\documentclass[preprint,3p,12pt,times]{elsarticle}

\usepackage{graphicx}
\usepackage{svg}
\usepackage{epsfig}
\usepackage{epstopdf}
\usepackage{amssymb}
\usepackage{amsthm}
\usepackage{bm}
\usepackage{placeins}
\usepackage{subcaption}

\usepackage[colorlinks=true,citecolor=blue,linkcolor=black]{hyperref}
\usepackage{color}
\usepackage{amsmath}
\usepackage{multirow}
\usepackage[justification=centering]{caption}
\usepackage{subcaption}
\usepackage{color}
\usepackage[algoruled,boxed,lined]{algorithm2e}

\usepackage{booktabs} 
\usepackage{lineno}
\usepackage{comment}

\journal{International Journal of Mechanical Sciences}

\begin{document}

\title{Multiscale modelling of thermally stressed superelastic polyimide}

\author[iitd]{Jerome Samuel S}
\author[iitkgp]{Puneet Kumar Patra}
\ead{puneet.patra@civil.iitkgp.ac.in}
\author[iitd]{Md Rushdie Ibne Islam\corref{coraut}}
\ead{rushdie@am.iitd.ac.in}

\cortext[coraut]{Corresponding Author}
\address[iitd]{Department of Applied Mechanics, Indian Institute of Technology Delhi, New Delhi 110016, India} 
\address[iitkgp]{Department of Civil Engineering, Indian Institute of Technology Kharagpur, West Bengal 721302, India}

\begin{abstract}
Many thermo-mechanical processes, such as thermal expansion and stress relaxation, originate at the atomistic scale. We develop a sequential multiscale approach to study thermally stressed superelastic polyimide to explore these effects. The continuum-scale smoothed particle hydrodynamics (SPH) model is coupled with atomistic molecular dynamics (MD) through constitutive modelling, where thermo-mechanical properties and equations of state are derived from MD simulations. The results are verified through benchmark problems of heat transfer. Finally, we analyse the insulating capabilities of superelastic polyimide by simulating the thermal response of an aluminium plate. The result shows a considerable reduction in the thermal stress, strain and temperature field development in the aluminium plate when superelastic polyimide is used as an insulator. The present work demonstrates the effectiveness of the multi-scale method in capturing thermo-mechanical interactions in superelastic polyimide.
\end{abstract}


\maketitle

\section{Introduction}
Thermal stresses, resulting from constrained thermal expansion or contraction in materials, are of significance in engineering systems subjected to uniform or non-uniform temperature loading. If not properly considered during the design phase, these stresses can lead to performance degradation, cracking, and even long-term failure. For example, thermal stresses can cause cracking, delamination, and performance degradation in semiconductors \cite{gong2024application}, depoling and a lowered Curie temperature in piezoelectric materials due to the pyroelectric effect \cite{patel2019ferroelectric}, thermal buckling and unwated deformations in structural elements \cite{pantousa2018thermal}, and instrument failure, structural deformation and microcracking in spacecrafts due to the exposure to extreme temperature variations and ionizing solar radiation \cite{wang2002radiation, dever2005degradation, abd2022radiation}. 

Thermal stress is commonly modelled by incorporating temperature-dependent material properties and thermally induced strains, which are treated as eigenstrains within the constitutive framework \cite{boley1997theory, nowacki2013thermoelasticity}. These thermal strains arise due to spatial or temporal temperature gradients and contribute to stress development even in the absence of external mechanical loads \cite{hetnarski2009thermal, lee2010postbuckling}. But, a proper understanding of thermoelastic phenomenon requires the use of multiscale analysis, as thermal stresses originate from the atomic-scale contraction or expansion of a material due to changes in the mean equilibrium positions of atoms or molecules as the temperature changes. There are two types of multiscale modelling: sequential and concurrent \cite{lu2004overview}. In sequential multiscaling, continuum scale models are developed using information from atomic scale models, with the two different length scales solved separately. Concurrent multiscaling, on the other hand, solves models of the two length scales in tandem, making the process more complex. Attempts to develop multiscale frameworks include coupling molecular dynamics (MD) with mesh-based finite element methods (FEM) \cite{lee2013multiscale, yamakov2008new}, meshfree methods \cite{gu2006concurrent, wang2009multiscale}, or combining coarse-grained continuum thermodynamics models with fine-scale non-equilibrium MD simulations \cite{li2010multiscale}.

In this work, we improve upon the novel sequential multiscale modelling framework proposed recently \cite{bhattacharyya2022multiscale} to incorporate thermal stress phenomena. The proposed approach links the atomic and continuum scales by coupling MD simulations with continuum-level smoothed particle hydrodynamics (SPH). This is achieved by extracting the key material properties -- stress-strain behaviour, bulk modulus, equation of state, thermal conductivity, and coefficient of thermal expansion -- directly from MD simulations, and using them subsequently for the SPH-based coupled thermo-mechanical continuum elasticity model. The conventional SPH is supplemented with corrections to mitigate the issues of instability near shock waves \cite{monaghan1983shock}, tensile instability \cite{monaghan2000sph}, particle interpenetration without sufficient dissipation, and inconsistent zeroth- and first-order accuracy at boundaries \cite{chen1999corrective}. The two methods have been chosen following three main considerations: (i) both MD and SPH rely on particle-based descriptions of matter, which naturally facilitates the transition between length scales without requiring complex interfacing, (ii) they have been shown to produce consistent predictions for both liquids and solids; isomorphic trajectories are obtained for fluid mechanics problems upon tuning the parameters properly \cite{hoover2006smooth}; for solids, consistent results at the two length scales can be obtained \cite{islam2022equivalence, ganesh2022multiscale}, and (iii) being mesh-free, SPH has inherent advantages over other techniques in problems involving large deformations \cite{ganesh2022pseudo}. 

The utility of the proposed modelling framework is demonstrated using aluminium-superelastic polyimide \cite{cheng2021super} film, a material relevant to the space industry. Unlike Kapton, which suffers from poor mechanical and thermal stability in harsh space conditions (for example, oxygen/nitrogen collisions, UV radiation, micrometeoroid impacts), superelastic polyimide offers better mechanical stability and comparable thermal properties, making it ideal as a space blanket and insulator for heat-sensitive equipment. While Kapton has been extensively investigated through reactive atomic modeling \cite{rahnamoun2014reactive}, steered MD simulations \cite{min2017computational}, including its degradation on collision with atomic oxygen \cite{rahnamoun2014reactive}, damage due to electron beam irradiation \cite{rahnamoun2019chemical}, water absorption \cite{marque2008molecular}, and pyrolysis \cite{lu2015reaxff}, superelastic polyimide is less explored despite its ability to endure high elastic strain and no loss of resilience even after sudden temperature jumps. Due to the unavailability of thermal conductivity and other key material properties at the macroscale for superelastic polyimide, a sequential multiscale modelling approach becomes essential. The sequential multiscale framework adopted here can systematically derive these effective properties from atomic-scale simulations. This approach ensures that the macroscopic behaviour accurately reflects the underlying microscale physics, ensuring a reliable and predictive modelling of the superelastic polyimide. After validating our SPH model with benchmark heat transfer problems, we analyse the insulating capabilities of superelastic polyimide by simulating the thermal response on aluminium-superelastic polyimide film within the multiscale framework. A considerable reduction in thermal stress, strain and temperature field is observed in the aluminium plate in the presence of superelastic polyimide film, showing that the sequential multiscale MD-SPH model can effectively capture thermal stresses in realistic scenarios.

This manuscript is organised as follows: Section 2 details the MD simulations adopted to estimate the different material properties of superelastic polyimide. The SPH methodology with thermo-mechanical coupling is described in Section 3, and subsequently, our validation results are discussed in Section 4. In Section 5, we perform numerical simulations to study the effects of insulation on aluminium plates. We conclude this manuscript with our conclusions and future directions.

\section{MD Simulation of Superelastic Polyimide}
Throughout this work, we simulate the superelastic polyimide in different MD ensembles, with NPT (constant temperature and pressure) environment being the most common, using the free-to-use software LAMMPS \cite{LAMMPS}. The isotropic formulation of the Nos\'e-Hoover barostat \cite{nose1984unified, hoover1985canonical} as modified by Martyna and coworkers \cite{martyna1994constant} and implemented in LAMMPS is used in the present study.  Consider a 3-dimensional system of $N$ particles. If $\mathbf{p}_i$ is the momentum vector of the $i^{th}$ particle of mass $m_i$ and $\Phi(\mathbf{r_1,r_2, \ldots, r_N})$ the coordinate-dependent potential energy of the system, then the NPT equations take the form: 
\begin{equation}
\begin{array}{rl}
        \dot{\mathbf{r}_i} & = \dfrac{\mathbf{p}_i}{m_i} + \dfrac{p_\epsilon}{W}\mathbf{r}_i, \\
        \dot{\mathbf{p}_i}  &= -\dfrac{\partial \Phi}{\partial \mathbf{r}_i} - \left[\left(1+\dfrac{1}{N}\right)\dfrac{p_\epsilon}{W} + \dfrac{p_\xi}{Q}\right]\mathbf{p}_i, \\
        \dot{V} &= \dfrac{3 V p_\epsilon}{W} \\
        \dot{p_\epsilon} & =  3V(P- P_0) +\dfrac{1}{N} \sum\limits_{i=1}^N \dfrac{\mathbf{p}_i \cdot \mathbf{p}_i}{m_i} - \dfrac{p_\xi}{Q}p_\epsilon \\
        \dot{p_\xi} & = \sum\limits_{i=1}^N \dfrac{\mathbf{p}_i \cdot \mathbf{p}_i}{m_i} + \dfrac{p_\epsilon^2}{W} - (3N+1)k_BT_0. \\
    \end{array}
    \label{eq:two}    
    \end{equation}
Here, $p_\epsilon$ and $p_\xi$ are the instantaneous barostat and thermostat variables, respectively, $W$ and $Q$ are the fictitious masses associated with the barostat and thermostat variables, respectively. Variables $P$, $T$ and $V$, respectively, denote the instantaneous pressure, temperature and volume of the system, which is subjected to an external isotropic pressure of $P_0$ and a temperature equalling $T_0$. The instantaneous isotropic pressure is calculated as:
\begin{equation}
    P = \dfrac{1}{3V} \left[ \sum\limits_{i=1}^N \dfrac{\mathbf{p}_i \cdot \mathbf{p}_i}{m_i} - \sum\limits_{i=1}^N \mathbf{r}_i \cdot \dfrac{\partial \Phi}{\partial \mathbf{r}_i} -3V \dfrac{\partial \Phi}{\partial V} \right]. 
\end{equation}
The constant energy ensemble (NVE) is obtained by considering $p_\epsilon$, $p_\zeta$ and their time derivatives to be equal to zero, while the constant temperature ensemble (NVT) is obtained if $p_\epsilon = \dot{p}_\epsilon=0$. All simulations assume periodicity in three directions and a timestep of 0.1 fs. We next detail the steps involved in modelling the initial geometry of the superelastic polyimide along with describing their interatomic interactions.

\subsection{Modelling Initial Geometry and Interatomic Interactions}
\begin{figure}[hbtp!] 
    \begin{subfigure}[b]{0.50\textwidth}
        \centering
        \includegraphics[width=\textwidth,trim={0 0 0 0}, clip]{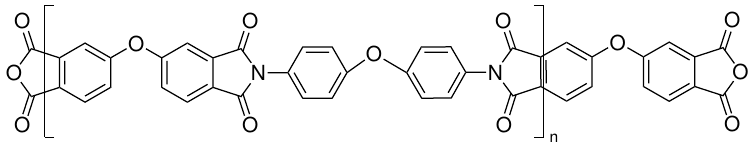}
        \caption{The molecular chain of pristine Kapton}
        \label{fig_2a}
    \end{subfigure}
    \begin{subfigure}[b]{0.50\textwidth}
        \centering
        \includegraphics[scale=0.7,trim={0 0 0 0}, clip]{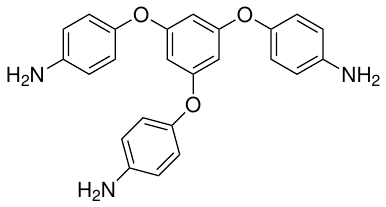}
        \caption{Molecular structure of cross-linker}
        \label{fig_3a}
    \end{subfigure}
    \hfill

    \vspace{5mm}
    \begin{subfigure}[b]{1.0\textwidth}
        \centering
        \includegraphics[width=\textwidth,trim={0 0 0 0}, clip]{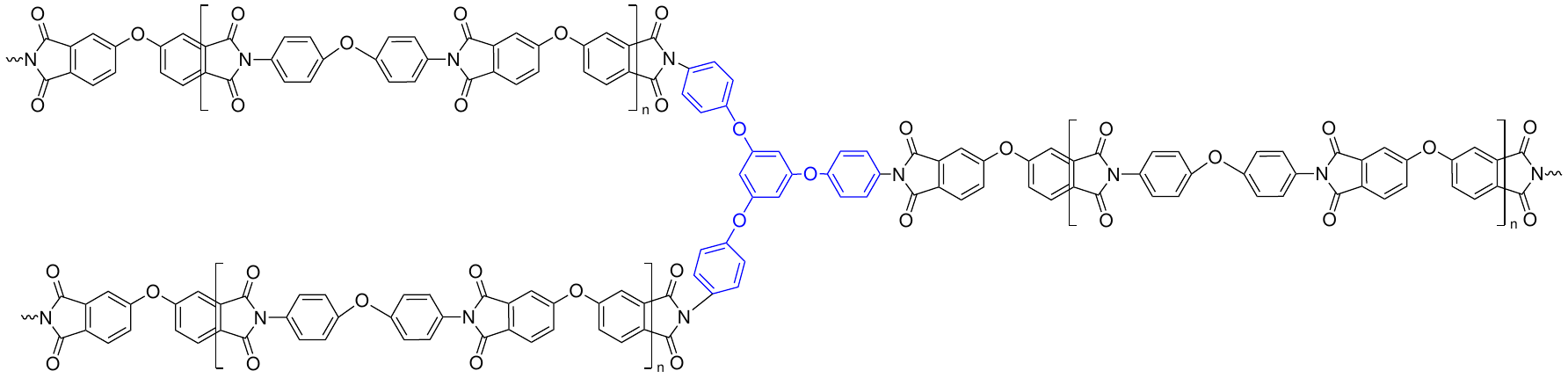}
        \caption{Molecular structure of superelastic polyimide}
        \label{fig_4a}
    \end{subfigure}
    
    \caption{(a) Ball and stick figure of the molecular structure of Pristine Kapton, (b) the central crosslinker necessary for the structure of superelastic polyimide and (c) molecular structure of superelastic polyimide.}
    \label{fig_1}
\end{figure}

The superelastic polyimide investigated in the present study is a cross-linked variant of Kapton \cite{cheng2021super}. The molecular chain of pristine Kapton, the cross-linker and the superelastic polyimide are shown in Figure \ref{fig_1}. The structural topology of the unit cell comprising superelastic polyimide, with three Kapton chains linked on each branch of the cross-linking agent, is first constructed. The unit cell, having 396 atoms, is then replicated 8 times each along the $x$, $y$ and $z$ directions so that a supercell of 202,752 atoms is created.

Kapton and its variants have been typically modelled using two interaction potentials - ReaxFF \cite{van2001reaxff}, and INTERFACE force field \cite{heinz2013thermodynamically}. The ReaxFF uses the bond-order mechanism with polarizable charge descriptions for reactive and non-reactive interactions between the different atoms \cite{senftle2016reaxff}. Unlike the ReaxFF, the INTERFACE force field cannot directly capture bond creation on the fly. So, in the present work, we have used the ReaxFF potential, in which the total potential energy of the system is given by:
\begin{equation}
    {\Phi} = {E_{bond}}+{E_{over}}+{E_{under}}+{E_{val}}+{E_{pen}}+{E_{tors}}+{E_{conj}}+{E_{vdW}}+{E_{Coulomb}}
    \label{eq:three}
\end{equation}
The bond energy, $E_{bond}$, of all the bonds is calculated dynamically through bond order. The bond orders higher (lower) than the equilibrium bond order are accounted by $E_{over}$ and $E_{under}$, respectively. ${E_{val}}$ represents the energy associated with the deviation of bond angles from their equilibrium value. The prevention of uncommon/energetically unfavourable configurations from occuring is through the penalty energy, $E_{pen}$. $E_{tors}$ represents the energy associated with the torsional angle, also referred to as the dihedral angle.  $E_{conj}$ represents the conjugation effect. The non-bonded interactions include the van der Waals interaction, $E_{vdWs}$, and Coulombic interactions, $E_{Coulomb}$, between two charged particles. Readers seeking further information are referred to the article \cite{van2001reaxff}.

\begin{figure}[hbtp!]
    \centering
    \begin{subfigure}{0.45\linewidth}
        \centering
        \includegraphics[width=\textwidth,trim={0 0 0 0}, clip]{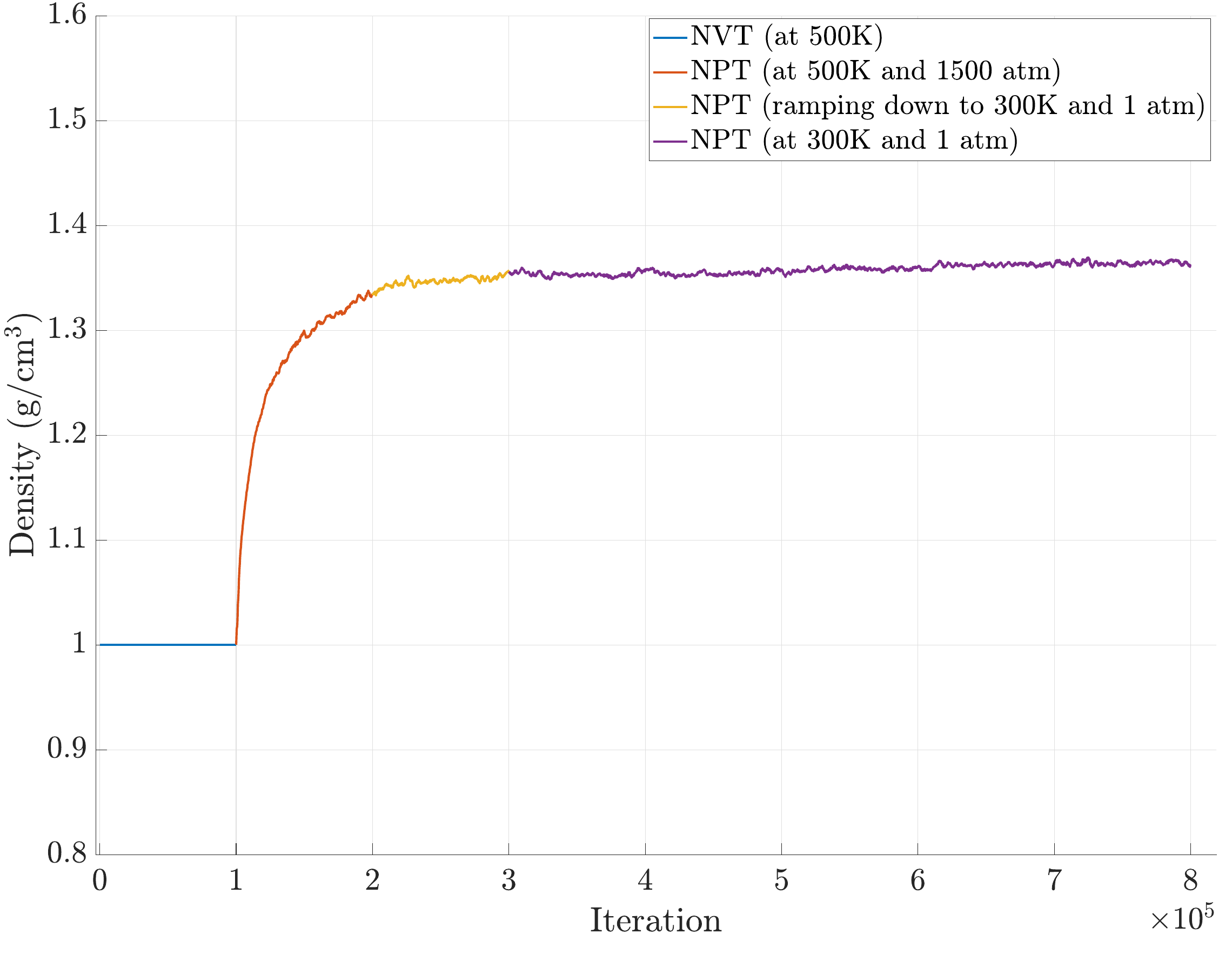}
        \caption{Initial equilibration runs to obtain the converged density of superelastic polyimide chain. }
        \label{fig_density_convergence}
    \end{subfigure}
    \begin{subfigure}{0.48\linewidth}
        \centering
        \includegraphics[width=\textwidth,trim={5 0 0 0}, clip]{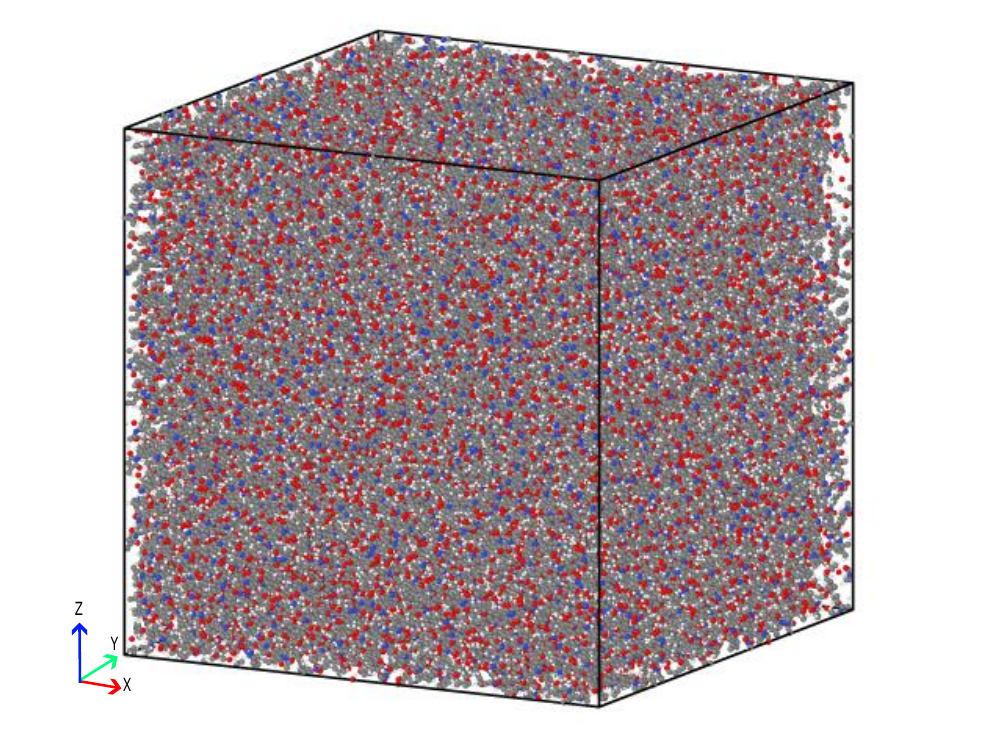}
        \caption{Resulting converged structure post-equilibration.}
        \label{fig_kapton_structure}
    \end{subfigure}

    \caption{Equilibration of superelastic polyimide chain. (a) Density convergence plot. (b) Converged structure visualised in OVITO.}
    \label{fig_equilibration_results}
\end{figure}

The supercell is first subjected to energy minimisation through conjugate gradient, and then undergoes equilibration in a series of steps. The timestep is set to 0.1 femtoseconds (fs), and the simulation begins with equilibration in the NVT ensemble at 500K for 100,000 iterations. This is followed by a series of NPT ensemble equilibration stages. First, the simulation box is equilibrated at 500K and 1500 atm for 100,000 iterations. Next, the temperature is gradually reduced to 300K and the pressure to 1 atm over another 100,000 iterations. Finally, the system is equilibrated at 300K and 1 atm for 500,000 iterations to ensure stability at room temperature and pressure. The equilibration process is validated by monitoring the density evolution, as shown in \autoref{fig_density_convergence}. The density converges to a value of $1.36 g/cm^3$. The converged density agrees well with that reported density for pristine Kapton in literature 1.40 g/cm$^3$ \cite{zhao2022dependence}.

\subsection{Calculating Elastic Properties of Superelastic Polyimide}
The simulation box obtained post equilibration is stretched uniaxially along the $x$ direction by subjecting it to an engineering strain rate of $0.000005$/fs. The coordinates of all atoms are remapped accordingly, ensuring homogeneous deformation according to the Cauchy-Born rule. This is performed in an NVE ensemble for 100,000 steps. We term these runs as production runs, and throughout them, the virial stress, defined in Eq. \eqref{eq:four}, is calculated and monitored.

\begin{equation}
\sigma_v^{\alpha \beta} = -\frac{1}{V} \sum_i \left( (v_i^\alpha - \langle v_i \rangle^\alpha)(v_i^\beta - \langle v_i \rangle^\beta) \right) + \sum_{j > i} F_c^{\alpha, ij} x_{ij}^\beta
\label{eq:four}
\end{equation}
In Eq. \eqref{eq:four}, the $\sigma_v^{\alpha \beta}$ represents the virial stress tensor's $\alpha$ and $\beta$ component. The average velocity vector in the neighbourhood of atom $i$ in the direction $\alpha$ is given by $\langle v_i \rangle^\alpha$, whereas $v_i^\beta$ is the instantaneous velocity of atom $i$ along $\beta$ direction. $F_c^{\alpha, ij}$ is the force exerted by atom $j$ on $i$ along $\alpha$ direction, while $x_{ij}^\beta$ represents the $\beta^{th}$ component of the position vectors of $i$ and $j$.

\begin{figure}[hbtp!]
    \centering
    \begin{subfigure}[b]{0.49\linewidth}
        \centering
        \includegraphics[width=\linewidth]{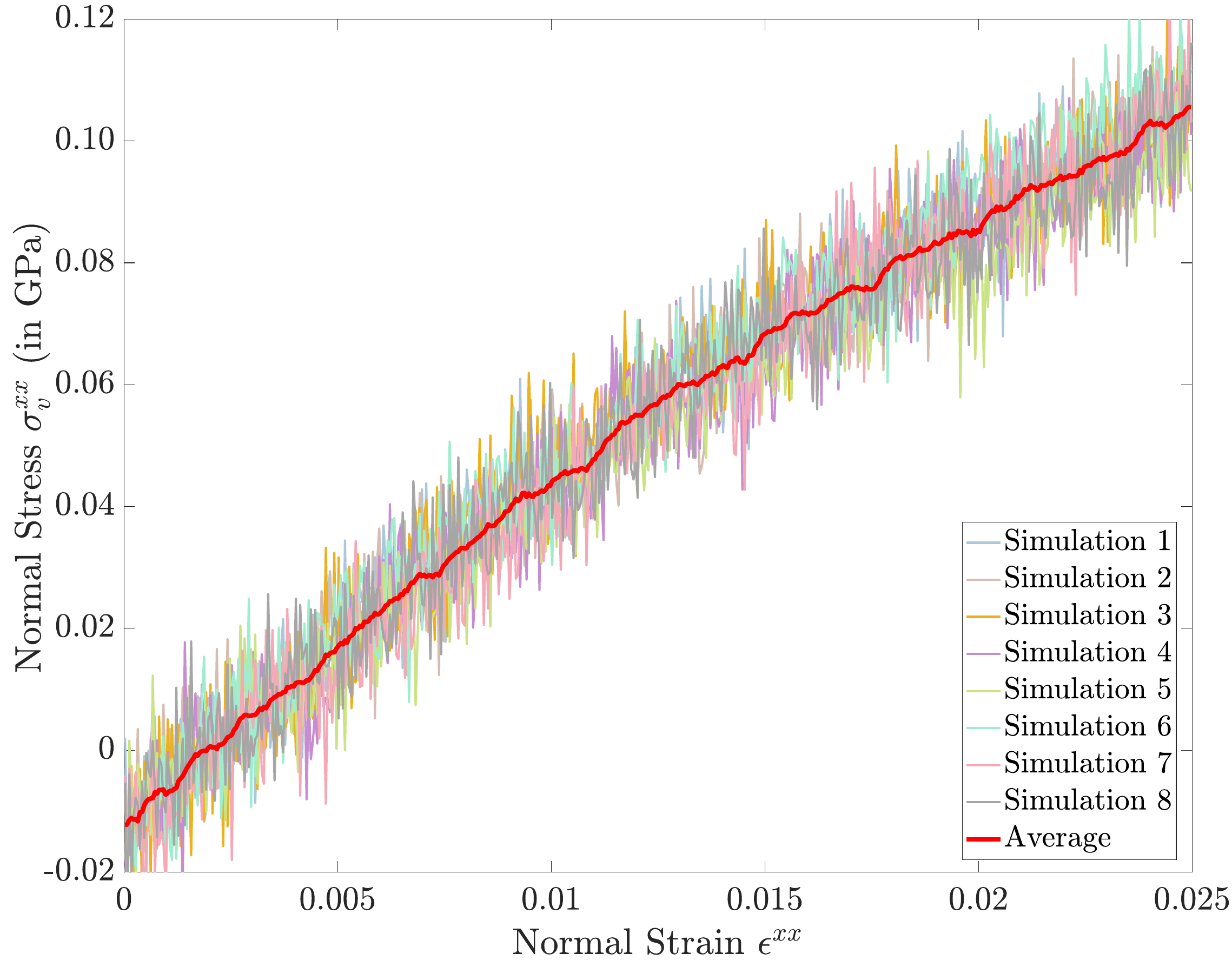}
        \label{fig_elasticmod}
        \caption{Normal Stress, $\sigma^{xx}_v$, vs. normal strain, $\epsilon^{xx}$, for each individual simulation and their average.}
    \end{subfigure}
    \hfill
    \begin{subfigure}[b]{0.49\linewidth}
        \centering
        \includegraphics[width=\linewidth]{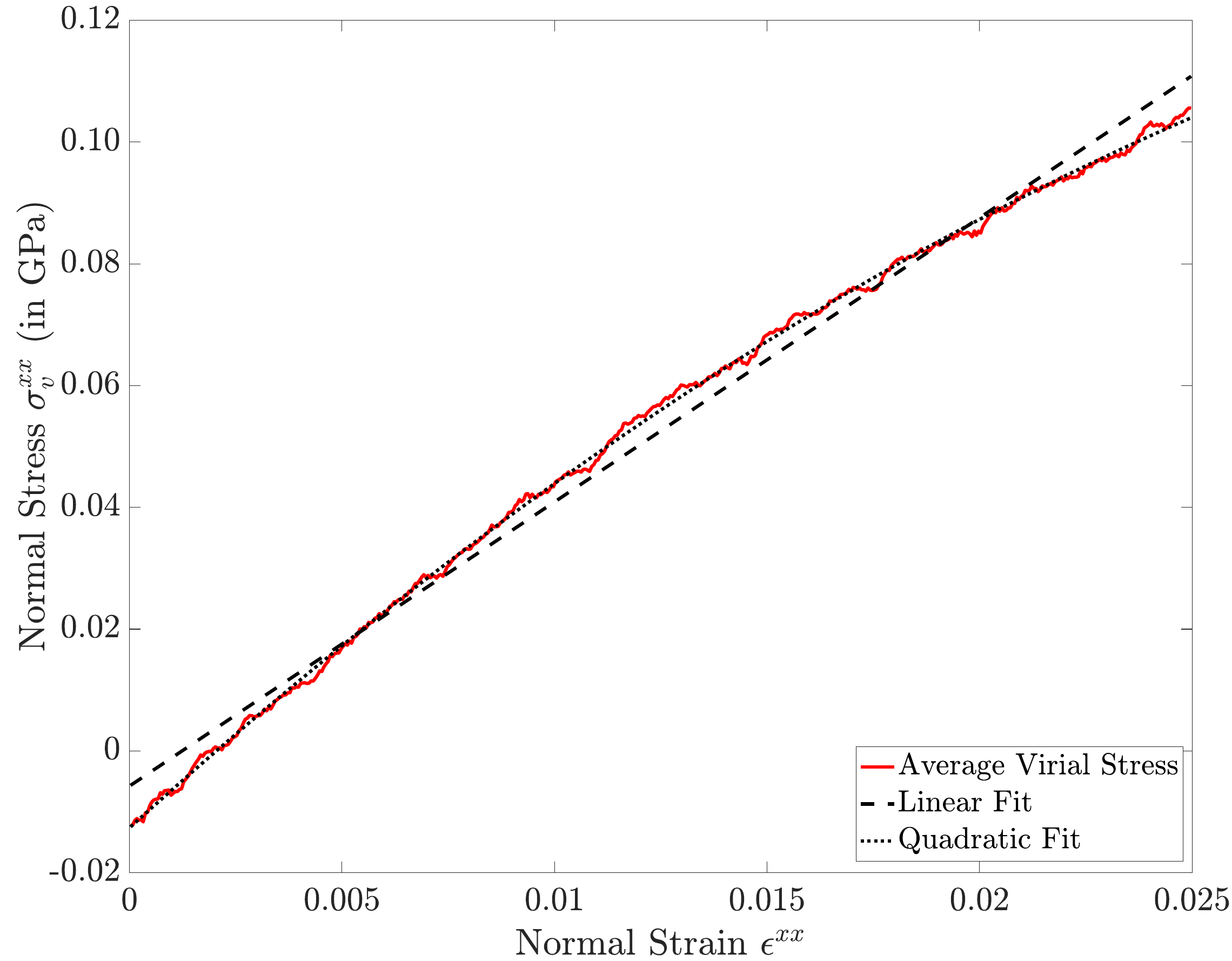}
        \label{fig_elasticmod2}
        \caption{Linear and quadratic fit of the average virial stress vs. strain curve. The quadratic fit shows a better match.}
    \end{subfigure}
    \caption{Instantaneous variation of normal stress, $\sigma^{xx}_v$, with normal strain, $\epsilon^{xx}$ across the 8 simulations along with its average, linear and quadratic fit. The variation of $\sigma^{xx}_v$ with $\epsilon^{xx}$ is best explained by a quadratic equation.}
    \label{fig_emodplot}
\end{figure}
The trajectory-dependent results are accounted for by considering eight independent production runs. These runs differ from each other in their initial velocity distribution, which is sampled from a Gaussian distribution with a different random seed at the start of the production runs. The stress-strain plot for each run can be seen in \autoref{fig_emodplot}(a) along with the average of the eight runs. It can be clearly observed from \autoref{fig_emodplot}(b) that the stress-strain curve has a nonlinear behaviour. The averaged stress, $\langle \sigma^{xx}_v \rangle$ varies quadratically with strain, $\epsilon^{xx}$, as given in Eq. \eqref{eq:five}.
\begin{equation}
    \langle \sigma^{xx}_v \rangle =-66.7\left(\epsilon^{xx}\right)^2+6.39\epsilon^{xx}-0.0128 \ \ \ \ (\text{in GPa})
    \label{eq:five}
\end{equation}
Thus, the elastic modulus, $E$, is strain-dependent, and is used for modelling the nonlinear behaviour in SPH. The strain-independent elastic modulus is obtained as $\epsilon^{xx} \rightarrow$ 0; the value $6.4$ GPa is obtained, which is also used in the SPH simulations later.

\begin{figure}[hbtp!]
    \centering
    \begin{subfigure}[b]{0.49\linewidth}
        \centering
        \includegraphics[width=\linewidth]{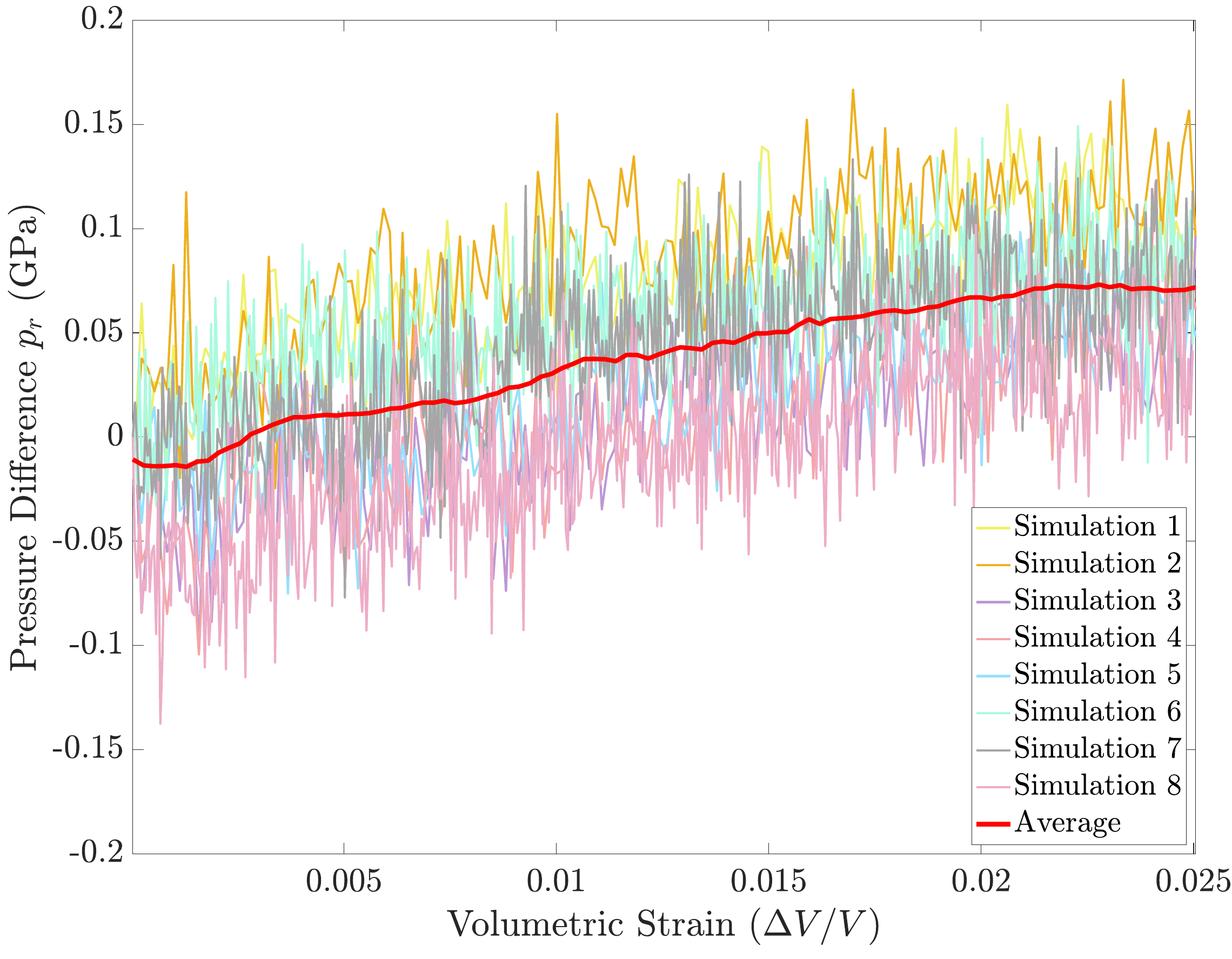}
        \label{fig_Bmod1}
    \end{subfigure}
    \hfill
    \begin{subfigure}[b]{0.49\linewidth}
        \centering
        \includegraphics[width=\linewidth]{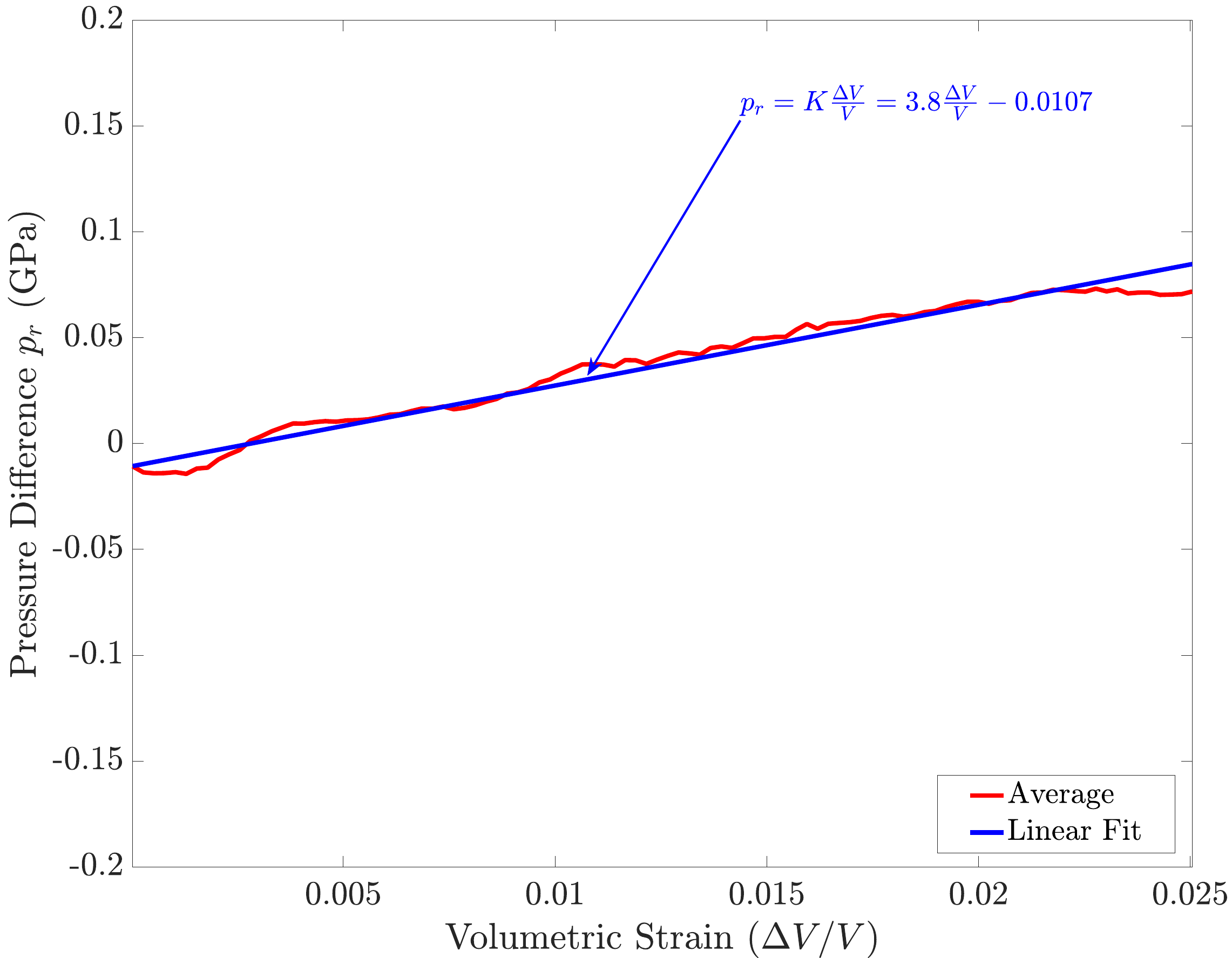}
        \label{fig_Bmod2}
    \end{subfigure}
    \caption{(Left) Hydrostatic pressure fluctuations during bulk modulus calculations, obtained from eight repeated simulations with identical setups. (Right) The averaged pressure is fitted with a linear curve, showing that, except for the endpoints, the hydrostatic pressure difference exhibits a linear trend with the volumetric strain ($\frac{\Delta V}{V}$). This validates the assumption of linear behaviour in the equation of state for SPH calculations.}
    \label{fig_Bmodplot}
\end{figure}

The bulk modulus is calculated following a similar approach. The simulation box is deformed along all three axes by an engineering strain rate of $0.000005$/fs up to an increase in the volume of $2.5\%$ as illustrated in \autoref{fig_Bmodplot}. The curve clearly exhibits a linear trend which is expressed in Eq. \eqref{eq:bmod}, where ${p_r}$ denotes the change in hydrostatic pressure, $\frac{\Delta V}{V}$ signifies the volume change ratio, and $K$ represents the bulk modulus, which is calculated to be $3.8$ GPa.

\begin{equation}
    p_r= K\frac{\Delta V}{V}= 3.8\frac{\Delta V}{V}-0.0107 \ \ (\text{in GPa})
    \label{eq:bmod}
\end{equation}
SPH incorporates a hydrostatic pressure term within its governing equations. Additionally, an equation of state that connects hydrostatic pressure with instantaneous density (${\rho}$) and initial density ({$\rho_o$}) is essential. Having now established a relationship between $p_r$ and $V$, Eq. \eqref{eq:bmod} can be altered to derive the linear equation of state represented by Eq. \eqref{eq:EOS}.
\begin{equation}
    {p_r}= K\left(\frac{\rho}{\rho_o}-1\right)
    \label{eq:EOS}
\end{equation}

\subsection{Calculating Thermal Properties of Superelastic Polyimide}
The thermal conductivity tensor $\kappa$ is the material property that measures the ability to transfer heat diffusively and is governed by Fourier's law: 
\begin{equation}
    Q= -{\kappa}{\nabla T},
    \label{eq:Fourier}
\end{equation}
where, $Q$ is the heat flux vector and $\nabla T$ is the spatial gradient of temperature. We use the M\"uller-Plathe method \cite{muller1997simple} within the framework of reverse non-equilibrium molecular dynamics (rNEMD) to calculate $\kappa$. In this approach, a heat flux is imposed within the system by exchanging the kinetic energy of two particles located in two different regions of the simulation domain, and the resulting temperature gradient is found.

The thermal conductivity is calculated by equilibrating the simulation box to a temperature of 300K and a pressure of 1 atm. The regions designated as ``hot'' and ``cold'', each of length 4$\AA$ in the z-direction, are established with a temperature of 350K in the hot region and 250K in the cold region. Upon achieving statistical convergence of the temperature difference, the cumulative heat energy added and removed is averaged. The coefficient of thermal conductivity is derived from the Eq. \eqref{eq:TC_calculation_equation}: 
\begin{equation}
    \kappa= \frac{\frac{Q_c}{(2A \Delta t) }}{\frac{\Delta T}{L_z}}
    \label{eq:TC_calculation_equation}
\end{equation}
In this equation, $Q_c$ represents the averaged cumulative heat energy, $A$ denotes the area perpendicular to the direction of heat flux, $\Delta t$ indicates the simulation time, $\Delta T$ signifies the temperature difference between the hot and cold regions, which is 100K in this instance, and $L_z$ refers to the distance separating the hot and cold regions. The cross-sectional area is multiplied by a factor of two to account for the influence of the periodic boundary condition. 
Following the convergence study (not shown here), $\kappa=$ 0.32 W/mK is used later in the SPH simulations.

\begin{figure}[hbtp!]
    \centering
    \includegraphics[width=0.5\linewidth]{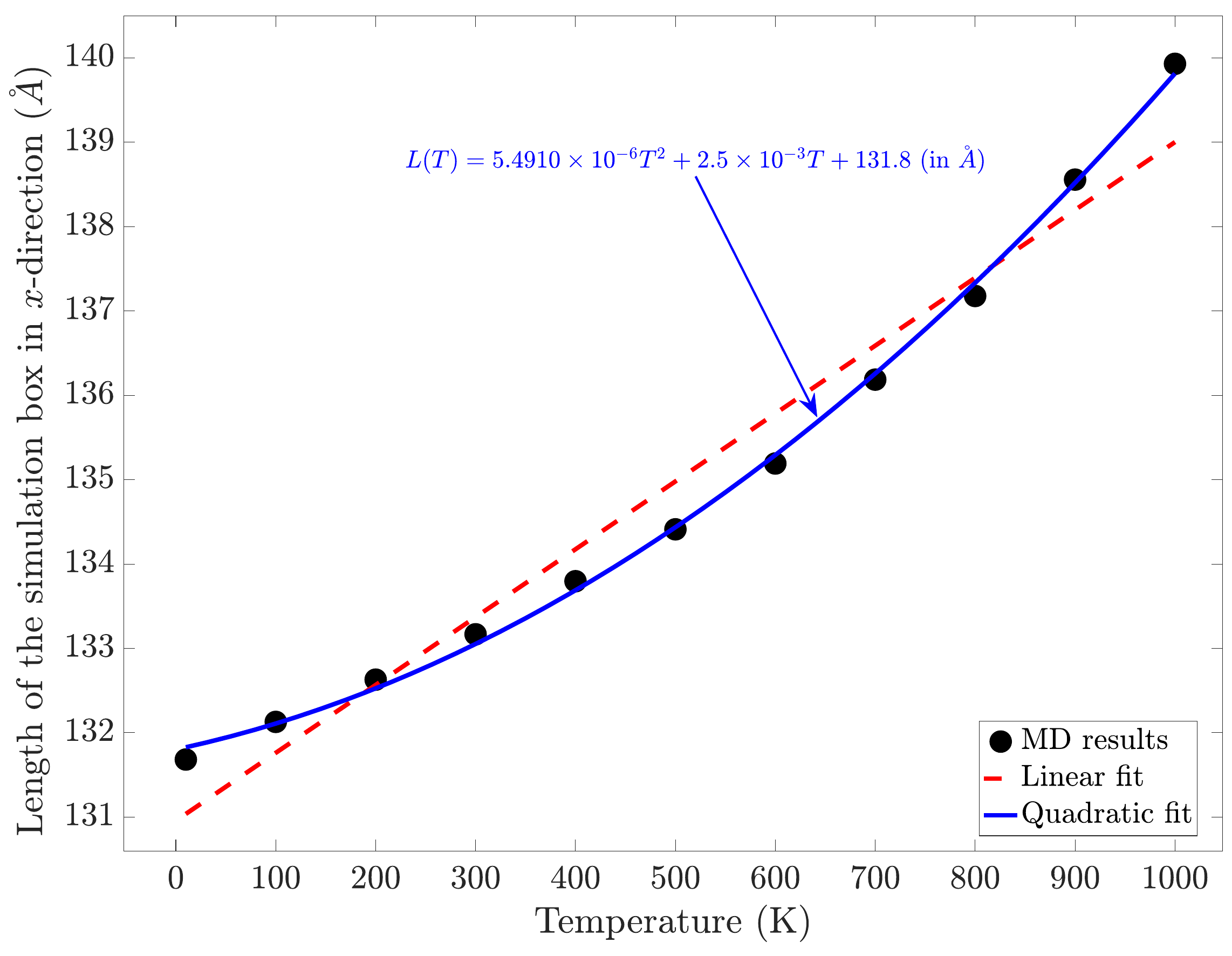}
    \caption{Plot illustrating the variation of the simulation box length (L) with temperature (T), along with linear and quadratic fits. The quadratic fit shows better alignment with the data and simplifies the determination of the coefficient of thermal expansion, an essential thermal field parameter in the SPH study. Without this empirical fit, an MD simulation would need to be conducted for every combination of $T_0$ and $T$, which is computationally expensive.}
    \label{fig_thermal_expansion_coefficient_plot}
\end{figure}

The thermal expansion coefficient ($\hat{\alpha}$) is determined using a relatively simple methodology \cite{pishkenari2016molecular}. The simulation box is set to an equilibrium state at a temperature of $T_0=300$K and a pressure of $P_0=1$atm using the NPT ensemble. In the production run, the simulation box is equilibrated to a final temperature denoted as $T$. The procedure is executed for a series of distinct $T$ values, ranging from 10K to 100K with an increment of 10K, utilising a time-step size of 0.1fs and a total of 100,000 iterations. The final temperature ($T$) and final length ($L$) are represented in the plot shown in \autoref{fig_thermal_expansion_coefficient_plot}. The quadratic curve, as represented by Eq. \eqref{length_dependence_on_temp_eqn}, demonstrates a strong fit. The Eq. \eqref{length_dependence_on_temp_eqn} is utilised to determine the length of the simulation box at a specified temperature $T$. This length is subsequently applied in Eq. \eqref{thermal_expansion_coeff_eqn} to calculate the thermal expansion coefficient for the temperatures $T$ and $T_0$.

\begin{equation}
    L(T) = 5.4910 \times 10^{-6} T^2 + 2.5 \times 10^{-3} T + 131.8 \quad \, \text{(\AA)}
    \label{length_dependence_on_temp_eqn}
\end{equation}

\begin{equation}
    \hat{\alpha}(T) = \frac{1}{\Delta T} \left( \frac{L(T) - L(T_0)}{L(T_0)} \right)
    \label{thermal_expansion_coeff_eqn}
\end{equation}

\begin{table}
    \centering
\caption{Calculated mechanical and thermal properties of superelastic polyimide}
\label{tab:my_label}
    \begin{tabular}{|c|c|c|} \hline 
         Property&  Value& Units\\ \hline 
         Elastic modulus&  6.39& $GPa$\\ \hline 
         Bulk modulus&  3.8& $GPa$\\ \hline 
         Poisson ratio&  0.218& \\ \hline 
         Density&  1.36& $g/cm^3$\\ \hline 
         Thermal conductivity&  0.32& $W/mK$\\ \hline
 Coefficient of thermal expansion& $4.723 \times 10^{-5}$&$/K$\\\hline
    \end{tabular}

\end{table}

\section{SPH and thermo-mechanical coupling methodology} 
In this section, we discuss the proposed methodology for coupled thermoelastic analysis \cite{hu2017thermomechanically} using SPH, which was initially developed for solving astronomical problems \cite{gingold1977smoothed}. It is now being increasingly adopted for studying a wide range of fluid \cite{monaghan1994simulating} and solid mechanics \cite{libersky1993high} problems. In SPH, the material points in the computational domain are represented by a collection of particles that carry intrinsic material properties such as density and field properties such as velocity, acceleration, temperature, etc. Two significant steps are followed in SPH: kernel approximation that mimics the properties of a Dirac-delta function and particle approximation \cite{liu2010smoothed}, which approximates the field variables and their special derivatives through the kernel approximation; more details can be found in \cite{liu2003smoothed}. 

Here, we solve the conservation equations of continuity, momentum and energy coupled with the heat equation. The following form of the continuity equation in the updated Lagrangian description is used:
\begin{equation}
    \frac{d \rho}{d t}=-\rho \frac{\partial v^\beta}{\partial x^\beta},
\end{equation}
where $\rho$, $v^\beta$, $x^\beta$ and $\frac{d}{dt}$ are the particle mass per unit volume, $\beta$ component of the particle velocity and position and time derivative in the Lagrangian description, respectively. Using the kernel and particle approximation \cite{islam2022equivalence}, we arrive at the following discretised form of the continuity equation: 

\begin{equation}
    \frac{d \rho_i}{d t}=\sum_j m_j v_{i j}^\beta W_{i j, \beta},
\end{equation}
where $i$ is the particle of interest, $j$ denotes the neighbouring particles within a finite radius (neighbourhood particles), $m_j$ denotes the particle mass of particle $j$, $v^{\beta}_{ij} = v^{\beta}_i - v^{\beta}_j$ and $W_{ij,\beta}$ is the derivative of the kernel function. In this work, we have used Lucy's kernel function \cite{liu2003smoothed}. 

The following linear momentum equation is solved:
\begin{equation}
    \frac{d v^\alpha}{d t}=\frac{1}{\rho} \frac{\partial \sigma^{\alpha \beta}}{\partial x^\beta} + g^\alpha,
\end{equation}
where, $g^\alpha$ is the body force components and $\sigma^{\alpha \beta}$ is the Cauchy stress components calculated as \cite{hu2017thermomechanically}: 
\begin{equation}
    \sigma^{\alpha \beta} = S^{\alpha \beta} - \left[p_r  + 3 K \hat{\alpha} (T - T_0)\right] \delta^{\alpha \beta}.
\end{equation}
Here, $S^{\alpha \beta}$ is the deviatoric component of Cauchy stress, $p_r$ is the hydrostatic pressure calculated from the linear equation of state $p = K \left(\frac{\rho}{\rho_0} - 1\right)$, $\rho_0$ is the reference particle density, $K$ is the bulk modulus, $\hat{\alpha}$ is the coefficient of thermal expansion, $T$ is the temperature, $T_0$ is the reference temperature and $\delta^{\alpha \beta}$ is the Direc-delta function. The rate of the deviatoric component is updated using the following Jaumann stress rate: 
\begin{equation}
    \dot{S}^{\alpha \beta} = 2 \mu \left(\dot{\epsilon}^{\alpha \beta} - \frac{1}{3} \delta^{\alpha \beta} \dot{e}^{\gamma \gamma}\right) + S^{\alpha \gamma} \omega^{\beta \gamma} + S^{\gamma \beta} \omega^{\alpha \gamma},
\end{equation}
with $\mu$ being the modulus of shear, the strain rate $\dot{\epsilon}^{\alpha \beta}$ and rotation rate $\omega^{\alpha \beta}$ tensors as following:

\begin{align}
    \dot{\epsilon}^{\alpha \beta} &= \frac{1}{2} \left(l^{\alpha \beta} + l^{\beta \alpha}\right), \\
    \omega^{\alpha \beta} &= \frac{1}{2} \left(l^{\alpha \beta} - l^{\beta \alpha}\right),
\end{align}
where the velocity gradient, \(l^{\alpha \beta}\), is:
\begin{equation}
    l^{\alpha \beta} = -\sum_j \frac{m_j}{\rho_j} v_{ij}^{\alpha} W_{ij,\beta}.
\end{equation}
The particle approximation form of the momentum equation is given below:
\begin{equation}
    \frac{d v_i^\alpha}{d t}=\sum_j m_j\left(\frac{\sigma_i^{\alpha \beta}}{\rho_i^2}+\frac{\sigma_j^{\alpha \beta}}{\rho_j^2}-\pi_{i j} \delta^{\alpha \beta}-P_{i j}^a \delta^{\alpha \beta}\right) W_{i j, \beta} + g^{\alpha}_i.
\end{equation}
where $\pi_{ij}$ is the artificial viscosity to stabilise the simulation in the presence of any shock or jump function \cite{monaghan1983shock} and $P^a_{ij}$ is the artificial pressure correction to suppress tensile instability \cite{monaghan2000sph}. The following form of $P^a_{ij}$ is used here:

\begin{equation}
    P_{i j}^a=\gamma\left(\frac{\left|p_{r_i}\right|}{\rho_i^2}+\frac{\left|p_{r_j}\right|}{\rho_j^2}\right)\left[\frac{W\left(d_{i j}\right)}{W(\Delta p)}\right]^{\bar{n}} \text {, }
\end{equation}
where $\gamma$ is a parameter to tune the artificial pressure, $d_{ij}$ and $\Delta p$ are the current and initial particle spacing and $\bar{n}=W(0)/W(\Delta p)$. The artificial viscosity term $\pi_{ij}$ is expressed as:
\begin{equation}
\pi_{ij} =
\begin{cases} 
\frac{-\beta_1 \bar{C}_{ij} \mu_{ij} + \beta_2 \mu_{ij}^2}{\bar{\rho}_{ij}}, & \text{if } \sum_\alpha v_{ij}^\alpha x_{ij}^\alpha \leq 0, \\
0, & \text{otherwise}.
\end{cases}
\end{equation}
\begin{equation}
\quad \text{where} \quad
\mu_{ij} = \frac{\sum_\alpha h_{ij} x_{ij}^\alpha v_{ij}^\alpha}{\sum_\alpha \left(x_{ij}^\alpha\right)^2 + 0.01 h_{ij}^2}.
\end{equation}
The variables $\beta_1$ and $\beta_2$ are parameters that control the damping intensity of the artificial viscosity. While large values of $\beta_1$ and $\beta_2$ ensure unconditional damping, they render the simulation unrealistic. The average speed of sound and density between particles $i$ and $j$ are defined as $
\bar{C}_{ij} = 0.5 \left(C_i + C_j\right), \quad \text{and} \quad \bar{\rho}_{ij} = 0.5 \left(\rho_i + \rho_j\right).$

The internal energy equation takes the following form when coupled with the heat equation \cite{eringen1967mechanics}:

\begin{equation}\label{en}
    \frac{de}{dt}  =  \frac{\sigma^{\alpha\beta}}{\rho}\frac{\partial v^\alpha}{\partial x^\beta} - \frac{1}{\rho} \frac{\partial q^\alpha}{\partial x^\alpha} + \frac{1}{\rho} q_s \delta(\bm r - \bm R_s),
\end{equation}
where $e$ is the internal energy, $q^\alpha$ is the heat flux, $q_s$ is the source strength in the dimension power per unit volume, and $R_s$ is the position vector of the heat source. The internal energy can be expressed in the following form \cite{eringen1967mechanics}:

\begin{equation}
    e = c_v (T-T_0) + \frac{3K}{\rho_0} \hat{\alpha} T_0 \epsilon_{kk} + \frac{1}{2 \rho_0} \left(  \lambda + 2 \mu \right) \epsilon^2_{kk} + \frac{\mu}{\rho_0} \left[ (\epsilon_{kk})^2 + \epsilon^2_{kk} \right] + constant.
\end{equation}
where $c_v$ is the heat capacity at constant volume, $\lambda$ and $\mu$ are the Lame's parameters. Fourier's law of heat conduction for a material with constant thermal conductivity $k$ is given below:

\begin{equation}\label{con}
    q_\alpha = -k \frac{\partial T}{\partial x_\alpha}
\end{equation}
Combining Eqs. \ref{en} and \ref{con}, we get the coupled heat equation in linear thermoelasticity \cite{hu2017thermomechanically}:

\begin{equation}\label{temp}
    \rho c_v \frac{dT}{dt} = -3 K \hat{\alpha} T_0 \dot{\epsilon}_{kk} + k \Delta T + q_s \delta(\bm r - \bm R_s)
\end{equation}
The discrete form of Eq. \ref{temp} for the heat conduction and source term is \cite{cleary1999conduction, monaghan2005solidification}: 

\begin{equation}
    c_v \frac{d T_i}{dt} = \sum\limits_j \frac{m_j}{\rho_i \rho_j} \left( 3 K \hat{\alpha} T_0 v^\beta_{ij} + 2 k (T_i - T_j) \frac{x^\beta_{ij}}{||x_{ij}||^2}  \right) W_{ij,\beta} + \frac{1}{\rho_i} \sum\limits_j Q_s \zeta_s W (\bm r_j - \bm R_s),
\end{equation}
where $Q_s$ is the power of the heat source, and $\zeta_s$ is the normalising factor for the heat source \cite{monaghan2005solidification} to correctly account for the rate of change of thermal energy with the following form:

\begin{equation}
    \zeta_s = \frac{1}{\sum\limits_j \frac{m_j}{\rho_j} W (\bm r_j - \bm R_s)}.
\end{equation}

To ensure numerical stability, the timestep size must be small enough to allow stable time integration while being large enough to reduce computational cost. The minimum timestep size for time-integration in the thermal field, as described in \cite{monaghan2005smoothed}, is provided by \eqref{Thermal_time_integration}, while for the mechanical field, it is determined by the Courant-Friedrichs-Lewy (CFL) criterion \cite{shaw2009heuristic}, given in \eqref{Mechanical_time_integration}.

\begin{equation}
\Delta t \leq \hat{\beta} \rho c_v \frac{h^2}{\kappa} = 1.44 \hat{\beta} \rho c_v \frac{\Delta x^2}{\kappa},
\label{Thermal_time_integration}
\end{equation}

\begin{equation}
\Delta t \leq \left\{ \frac{c_s h}{c_i + |v_i|} \right\},
\label{Mechanical_time_integration}
\end{equation}

The parameter $\hat{\beta}$ in \eqref{Thermal_time_integration} is typically set to 1.5, and the {$c_s$} in \eqref{Mechanical_time_integration} is chosen as 0.3. The wave speed in the material, {$c_i$}, is defined as {$c_i$} = $\sqrt{\frac{E}{\rho_i}}$. Here, h represents the kernel's support radius.

\section{Numerical examples}
This section uses benchmark examples to present the thermo-mechanical coupled-SPH (TMC-SPH) code validation. Initially, the SPH code is validated through a one-dimensional and two-dimensional heat transfer scenario, excluding considerations of the mechanical field. Thermomechanical coupling is subsequently validated by examining thermal-induced deformation in a two-dimensional heat transfer context, incorporating both thermal and mechanical fields.

\subsection{Transient heat transfer in 1D}
\begin{figure}[hbtp!]
    \centering
    \includegraphics[width=0.5\linewidth]{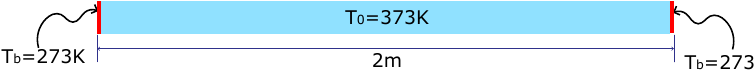}
    \caption{Initial and boundary conditions for 1D transient heat transfer example}
    \label{fig_1D_Initial_and_boundary_conditions}
\end{figure}

For the 1D transient heat transfer case \cite{}, a finite rod of length $L=2$ m, initially at temperature $T_0=373 \mathrm{K}$ is considered. The temperature of the boundary particles is kept at $T_B=273K$ at the beginning of the simulation as illustrated in \autoref{fig_1D_Initial_and_boundary_conditions}. The values of coefficient of thermal conductivity ($\kappa$), and specific heat capacity($c$) are $106~ \mathrm{W/mK}$ and $100~ \mathrm{J/kgK}$ respectively. The rod is discretised into 201 particles with an initial inter-particle spacing ($\Delta p$) of $0.01$ m and a smoothing length of $0.018$ m. The timestep size for integration of the thermal field is taken as $10^{-4}$ s. The analytical solution for this case can be expressed as \cite{zhou2024thermal}:
\begin{equation}
    T(t) = \frac{4(T_0-T_B)}{\pi} \sum_{n=0}^\infty \frac{1}{(2n+1)} 
    e^{\frac{-\kappa(2n+1)^2\pi^2t}{\rho c L^2}} 
    \sin\left(\frac{(2n+1)\pi x}{L}\right)
    \label{Analytical_soln_1D_heat_transfer}
\end{equation}

\begin{figure}[hbtp!]
    \centering
    \includegraphics[width=0.75\linewidth]{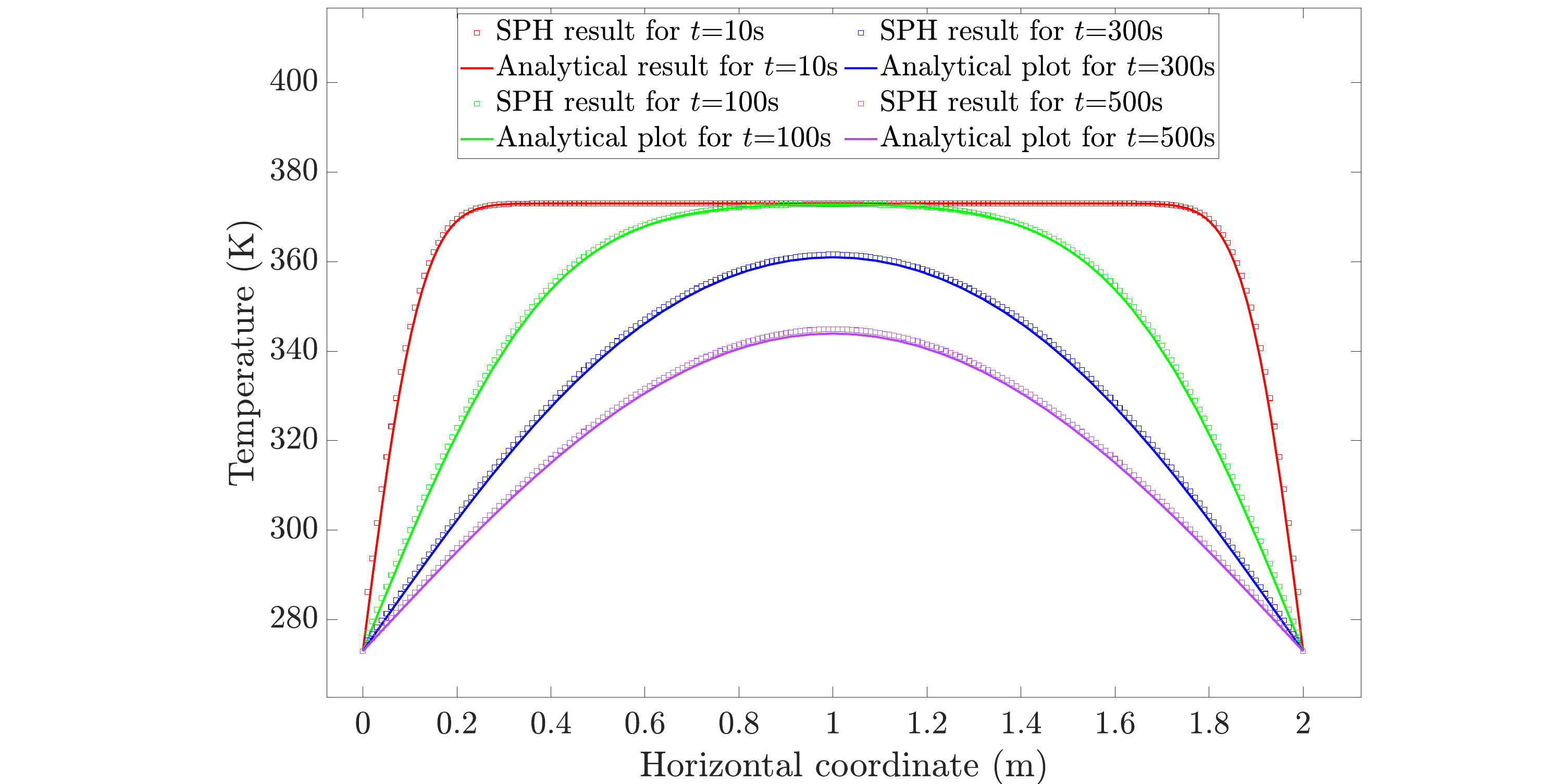}
    \caption{Temperature plot comparing SPH and analytical results at time 10s, 100s, 300s and 500s for 1D transient heat transfer problem }
    \label{fig_1D_transient_temp_vs_x_plot}
\end{figure}
The temperature variations obtained from the SPH simulation along the length of the bar over different times are presented in \autoref{fig_1D_transient_temp_vs_x_plot}, along with the analytical results for comparison. The simulated results align well with the analytical solution, validating the developed code for the one-dimensional heat transfer case.

\subsection{Transient heat transfer in 2D}
\begin{figure}[hbtp!]
    \centering
    \includegraphics[width=0.5\linewidth]{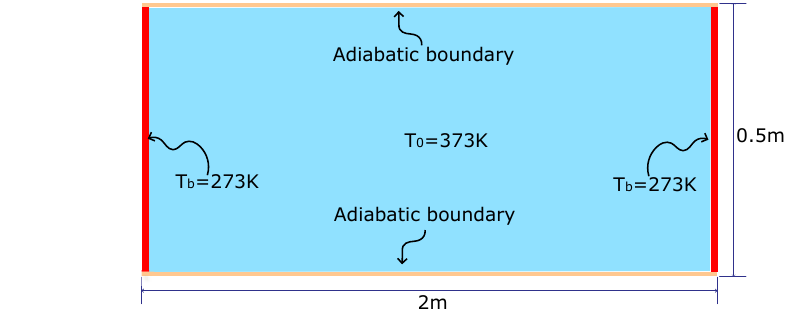}
    \caption{Initial and boundary conditions for the 2D transient heat transfer example}
    \label{fig_2D_initial_and_boundary_conditions}
\end{figure}

Next, we look into a two-dimensional transient heat transfer example. Here, the domain of interest is a rectangular plate $2$m wide with a height of $0.5$m (\autoref{fig_2D_initial_and_boundary_conditions}). The plate is initially at a temperature $T_0=375$K and the upper and lower edges are subject to adiabatic conditions (these boundaries do not exchange heat with their surroundings. The left and right edges of the plate are kept at $273$K temperature ($T_b$) as shown in \autoref{fig_2D_initial_and_boundary_conditions}. We have considered the following material parameters for the plate \cite{zhou2024thermal}: material density $\rho=2000~\mathrm{kg/m^3}$, Young's modulus $E=50$GPa and poissons ratio $\nu=0.3$. The specific heat capacity (c), coefficient of thermal expansion ($\hat{\alpha}$) and thermal conductivity ($\kappa$) are taken as $106~\mathrm{J/kgK}$, $10^{-5}/K$ and $100W/mK$ respectively. The plate is discretised into a regular Cartesian grid with 10,251 particles, with 201 particles in the x direction and 51 particles in the y direction. The initial inter-particle spacing ($\Delta p$) is 0.01m, and the support radius is of length 0.018m. The timestep size for integration is set to $10^{-3}$ s. This scenario has been analyzed in the works of \cite{mu2022coupled}, \cite{yan2017coupled}, 
\cite{zhou2024thermal}, among others. The initial and boundary conditions are expressed mathematically as follows:

\begin{align}
    T\big|_{t=0} &= T_0, \quad 0 \leq x \leq L, \\
    T\big|_{x=0} &= T_B, \quad T\big|_{x=L} = T_B, \quad t \geq 0
\end{align}

We validate the present SPH results by comparing the temperature and thermal stress data with analytical solutions for the temperature ($T$) and thermal stress ($\sigma^{TH}$):

\begin{equation}
    T(t) = \frac{4(T_0-T_B)}{\pi} \sum_{n=0}^\infty \frac{1}{(2n+1)} 
    e^{\frac{-\kappa(2n+1)^2\pi^2t}{\rho c L^2}} 
    \sin\left(\frac{(2n+1)\pi x}{L}\right)
    \label{Analytical_soln_2D_heat_transfer}
\end{equation}

\begin{equation}
    \sigma^{TH}(t)=-3K\hat{\alpha}[T-(T_0-T_B)]
    \label{Analytical_soln_for_thermal_stress_2D_heat_transfer}
\end{equation}

\begin{figure}[hbtp!]
    \centering
    \begin{subfigure}[b]{0.49\linewidth}
        \centering
        \includegraphics[width=\linewidth]{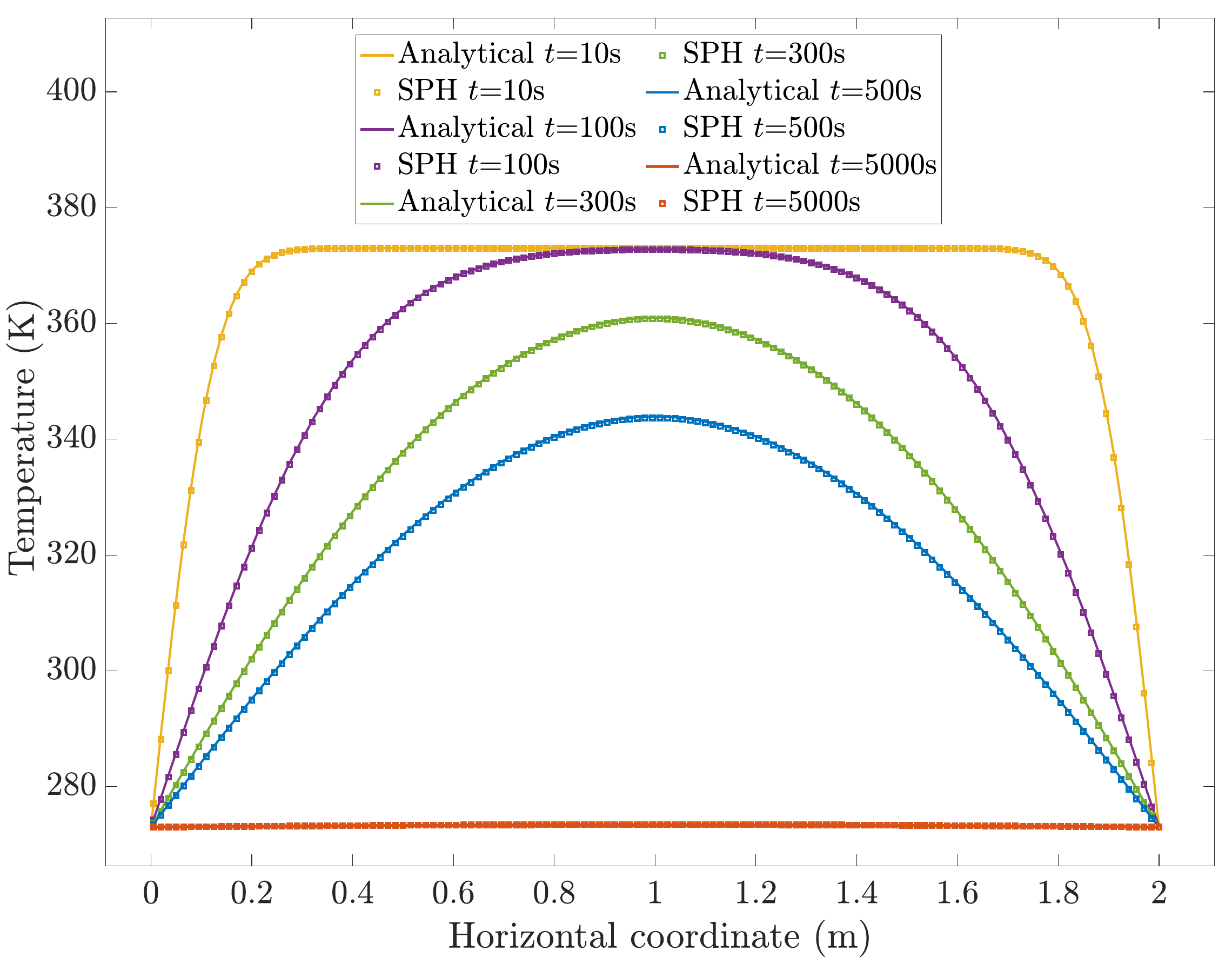}
        \label{fig_tempvspos2D}
    \end{subfigure}
    \hfill
    \begin{subfigure}[b]{0.49\linewidth}
        \centering
        \includegraphics[width=\linewidth]{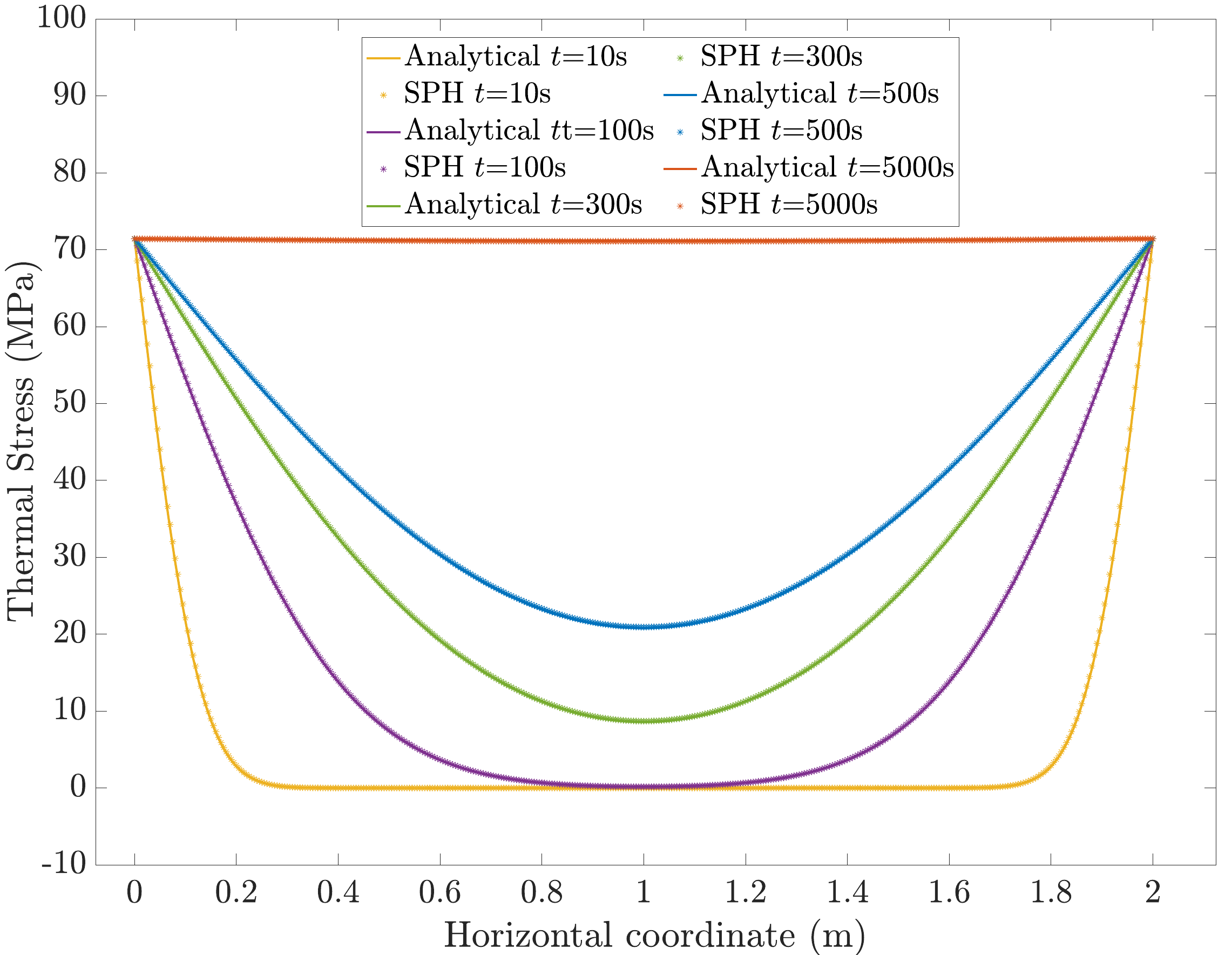}
        \label{fig_tstress_vs_position_for_2D}
    \end{subfigure}
    \caption{Temperature and thermal stress plots comparing SPH and analytical results at times 10s, 100s, 300s and 500s.}
    \label{fig_temp_and_stress_2D}
\end{figure}

The temperature and thermal stress distributions obtained from our SPH simulations are illustrated in \autoref{fig_temp_and_stress_2D}. Initially, the temperature gradient is higher between successive layers, facilitating quicker heat flux flow. This results in high thermal stress being induced quickly in the boundaries, as represented by the temperature and thermal stress distributions at 10 seconds and 100s. As the simulation proceeds, the heat transfer slows because of the exponential nature of temperature dependence with time, as expressed by \eqref{Analytical_soln_2D_heat_transfer}. It is also important to note that the necessary condition of symmetry of temperature and thermal stress field about the central vertical axis, owing to the symmetry of initial and boundary conditions, is satisfied. 

\subsection{Thermal-induced deformation in 2D}
In this numerical example, we compare the mechanical deformation caused by thermal loading to validate the thermomechanical coupling strategy in SPH. A key challenge in thermo-mechanical coupling using particle-based methods is the significant disparity in the time steps between the thermal and mechanical fields during numerical integration. It arises because the timestep for the thermal field is governed by thermal diffusivity (see Eq. \eqref{Thermal_time_integration}), while the mechanical field is influenced by wave speed (see Eq. \eqref{Mechanical_time_integration}). Since the timestep is typically selected as the minimum of the thermal and mechanical field values, the time derivative of temperature becomes very small. As a result, this small value gets truncated, leading to a loss of precision as the error accumulates. \cite{d2017thermal}. This large disparity requires a multi-rate time integration approach to ensure that $\frac{dT}{dt}$ does not experience a loss in precision, as discussed in \cite{d2017thermal}. 

We, therefore, consider a square plate with mechanical and thermal parameters calibrated as proposed by Sun et al. \cite{sun2021pd} to avoid the multi-rate time integration approach. The dimensions of the plate are: $1~\mathrm{m}~\times~1~\mathrm{m}$ and has an an initial temperature of  $274$K. The left and right boundaries are kept at $273$K, while the top and bottom edges are subject to adiabatic conditions. We use the following material properties for our simulation: $E = 1$Pa, $\mu$ = 0.3, $\rho$ = $1~\mathrm{kg/m^3}$, $\kappa$ = $1~ \mathrm{W/mK}$, $\hat{\alpha}$ = $0.02 \mathrm{/K}$, and $c_v$ = $1~\mathrm{J/kgK}$. With these parameters, a time-step size of $10^{-5}$s is chosen. The simulation domain is discretised into 29,241 particles ($\Delta p =5.88~\mathrm{mm}$; $h =10.58~\mathrm{mm}$), with 171 particles uniformly distributed along both the horizontal and vertical directions. Both the parameters of artificial viscosity, i.e., $\beta_1$ and $\beta_2$, are set as 0.5.

\begin{figure}[hbtp!]
    \centering
    \begin{subfigure}[b]{0.49\linewidth}
        \centering
        \includegraphics[ width=\linewidth]{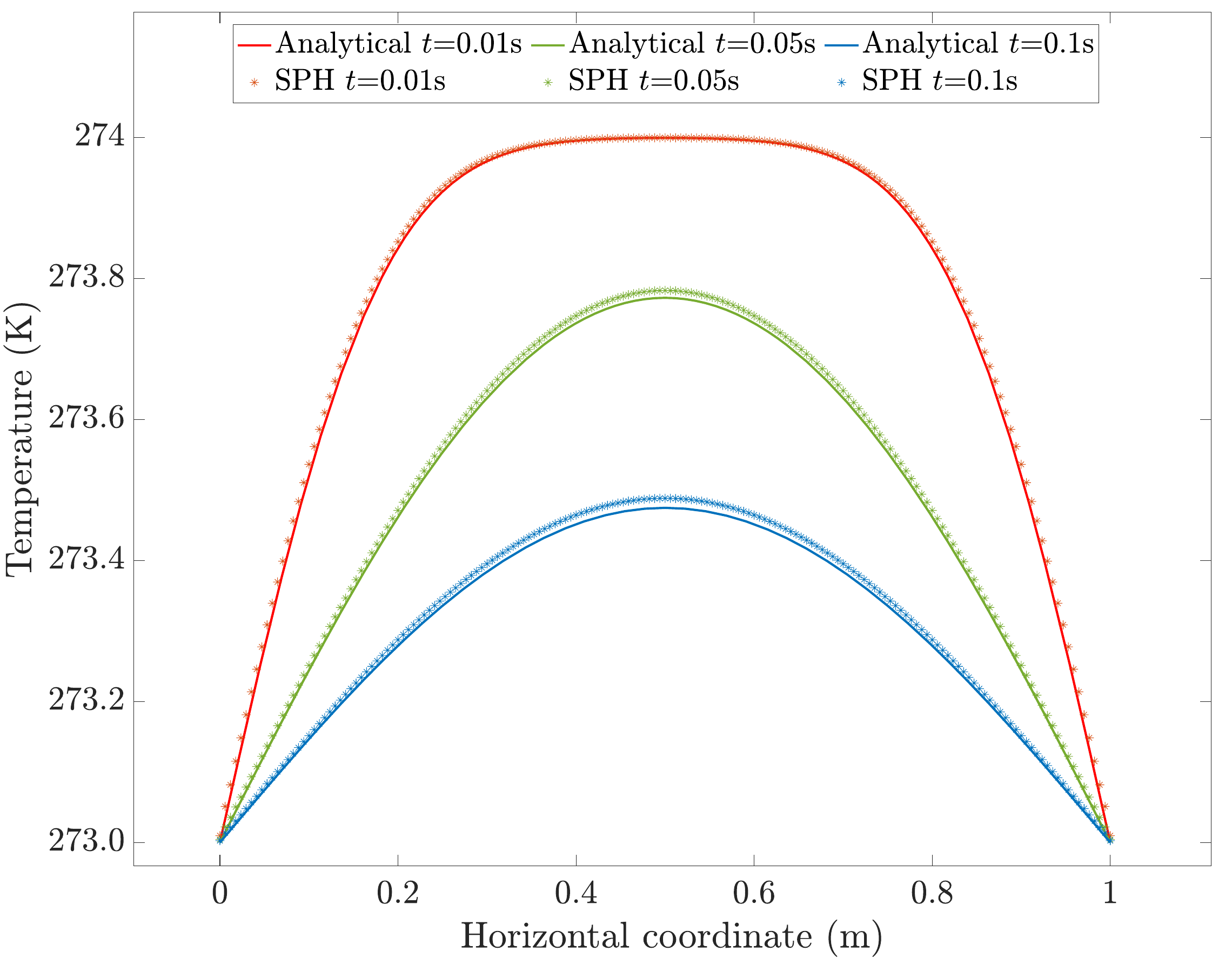}
        \label{fig_tempvspos2D_thermalinduceddeformation}
    \end{subfigure}
    \hfill
    \begin{subfigure}[b]{0.49\linewidth}
        \centering
        \includegraphics[width=\linewidth]{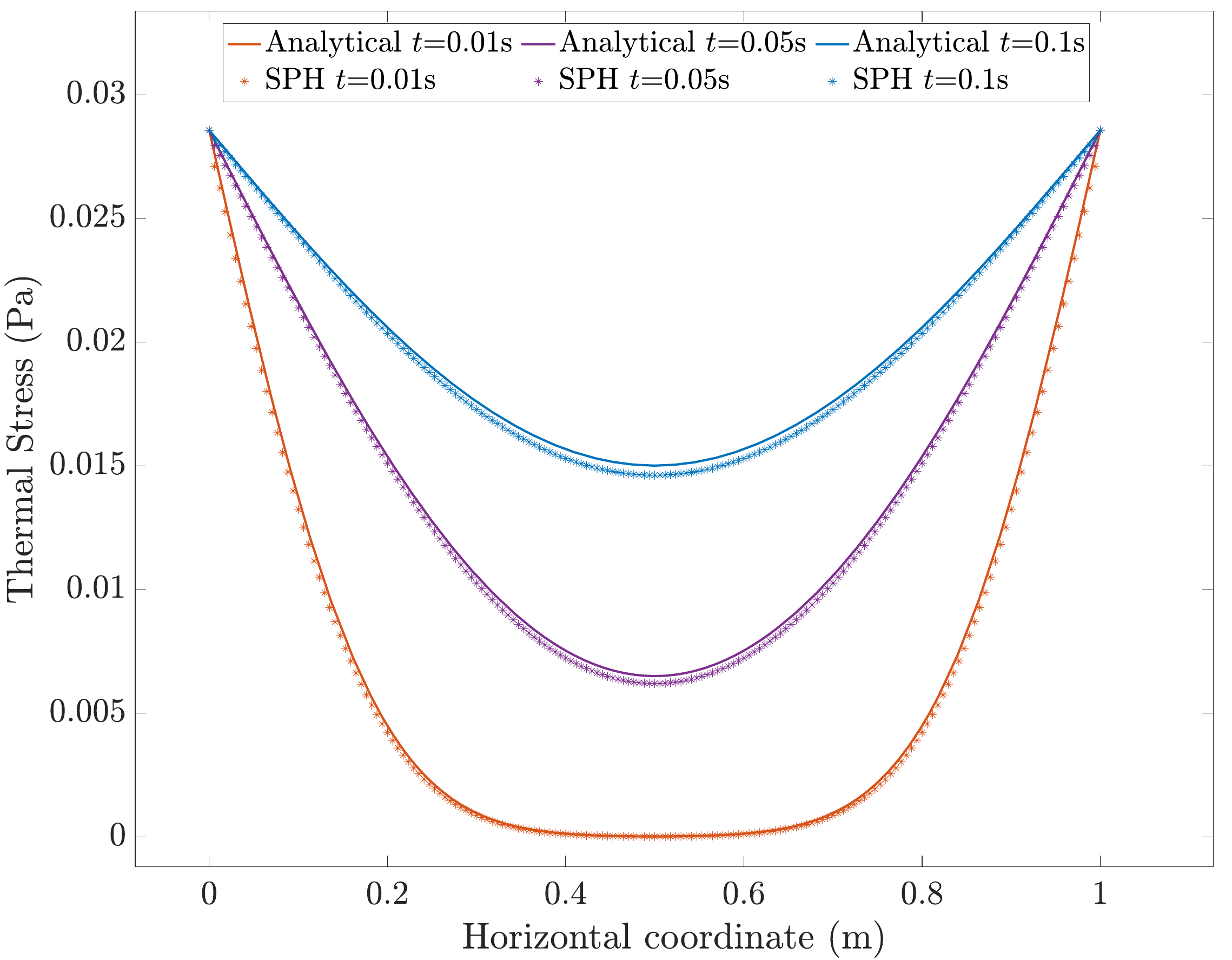}
        \label{fig_tstress2D_thermalinduceddeformation}
    \end{subfigure}
    \caption{Temperature and thermal stress plots comparing analytical results with SPH results. The SPH results show slight deviations because the analytical equations are derived from the uncoupled heat transfer equation, whereas the SPH results account for coupling with the mechanical field.}
    \label{fig_temp_and_stress_thermal_induced_deformation}
\end{figure}

\begin{figure}[hbtp!]
    \centering
    \begin{subfigure}[b]{0.49\textwidth}
        \centering
        \includegraphics[width=\textwidth]{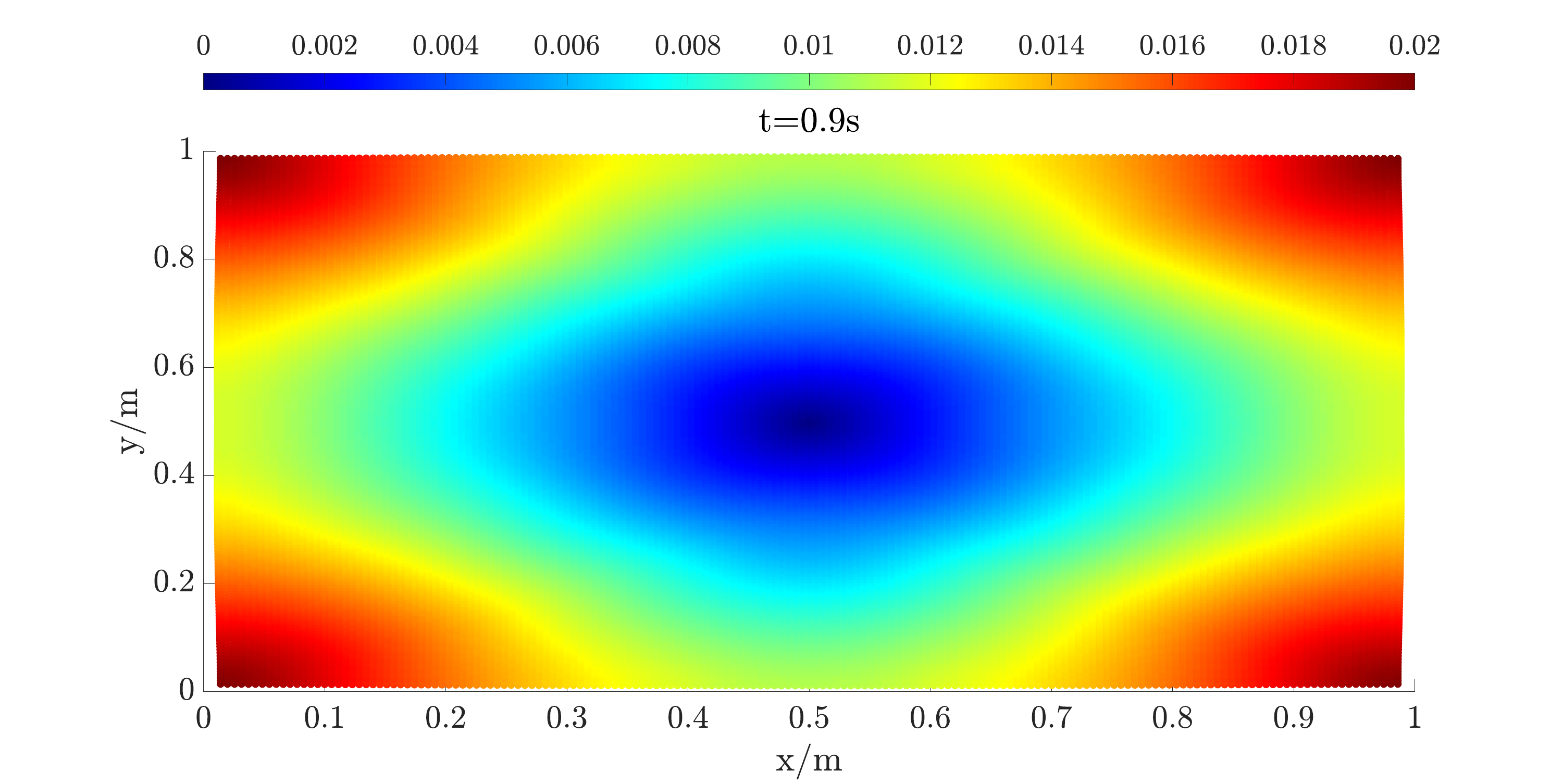}
        \label{fig_disp 0.9s}
    \end{subfigure}
   \hfill
    \begin{subfigure}[b]{0.49\textwidth}
        \centering
        \includegraphics[width=\textwidth]{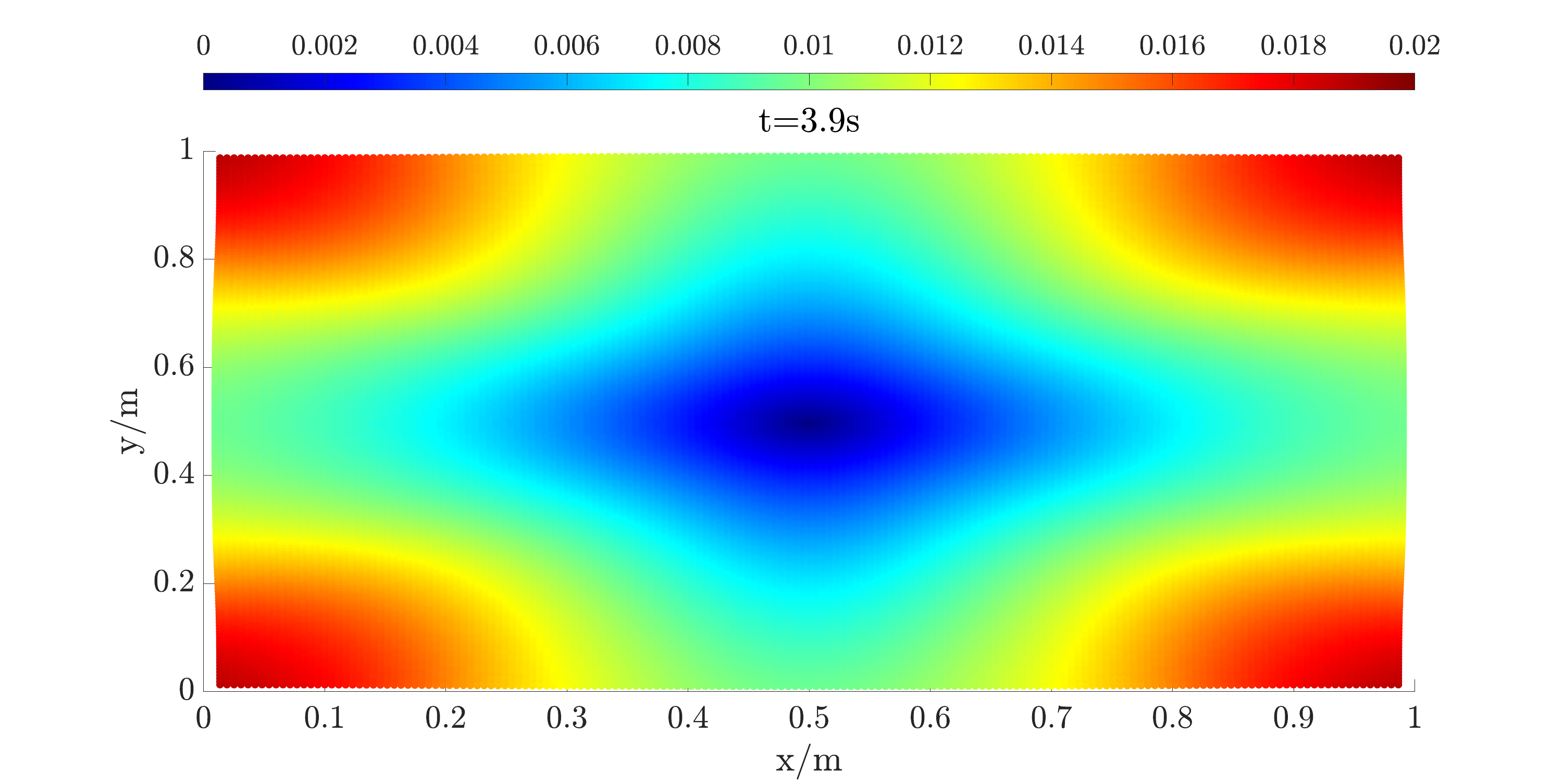}
        \label{fig_disp 3.9s}
    \end{subfigure}
    \hspace{0.003\textwidth} 
    \begin{subfigure}[b]{0.49\textwidth}
        \centering
        \includegraphics[width=\textwidth]{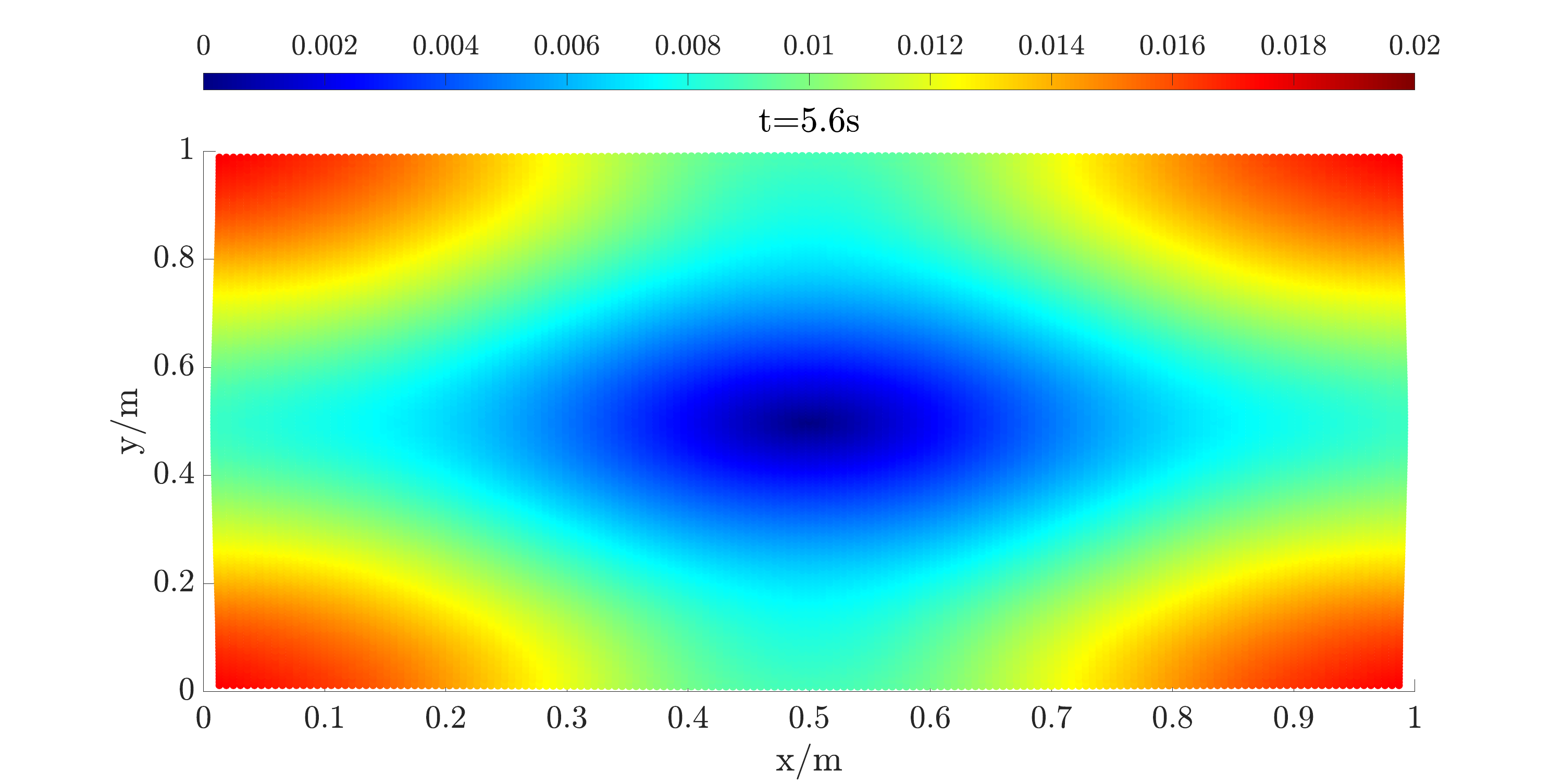}
        \label{fig_disp 5.6s}
    \end{subfigure}
    \caption{The nearly identical displacement (in meters) of SPH particles at 0.9s (Top left), 3.9s (Top right), and 5.6s (Bottom) demonstrates the periodic return of the displacement field to the same distribution. Exactly identical displacement distributions can be obtained by capturing data at even smaller intervals.}
    \label{fig_thermal_deformation_disp}
\end{figure}

\begin{figure}[hbtp!]
    \centering
    \begin{subfigure}[b]{0.4\linewidth}
        \centering
        \includegraphics[width=\linewidth]{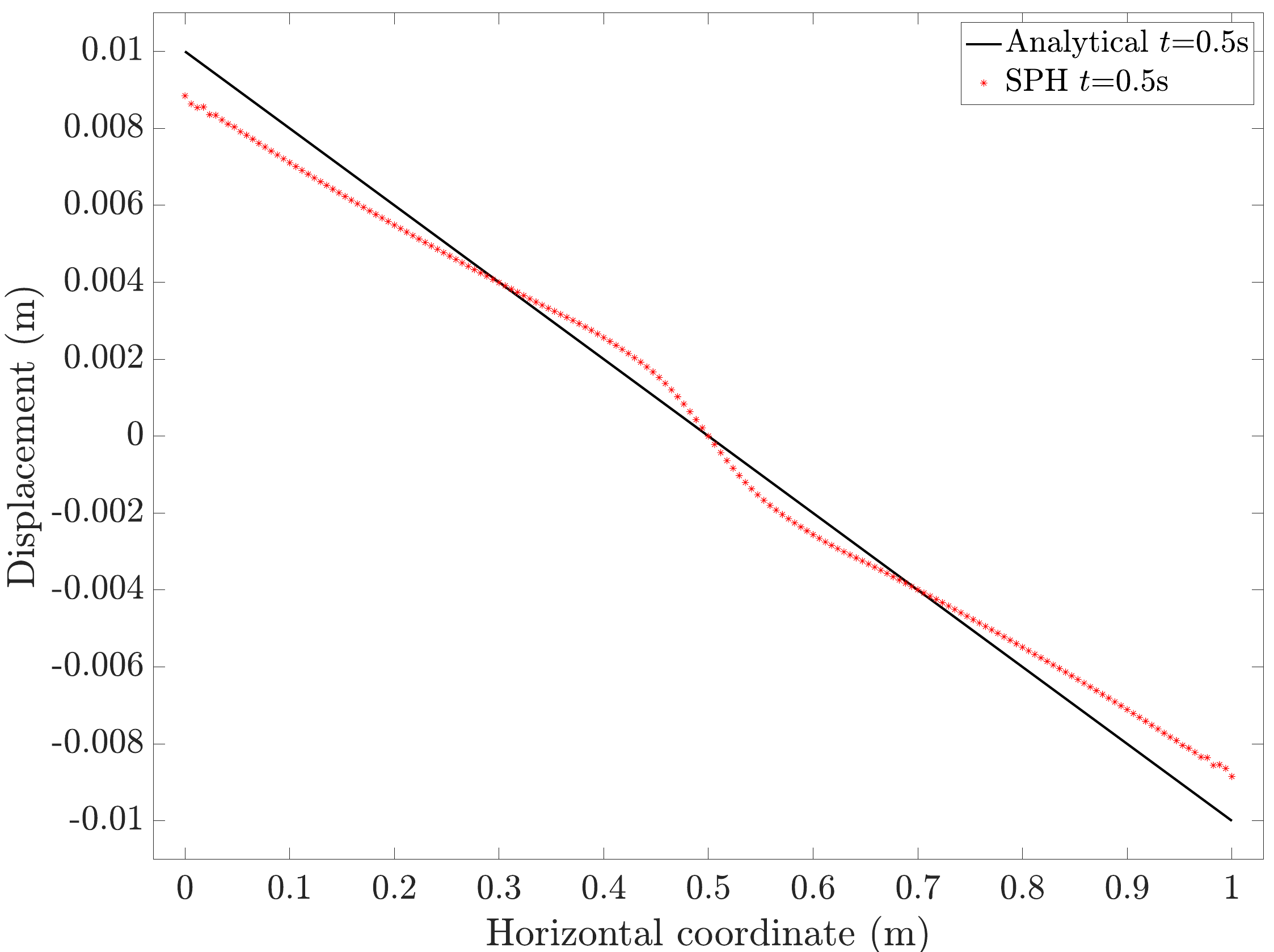}
        \caption{$t = 0.5$ s}
    \end{subfigure}
    \hfill
    \begin{subfigure}[b]{0.4\linewidth}
        \centering
        \includegraphics[width=\linewidth]{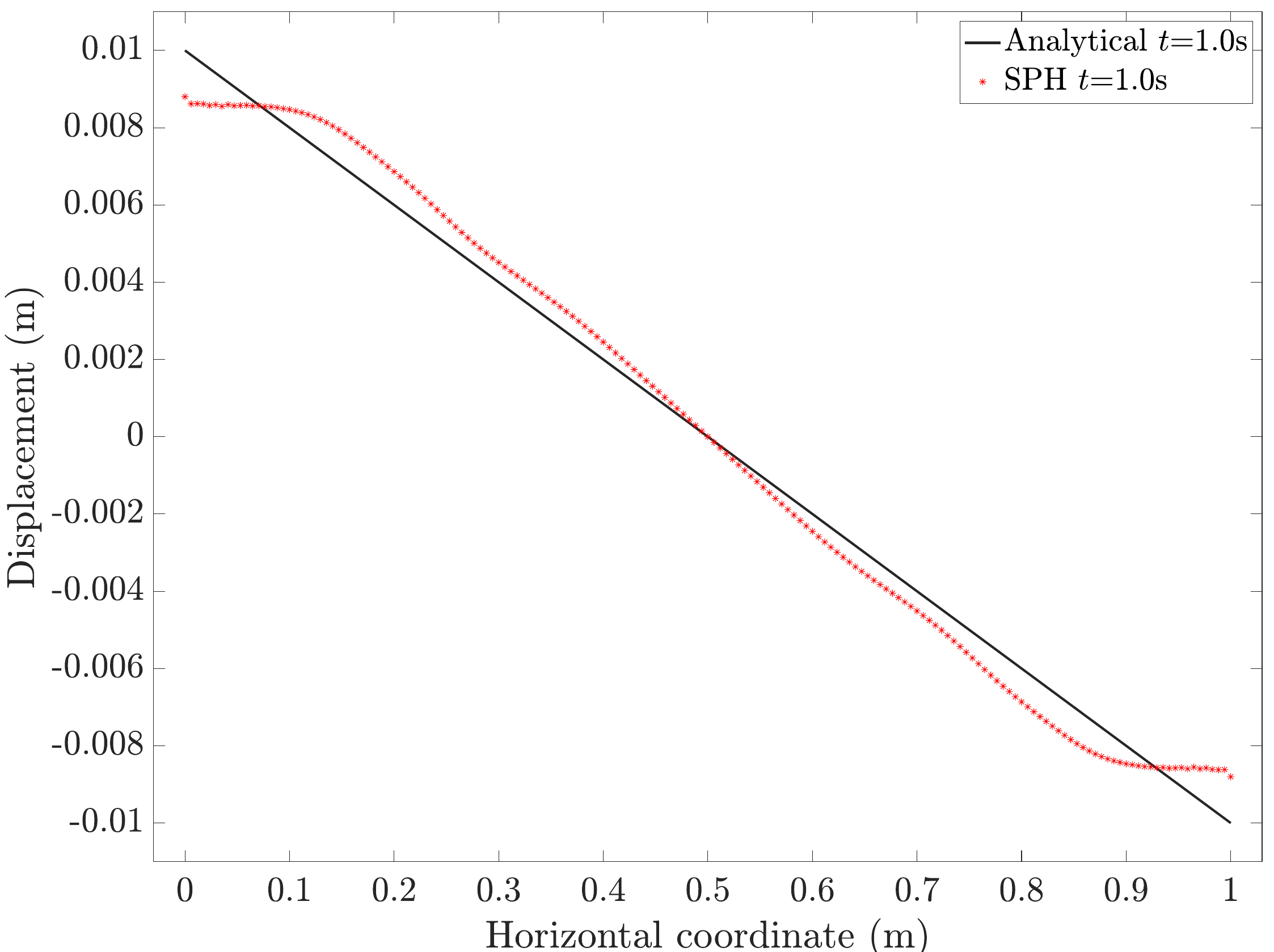}
        \caption{$t = 1.0$ s}
    \end{subfigure}

    \vspace{0.5em} 

    \begin{subfigure}[b]{0.4\linewidth}
        \centering
        \includegraphics[width=\linewidth]{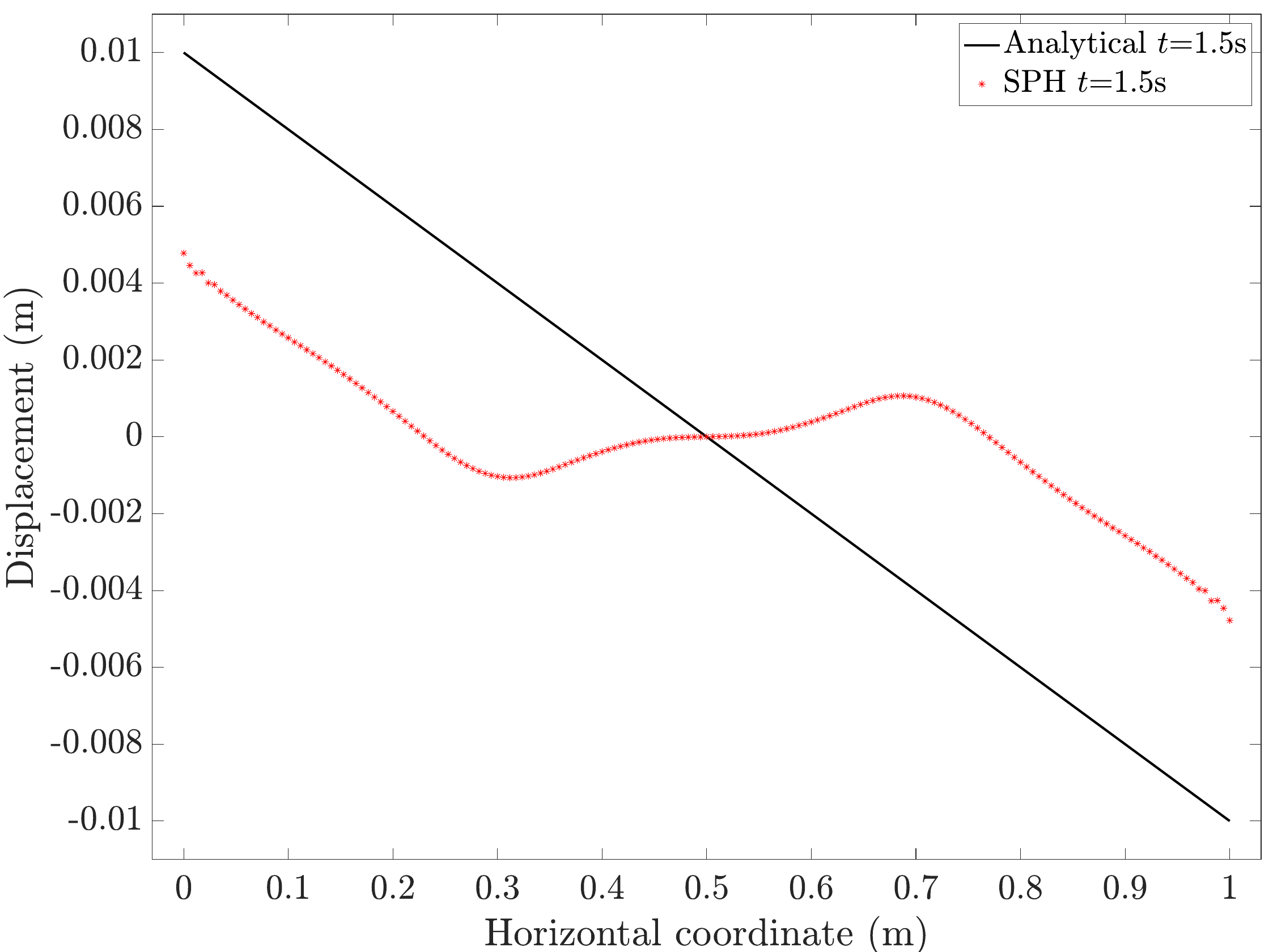}
        \caption{$t = 1.5$ s}
    \end{subfigure}
    \hfill
    \begin{subfigure}[b]{0.4\linewidth}
        \centering
        \includegraphics[width=\linewidth]{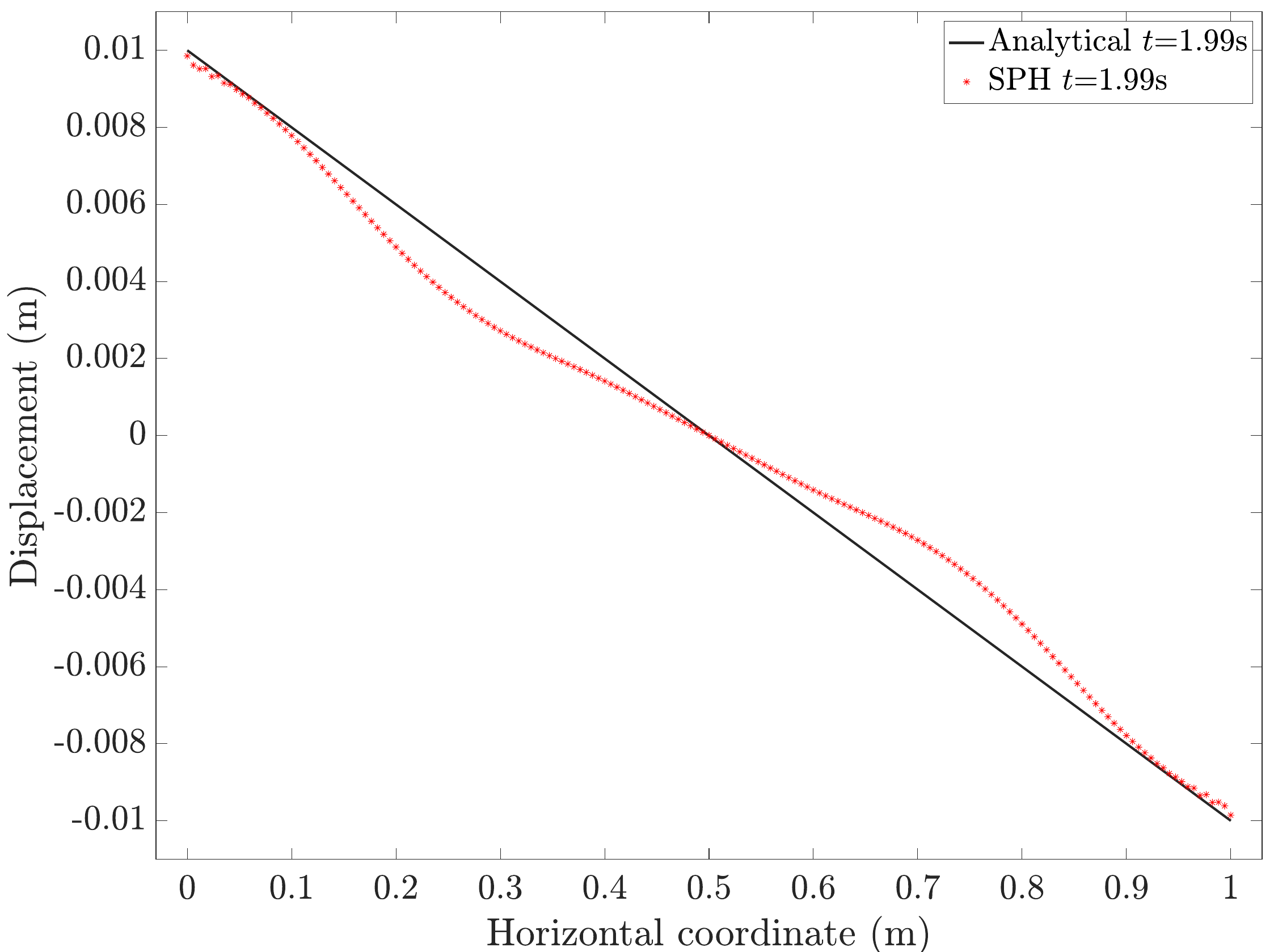}
        \caption{$t = 1.99$ s}
    \end{subfigure}

    \caption{Displacement profiles of SPH particles along the central horizontal axis at various time steps, compared with the analytical solution derived from thermal strain. The fluctuations observed in the SPH results arise because the displacement is computed from the total strain, which includes both elastic and thermal components. While thermal strain stabilises at steady state, the elastic strain continues to evolve, leading to the observed variations.}
    \label{fig_disp_snapshots_thermal_deformation}
\end{figure}

The distributions of thermal stress and temperature closely align with the analytical solutions given by Eqs. \eqref{Analytical_soln_2D_heat_transfer} and \eqref{Analytical_soln_for_thermal_stress_2D_heat_transfer}, as shown in \autoref{fig_temp_and_stress_thermal_induced_deformation}.  While the system approaches a near steady state at approximately 0.5 seconds, the particles continue to oscillate due to the sudden stress induced by the temperature gradient along with the very low elastic modulus (1 Pa) and high coefficient of thermal expansion (0.02/K). This behaviour is demonstrated in \autoref{fig_thermal_deformation_disp}. Although a steady state is reached, the particles continue to undergo minor oscillations, creating a displacement field that varies periodically with time. Notably, particles in and around the centre of the plate remain stationary. Displacement of SPH particles subjected to thermal loading compared against theoretical elastic displacement calculated from $\varepsilon = \alpha \Delta T$, under the assumption of linear elastic and static conditions, is presented in \autoref{fig_disp_snapshots_thermal_deformation}. The particles deviate slightly from the analytical curve due to the oscillations persisting in the system.

\subsection{Numerical simulation to study the effects of insulation}
Satellites, when orbiting, are subjected to cyclic thermal loading that results in the build-up of significant thermal stresses. Therefore, insulators are provided to alleviate the consequences of the thermal deformation. In this section, the multiscale model is employed to understand the insulating characteristics of the superelastic polyimide. The simulation is set up for the case similar to the thermal diffusion experienced by satellites in the Geostationary orbit (GEO). The study is divided into two subsections: (i) a thermomechanical study is conducted to understand how an aluminium plate would respond to thermal diffusion for 50 seconds, and (ii) the study is then repeated for an aluminium plate insulated with superelastic polyimide. However, the thermal deformation alone is considered for the super-elastic polyimide, as its low thermal conductivity renders internal mechanical deformation negligible.

\begin{figure}[hbtp!]
    \centering
    \begin{subfigure}[b]{0.48\linewidth}
        \centering
        \includegraphics[trim=0cm 2cm 0cm 0cm,clip, width=\linewidth]{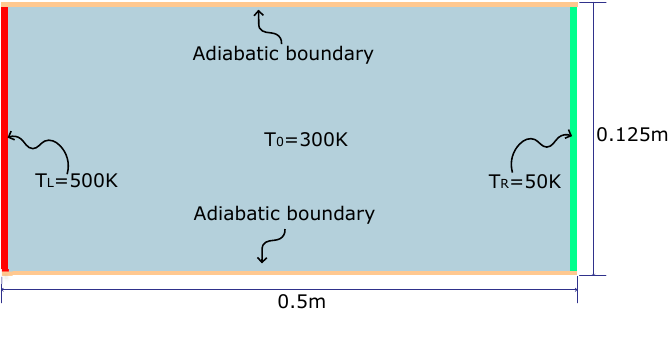}
        \caption{Initial and boundary conditions of the aluminium plate subject to thermal loading}
        \label{fig_Aluminium plate}
    \end{subfigure}
    \hfill
    \begin{subfigure}[b]{0.48\linewidth}
        \centering
        \includegraphics[width=\linewidth]{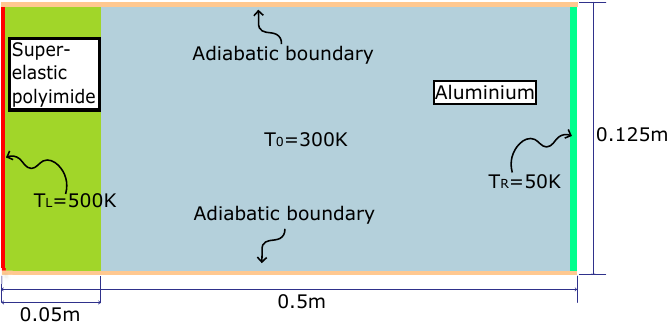}
        \caption{Initial and boundary conditions of the aluminium plate insulated by superelastic polyimide subjected to thermal loading}
        \label{fig_Aluminium-kapton plate}
    \end{subfigure}
\end{figure}



We consider a rectangular plate of dimensions 0.5m $\times$ 0.125m (\autoref{fig_Aluminium plate}). The density of aluminium is scaled (by a factor of $100$) to avoid the multi-rate time integration. This technique is called mass-scaling and is frequently used in explicit time integration schemes to address the problem of very small timestep size \cite{heinze2016systematic}. To keep thermal diffusivity ($k$) constant, that is, $k=\kappa/\rho c_v$=constant, $\kappa$ is also scaled by a factor of 100. The mechanical parameters used in the simulation as follows: Young's modulus ($E$) = 68.3 GPa, Poisson's ratio ($\mu$) = 0.34, density ($\rho$) = $268900.8 ~ \mathrm{kg/m^3}$ and yield stress ($\sigma_{Y}$) = 30 MPa. The thermal parameters are: thermal conductivity ($\kappa$) = $23800.5 ~ \mathrm{W/mK}$, coefficient of thermal expansion ($\hat{\alpha}$) = $23.5\times10^{-6} /\mathrm{K}$, and specific heat capacity ($c_v$) = $916.3 ~ \mathrm{J/kgK}$. The left boundary of the plate is fixed at 500 K, whereas the right boundary is fixed at 50 K. The other two sides have adiabatic constraints. \autoref{fig_Aluminium plate} details the initial and boundary conditions for the simulation. The response of the plate is assumed to be perfectly plastic. The setup is run for 50 seconds with a timestep size of $10^{-6}$ seconds.

Next, we analyse a similar scenario with superelastic polyimide insulating the plate from the high-temperature region. The same rectangular plate, measuring 0.5 m by 0.125 m, is considered, but the material configuration is modified: the first 0.05 m is superelastic polyimide, while the remainder is aluminium. The conditions are illustrated in \autoref{fig_Aluminium-kapton plate}. The mechanical and thermal properties of the superelastic polyimide are obtained from MD simulation as discussed before: Young's modulus ($E$) = 6.39 GPa, Poisson's ratio ($\mu$) = 0.218, density ($\rho$) = $1360.0~ \mathrm{kg/m^3}$, thermal conductivity ($\kappa$) = $0.32~\mathrm{W/mK}$, coefficient of thermal expansion ($\hat{\alpha}$) = $4.723 \times 10^{-5}/\mathrm{K}$, and specific heat capacity ($c$) = $1090.0  ~\mathrm{J/kgK}$.  We have ignored the mechanical stress developed in the superelastic polyimide as the thermal conductivity of the polyimide is very low (compared to Aluminum) to produce any significant mechanical deformation on the aluminium plate. The setup is run for 50 seconds with a timestep size of $10^{-6}$ seconds.


\begin{figure}[hbtp!]
    \centering
    \begin{subfigure}[b]{0.49\textwidth}
        \centering
        \includegraphics[width=\textwidth]{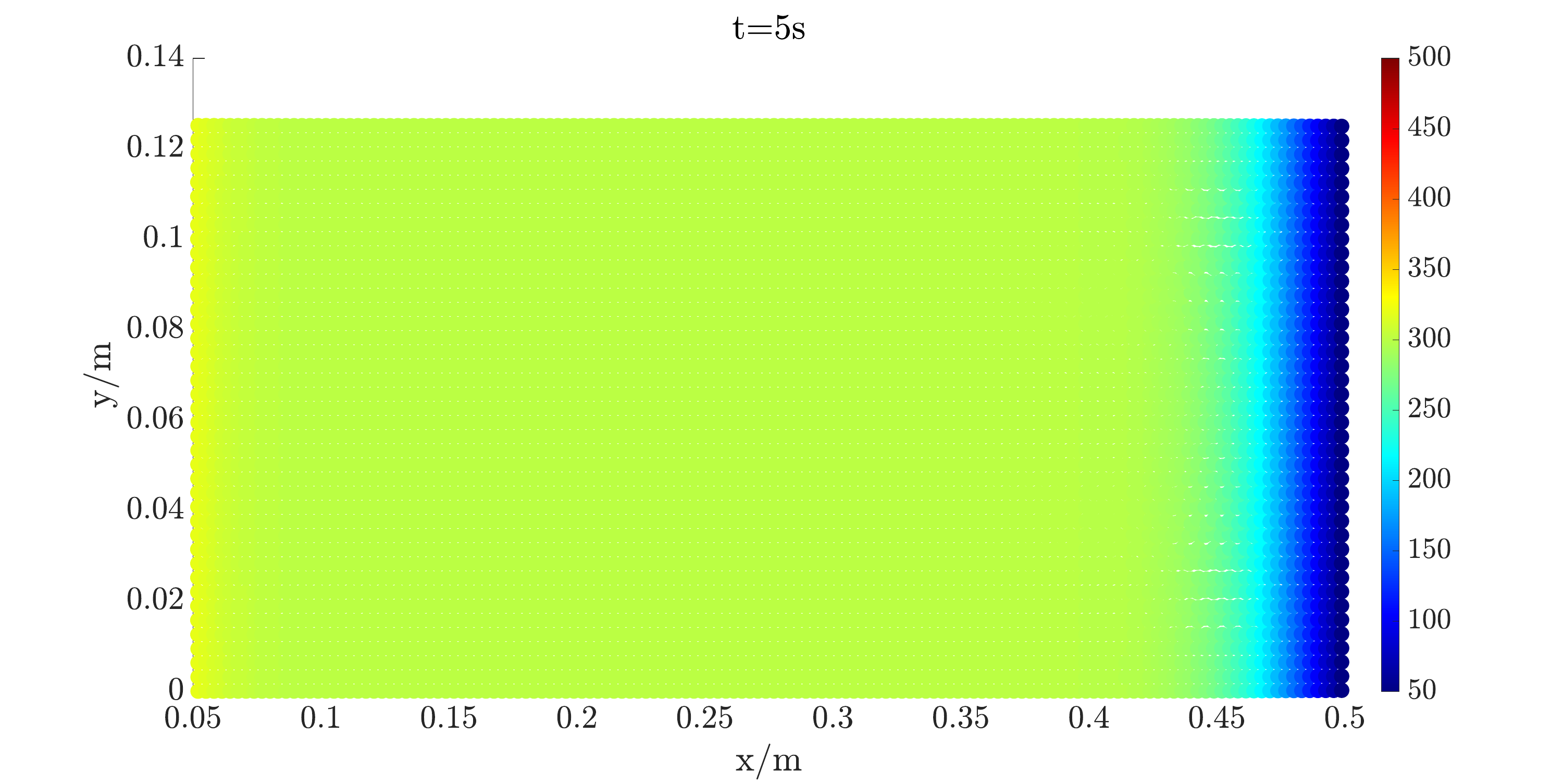}

    \end{subfigure}
   \hfill
    \begin{subfigure}[b]{0.49\textwidth}
        \centering
        \includegraphics[width=\textwidth]{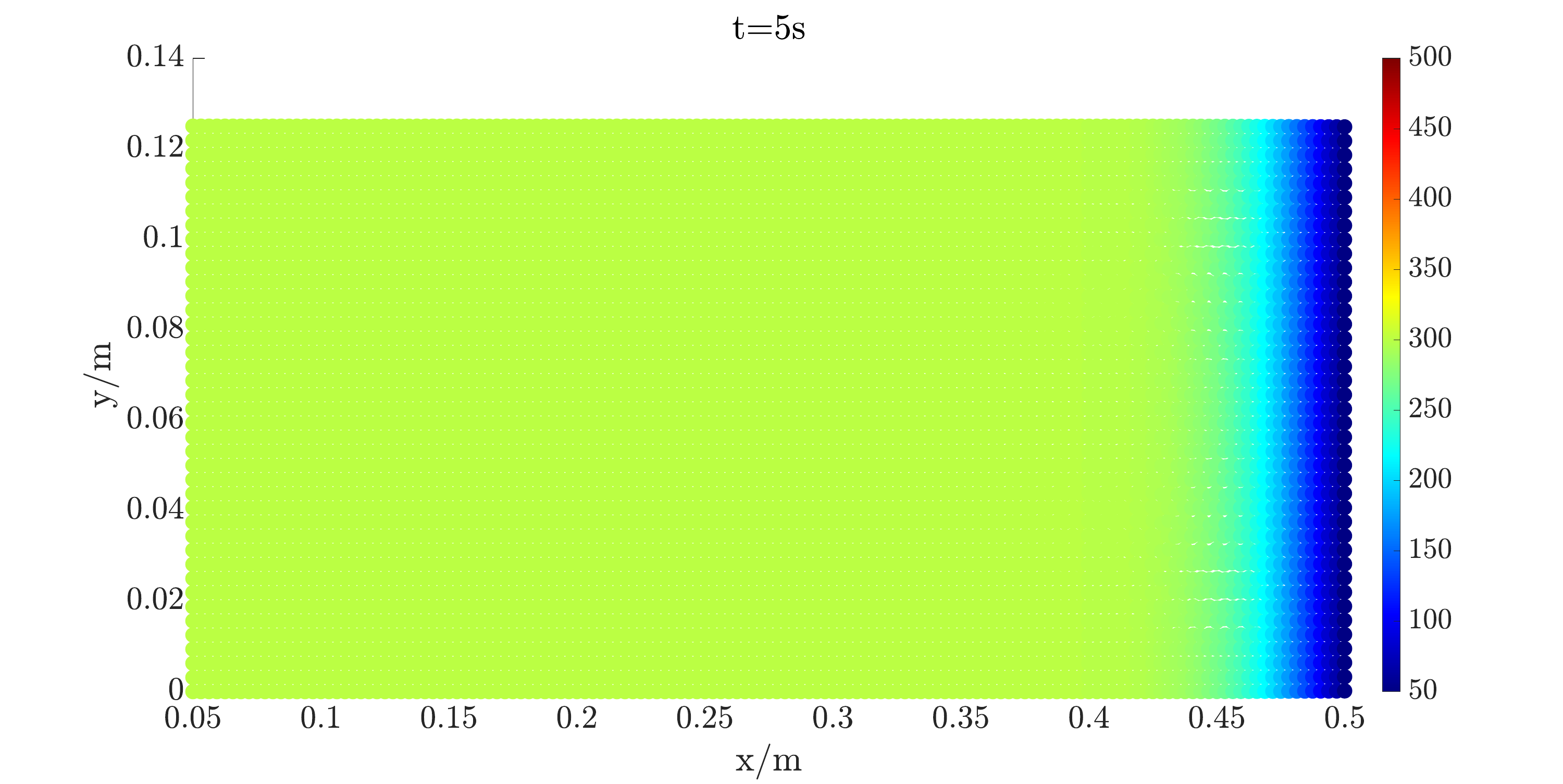}

    \end{subfigure}
    \hfill
    \begin{subfigure}[b]{0.49\textwidth}
        \centering
        \includegraphics[width=\textwidth]{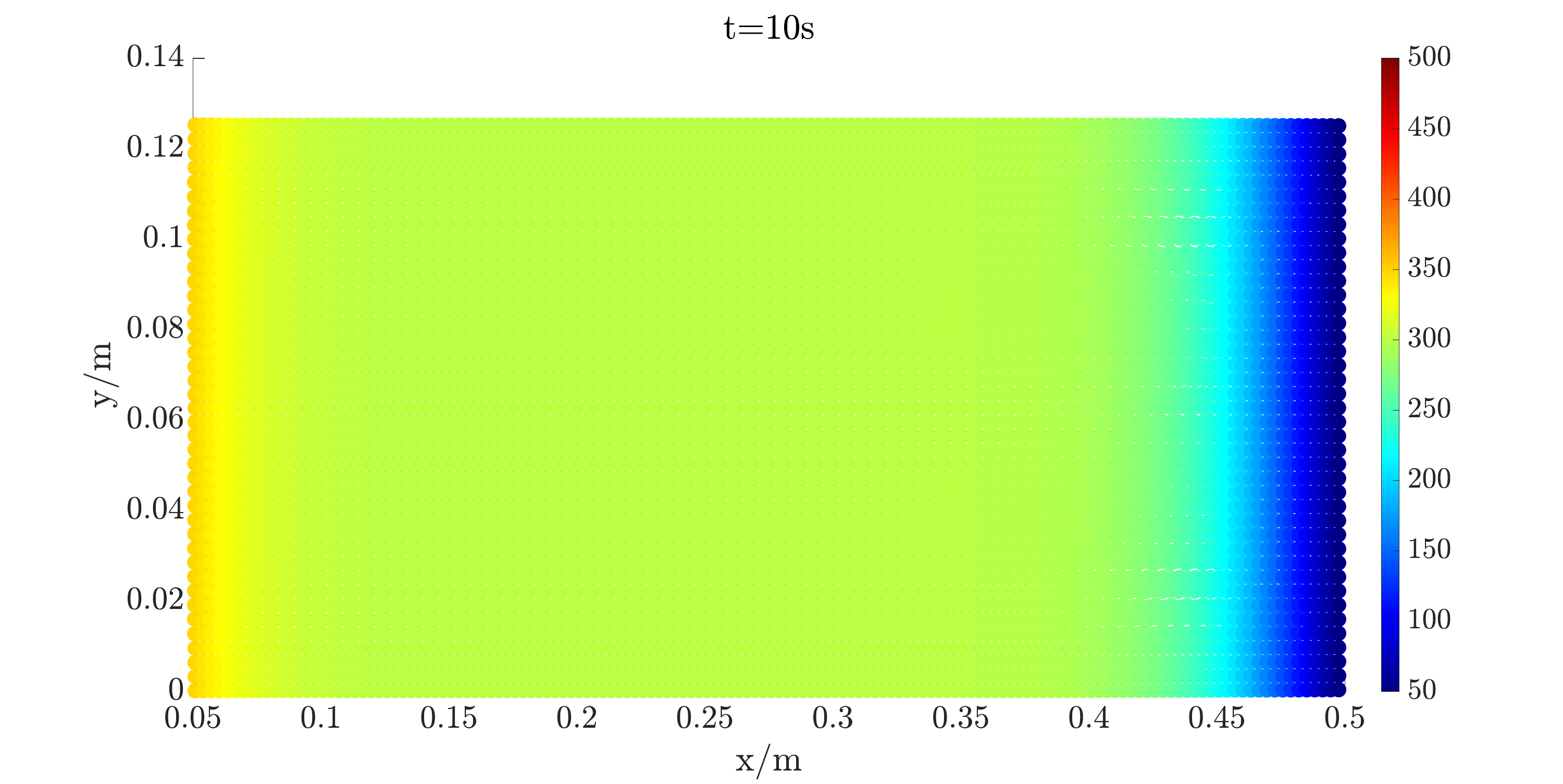}

    \end{subfigure}
    \hfill
    \begin{subfigure}[b]{0.49\textwidth}
        \centering
        \includegraphics[width=\textwidth]{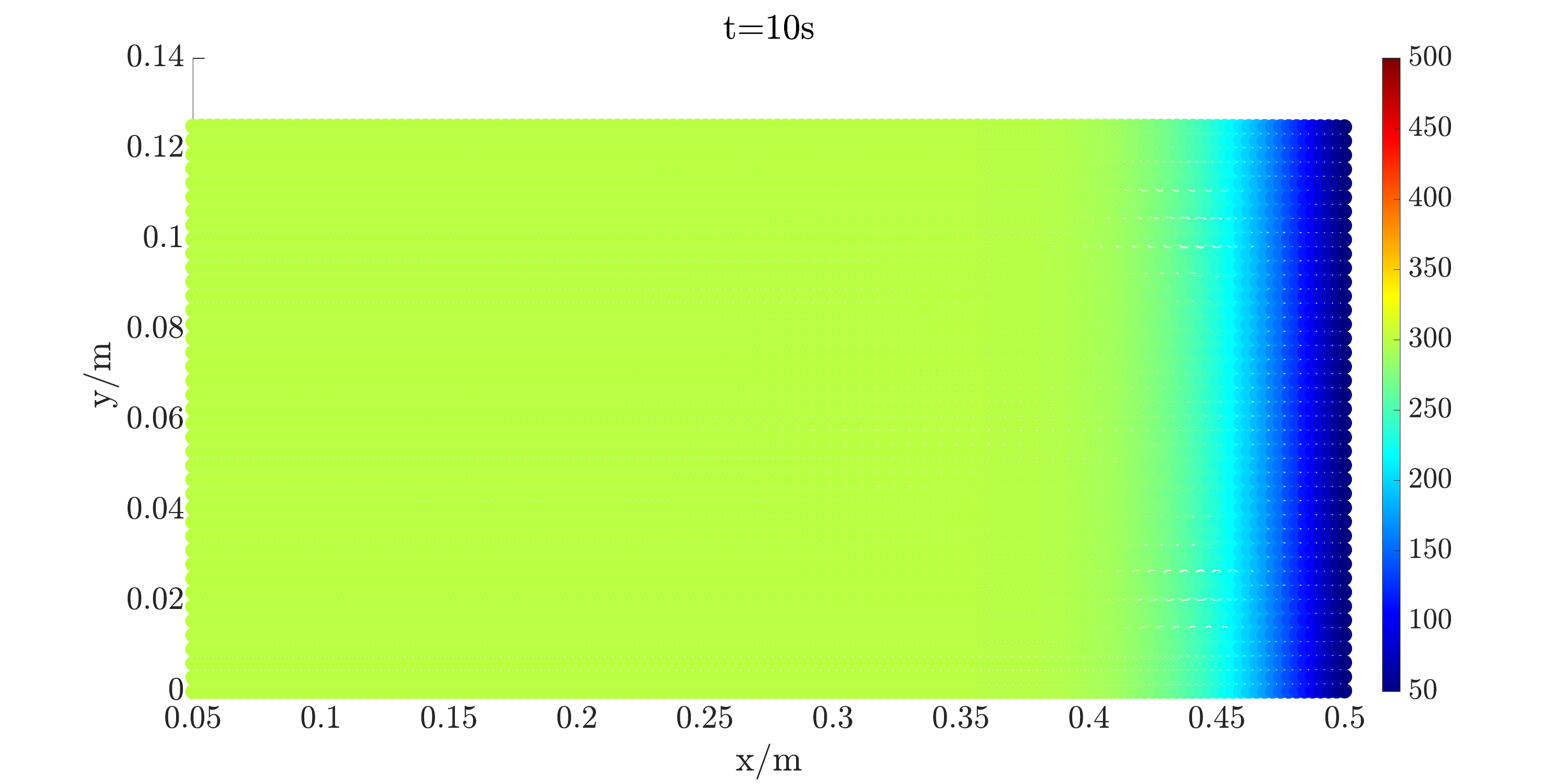}

    \end{subfigure}
    \hfill
    \begin{subfigure}[b]{0.49\textwidth}
        \centering
        \includegraphics[width=\textwidth]{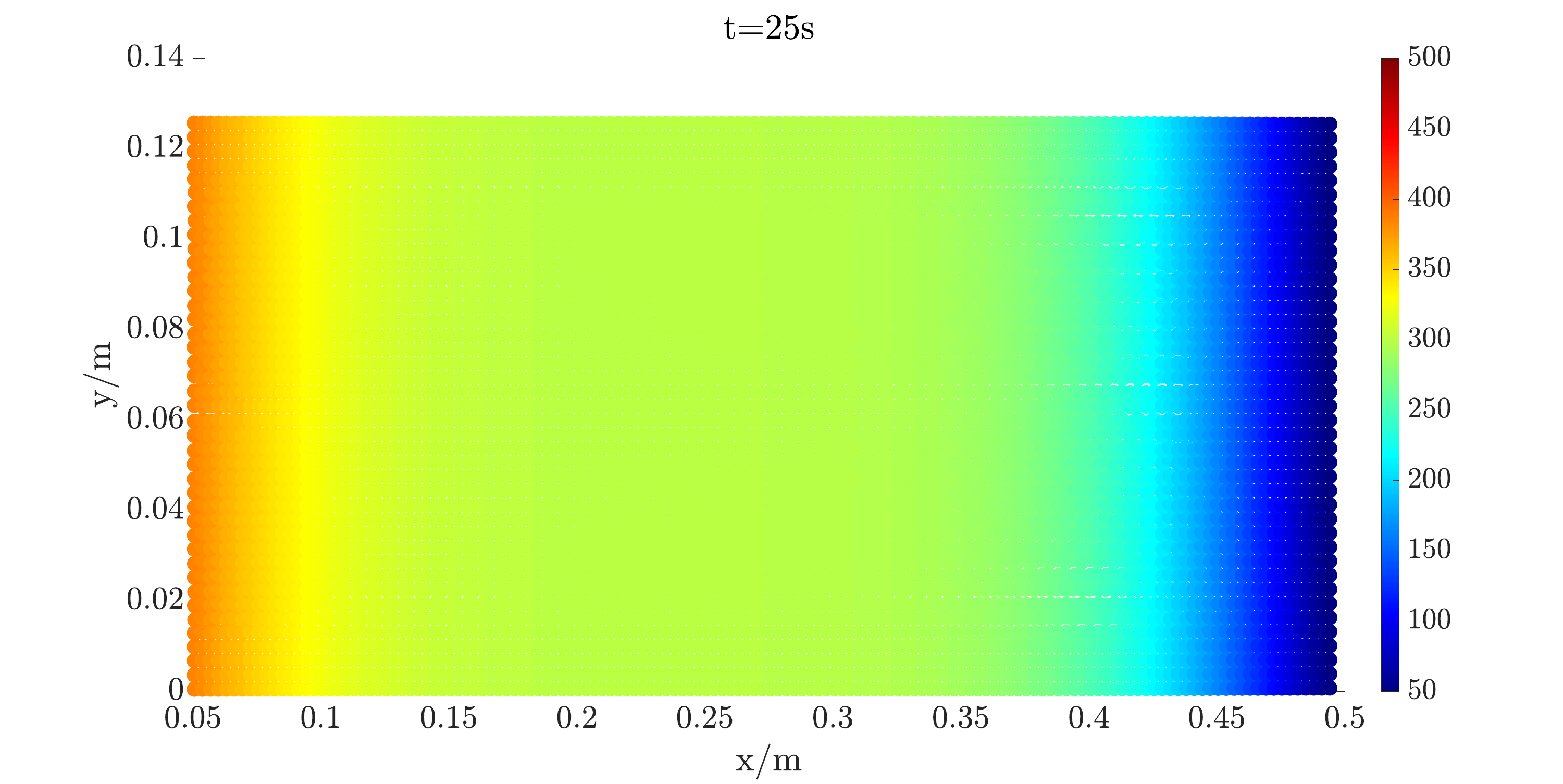}

    \end{subfigure}
    \hfill
    \begin{subfigure}[b]{0.49\textwidth}
        \centering
        \includegraphics[width=\textwidth]{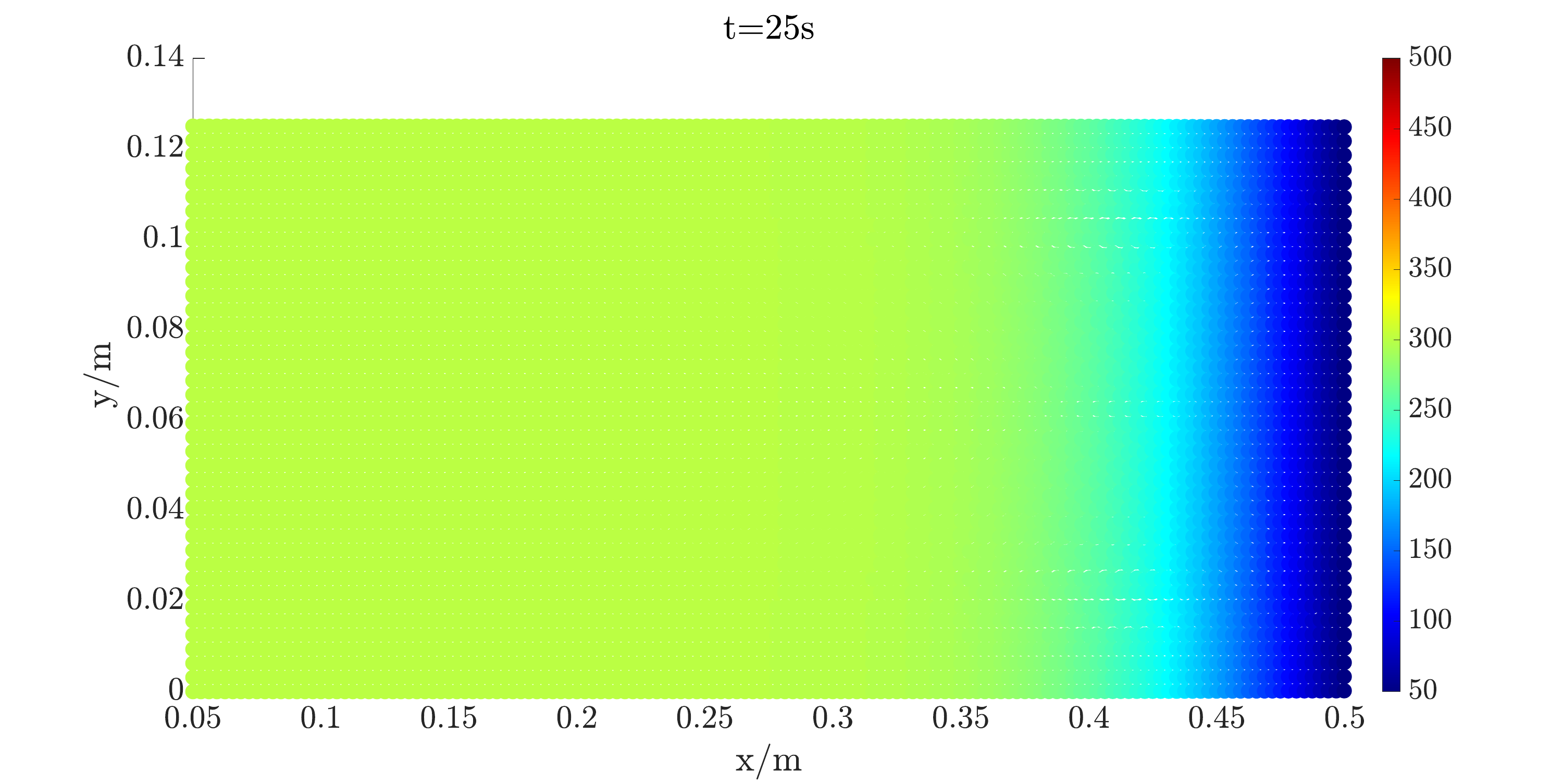}
  
    \end{subfigure}
    \hfill
    \begin{subfigure}[b]{0.49\textwidth}
        \centering
        \includegraphics[width=\textwidth]{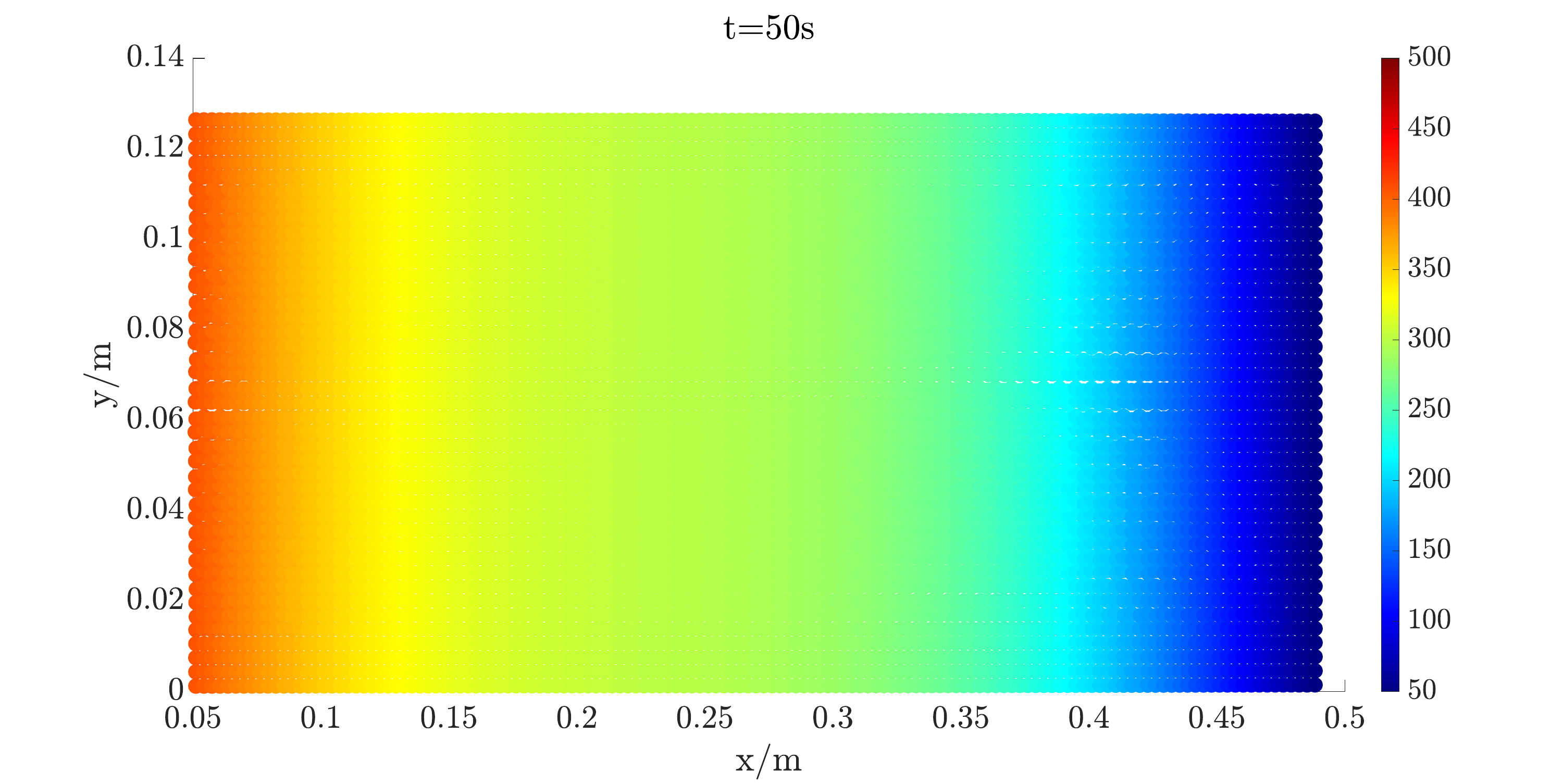}

    \end{subfigure}
    \hfill
    \begin{subfigure}[b]{0.49\textwidth}
        \centering
        \includegraphics[width=\textwidth]{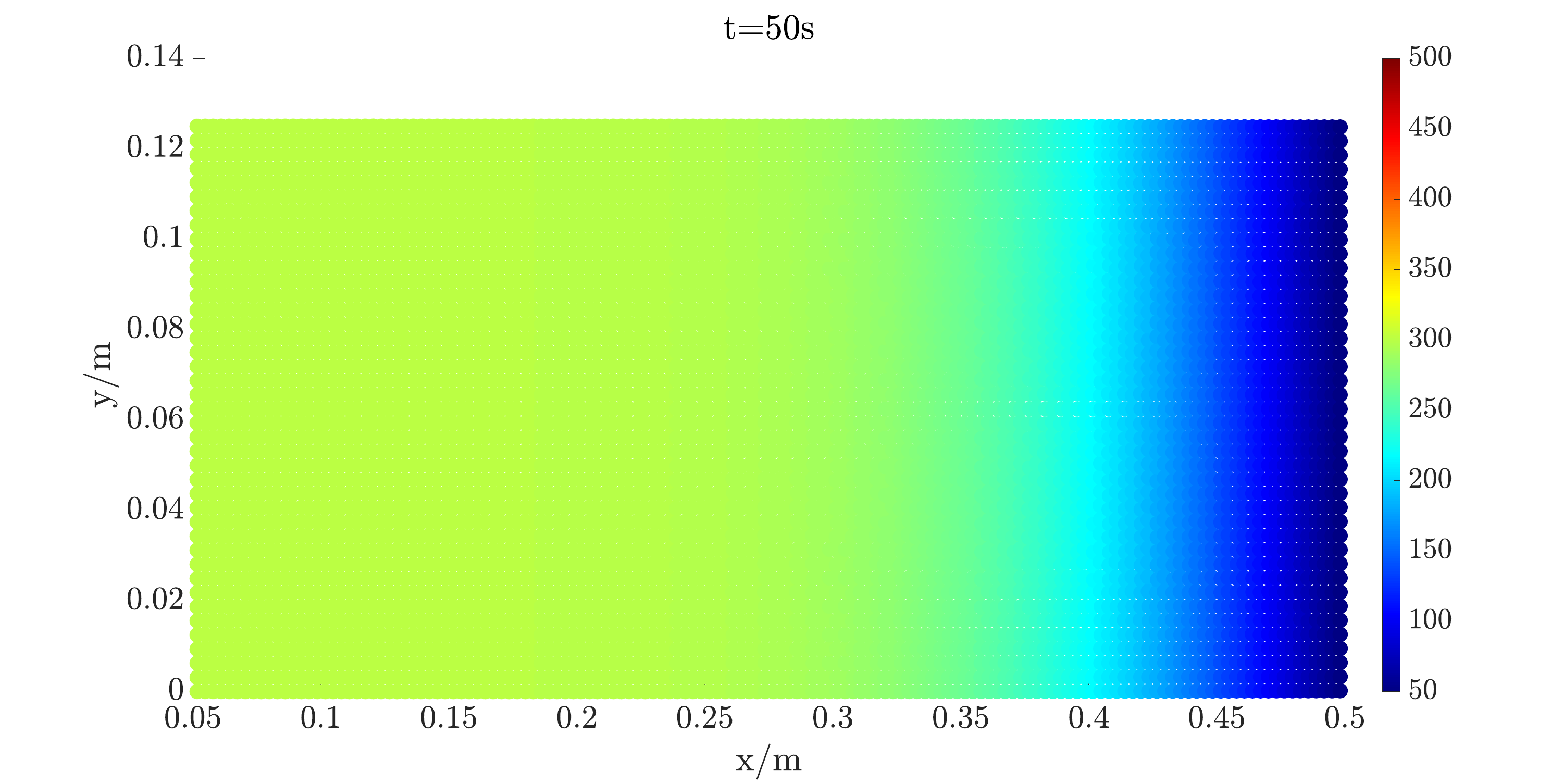}
  
    \end{subfigure}
    \caption{Temperature distribution results for an uninsulated aluminium plate (Left) and an aluminium plate insulated with superelastic polyimide (Right) at 5, 10, 25 and 50 seconds. The particles are shown at their initial positions, with the colour indicating the temperature (in K).}
    \label{fig_Temp_distribution_comparison_for_5_10_25_50}
\end{figure}

\begin{figure}[hbtp!]
    \centering

    \begin{subfigure}[b]{0.49\textwidth}
        \centering
        \includegraphics[width=\textwidth]{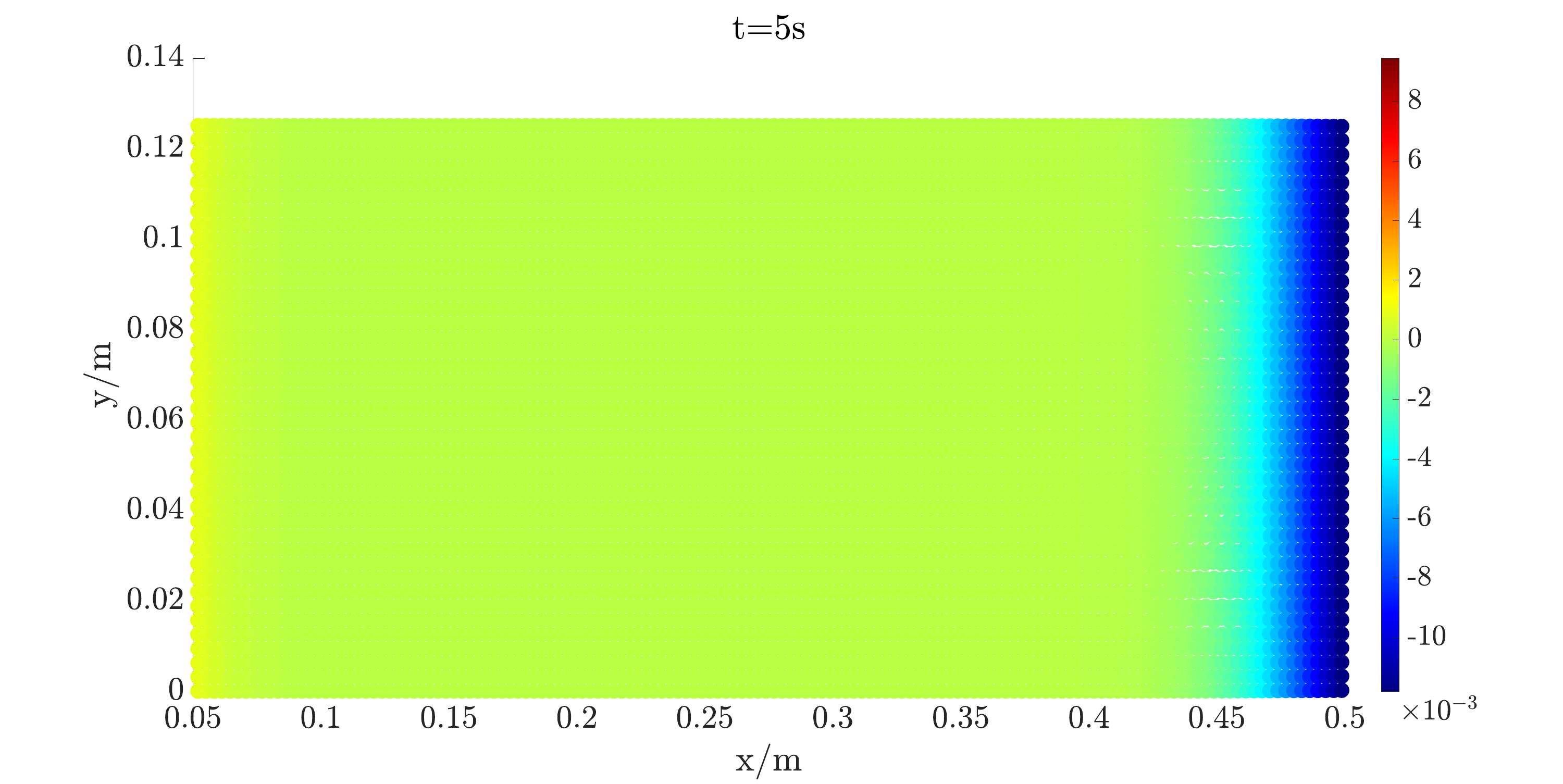}
        \label{fig_5s_alonly_tstrain}
    \end{subfigure}
    \hfill
    \begin{subfigure}[b]{0.49\textwidth}
        \centering
        \includegraphics[width=\textwidth]{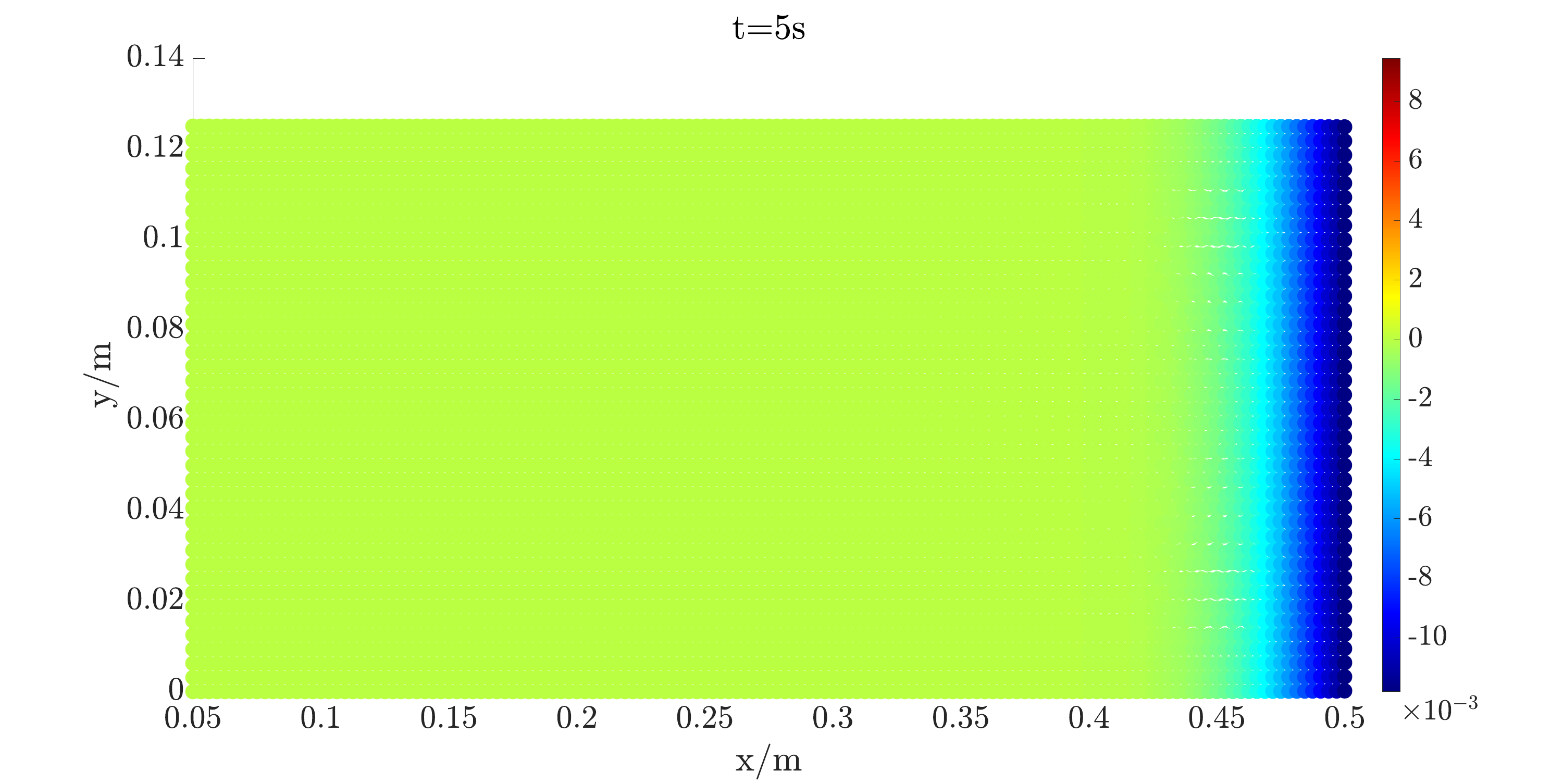}
        \label{fig_5s_kapal_tstrain}
    \end{subfigure}

    \vskip\baselineskip

    \begin{subfigure}[b]{0.49\textwidth}
        \centering
        \includegraphics[width=\textwidth]{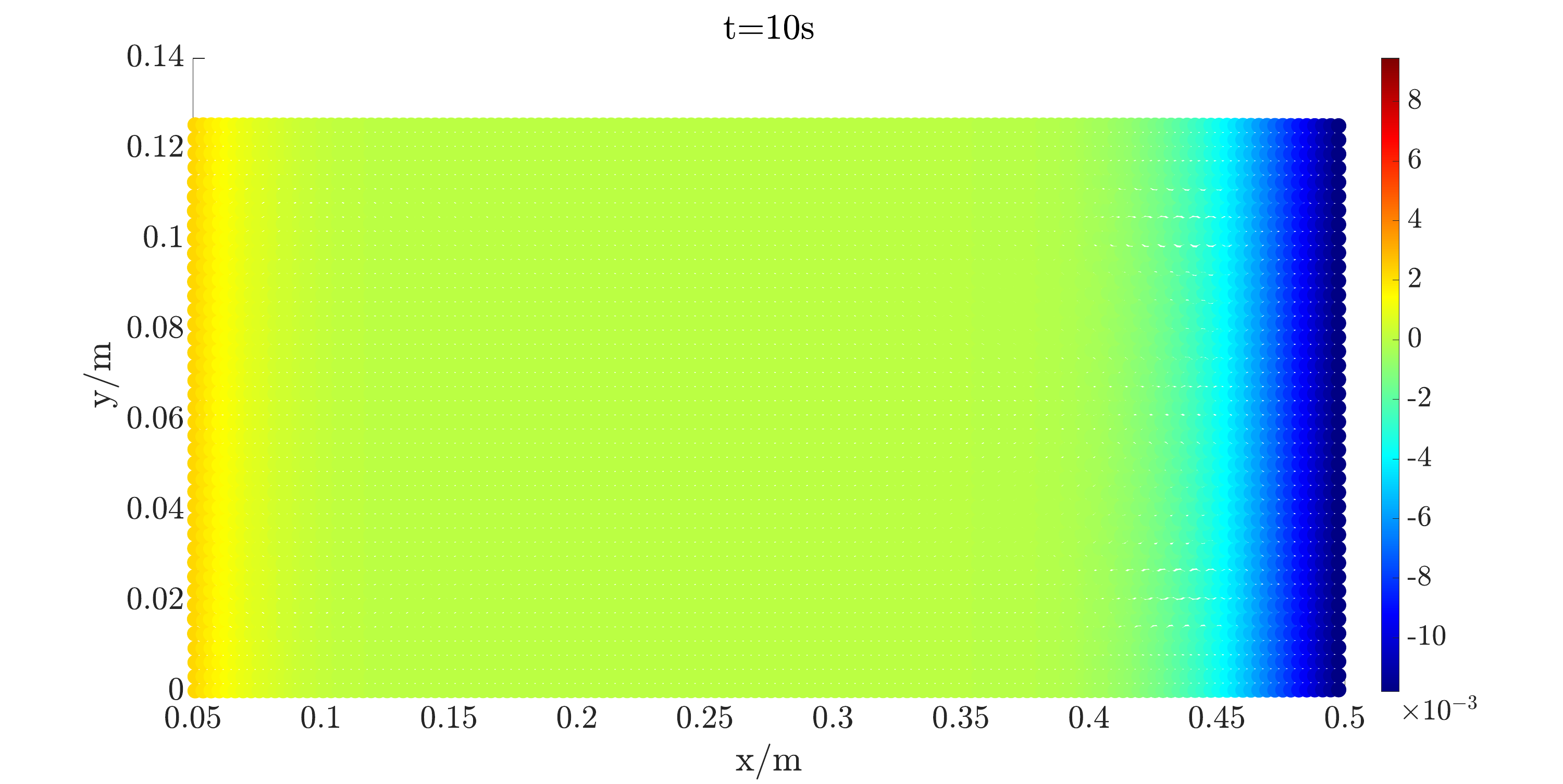}
        \label{fig_10s_alonly_tstrain}
    \end{subfigure}
    \hfill
    \begin{subfigure}[b]{0.49\textwidth}
        \centering
        \includegraphics[width=\textwidth]{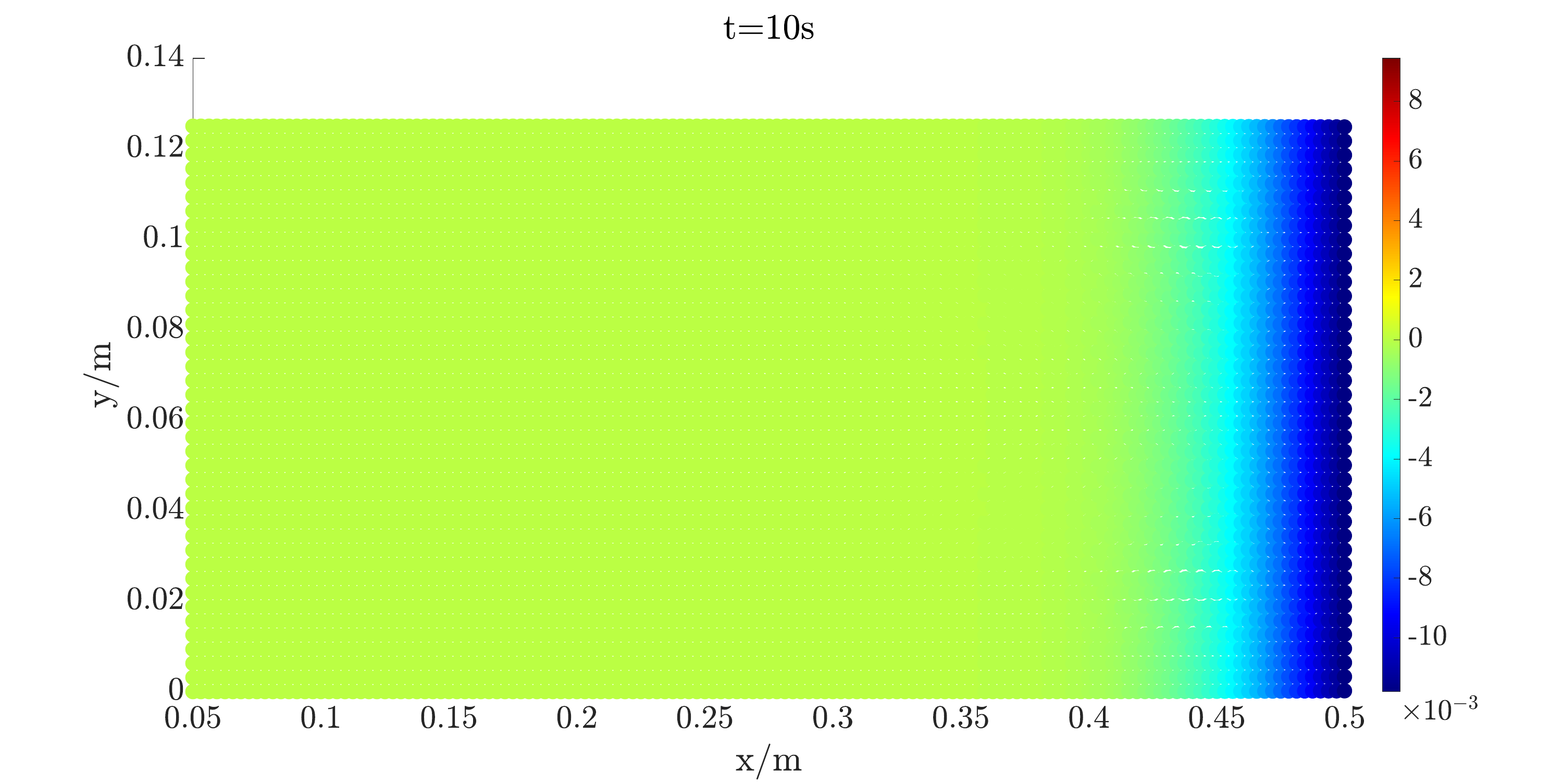}
        \label{fig_10s_kapal_tstrain}
    \end{subfigure}

    \vskip\baselineskip

    \begin{subfigure}[b]{0.49\textwidth}
        \centering
        \includegraphics[width=\textwidth]{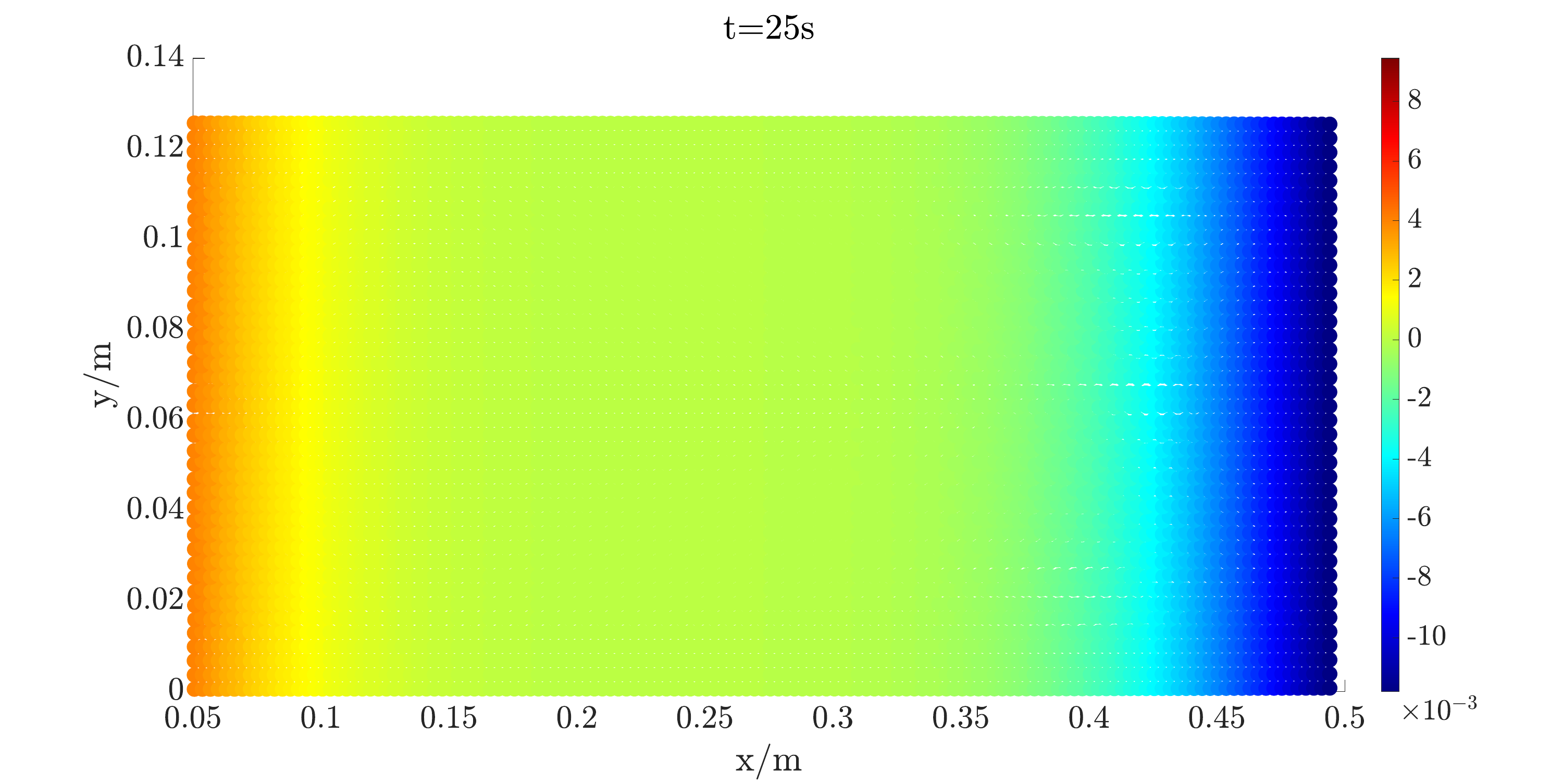}
        \label{fig_25s_alonly_tstrain}
    \end{subfigure}
    \hfill
    \begin{subfigure}[b]{0.49\textwidth}
        \centering
        \includegraphics[width=\textwidth]{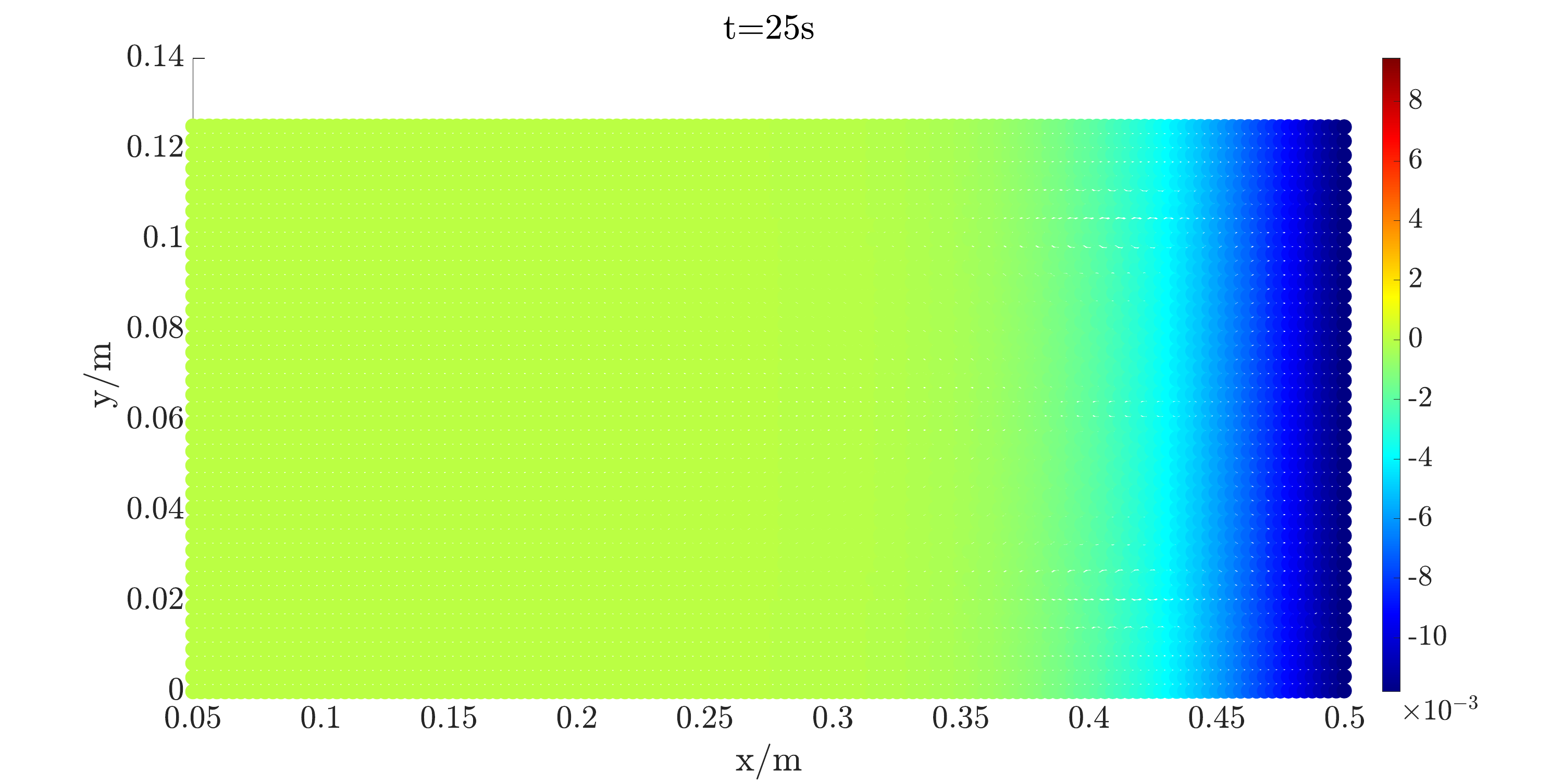}
        \label{fig_25s_kapal_tstrain}
    \end{subfigure}

    \vskip\baselineskip

    \begin{subfigure}[b]{0.49\textwidth}
        \centering
        \includegraphics[width=\textwidth]{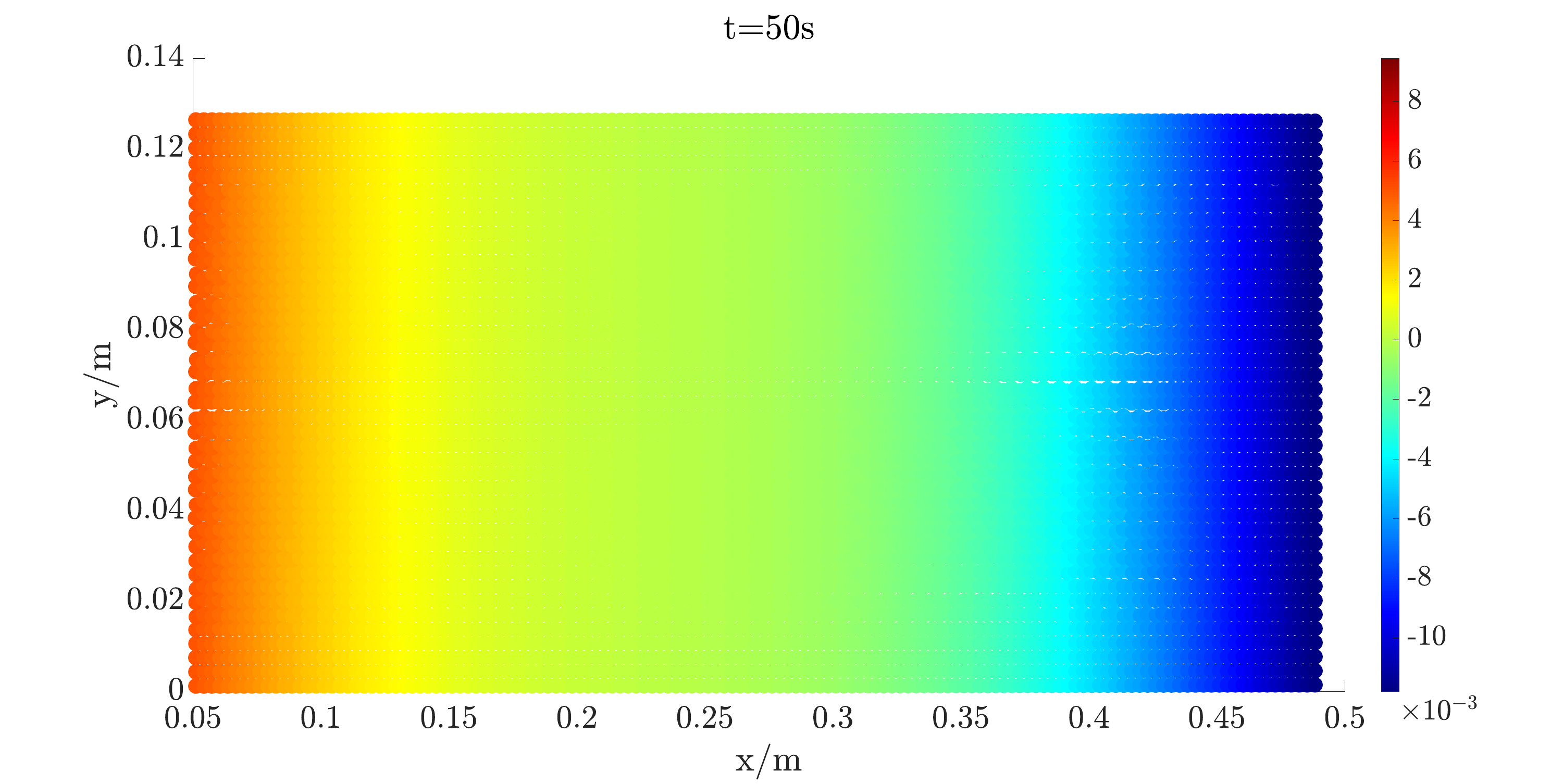}
        \label{fig_50s_alonly_tstrain}
    \end{subfigure}
    \hfill
    \begin{subfigure}[b]{0.49\textwidth}
        \centering
        \includegraphics[width=\textwidth]{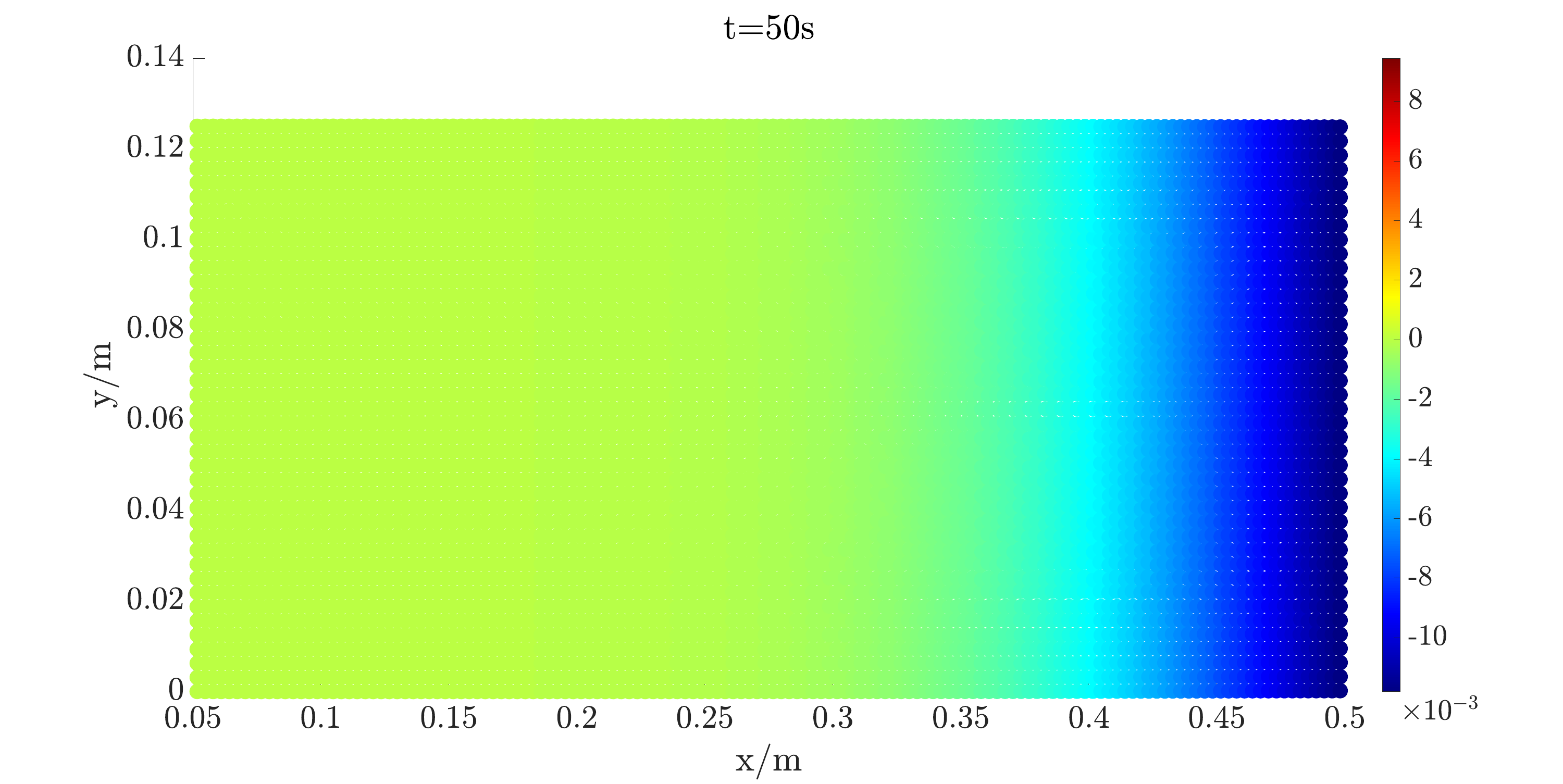}
        \label{fig_50s_kapal_tstrain}
    \end{subfigure}

    \caption{Thermal strain distribution for an uninsulated aluminium plate (Left) and an aluminium plate insulated with superelastic polyimide (Right) at 5, 10, 25, and 50 seconds.}
    \label{fig_tstrain_distribution_comparison}
\end{figure}

\begin{figure}[hbtp!]
    \centering
    \begin{subfigure}[b]{0.49\textwidth}
        \centering
        \includegraphics[width=\textwidth]{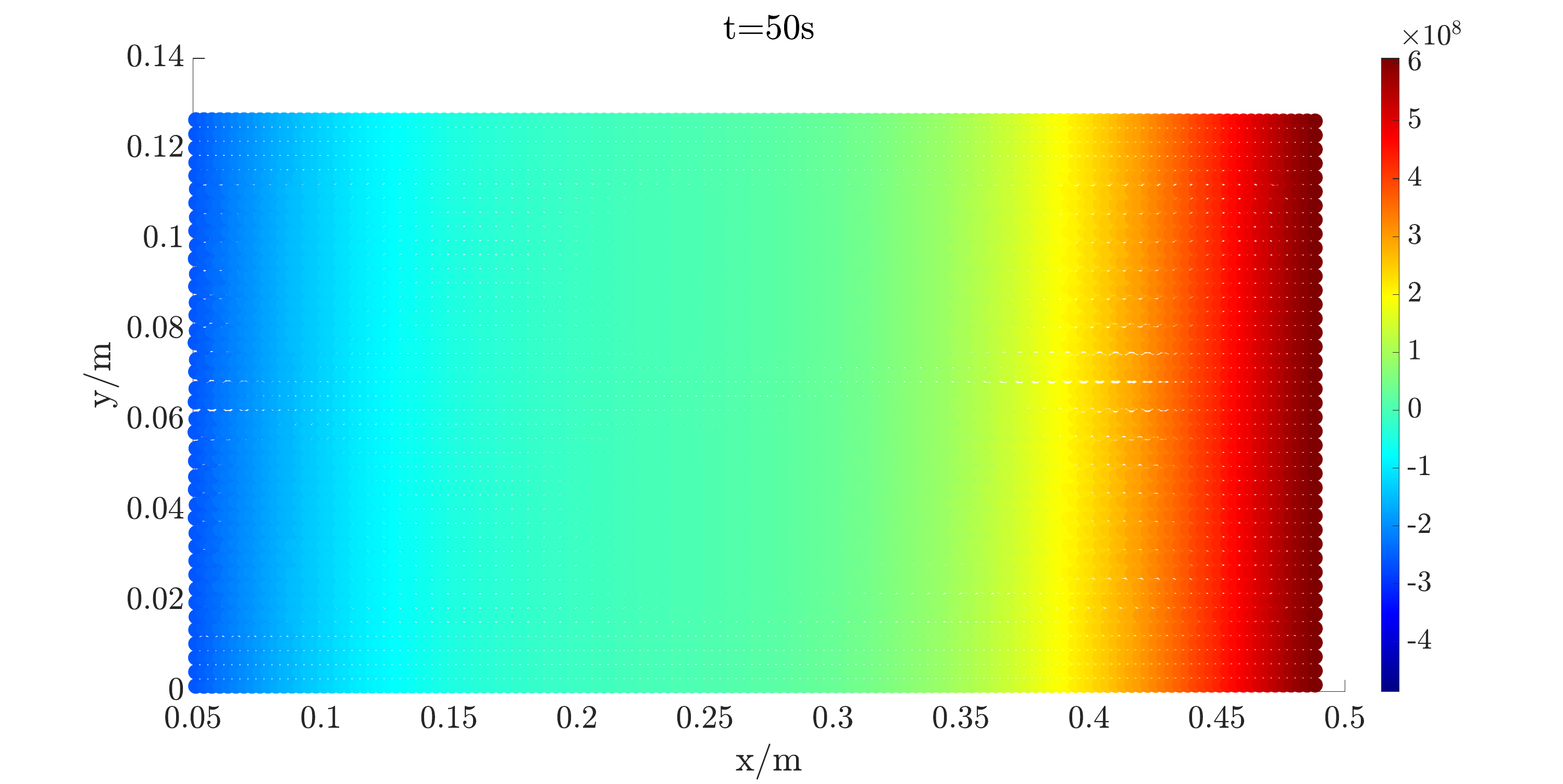}
    \end{subfigure}
    \hfill
    \begin{subfigure}[b]{0.49\textwidth}
        \centering
        \includegraphics[width=\textwidth]{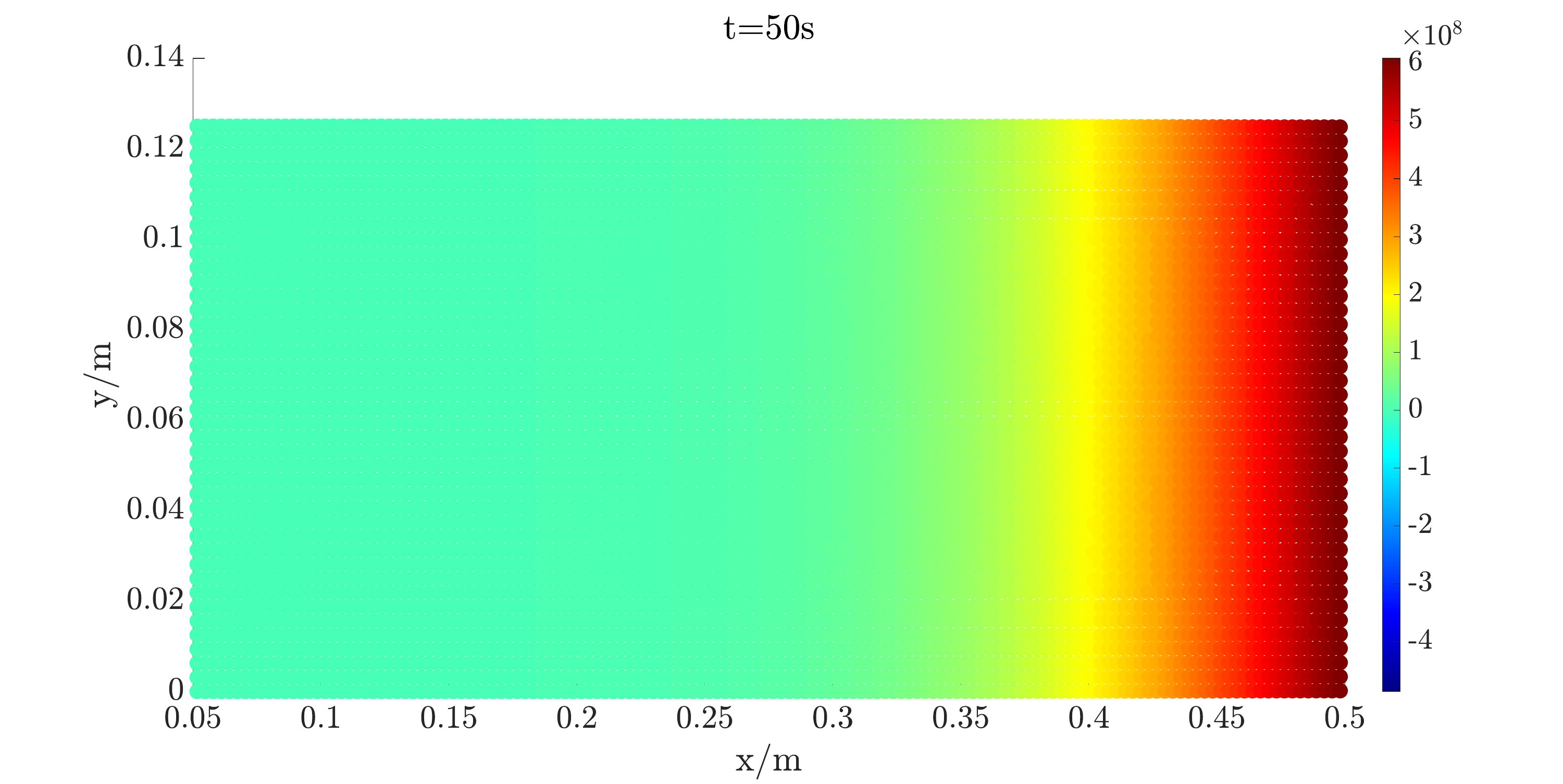}
    \end{subfigure}
    \caption{Thermal stress distribution (in Pa) for an uninsulated aluminium plate (Left) and an aluminium plate insulated with superelastic polyimide (Right) at 50 seconds. It can be observed that the left end of the insulated plate experiences very little thermal stress compared to the left end of the uninsulated plate.}
    \label{fig_tstress_distribution_comparison}
\end{figure}

The temperature evolution for the insulated and uninsulated aluminium plates are shown in \autoref{fig_Temp_distribution_comparison_for_5_10_25_50}. The insulating effect of superelastic polyimide is evident in these figures. Due to its very low thermal conductivity and high specific heat capacity, superelastic polyimide significantly slows down heat transfer compared to the uninsulated case, whereas the high thermal conductivity of aluminium allows rapid heat transfer.

In the horizontal direction, the plate consists of superelastic polyimide particles from 0 to 0.05 m, with the remainder composed of aluminium particles. It is noticeable that the right end of the plate, which is directly exposed to a sudden temperature drop, experiences high thermal stress and consequently a large and continuously growing displacement, portrayed clearly by the thermal strain distribution plot \autoref{fig_tstrain_distribution_comparison}. At 50 seconds, the thermal stress at the left end of the uninsulated aluminium plate reached approximately 480 MPa, calculated using the relation \( p_r = 3K\hat{\alpha}(T - T_0) \), illustrated by \autoref{fig_tstress_distribution_comparison}. In contrast, the thermal stress at the left end of the insulated plate remained significantly lower, staying well below \( 5000 \, \text{Pa} \). The thermal stress distribution at the right end of both the insulated and uninsulated plates was identical, as the right ends of both plates were uninsulated. 

\begin{figure}[hbtp!]
    \centering
    \begin{subfigure}[b]{0.49\textwidth}
        \centering
        \includegraphics[trim=3cm 0cm 4cm 0cm, clip,width=\textwidth]{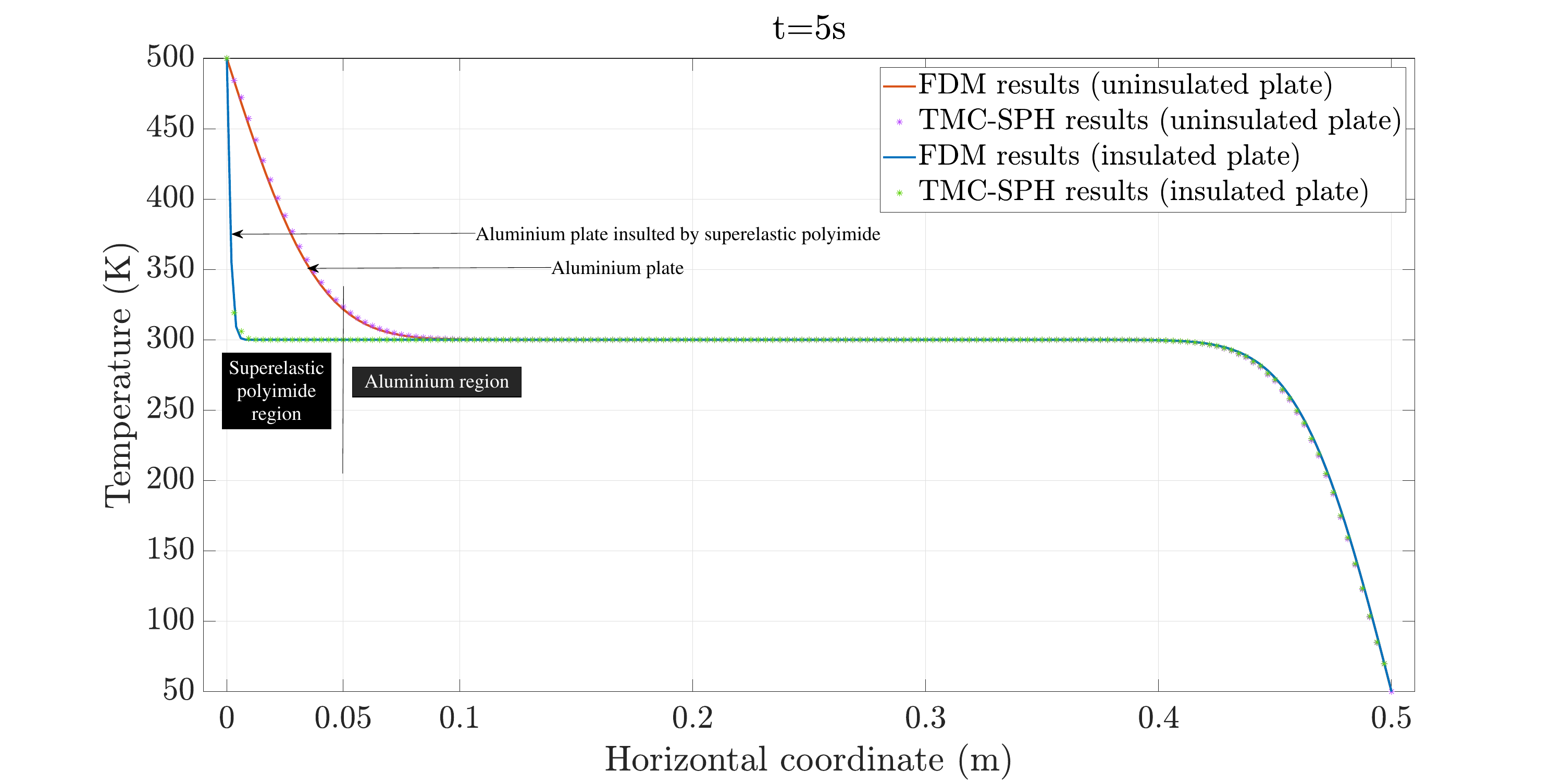}
        \label{fig_temp_5s}
    \end{subfigure}
   \hfill
    \begin{subfigure}[b]{0.49\textwidth}
        \centering
        \includegraphics[trim=3cm 0cm 4cm 0cm, clip,width=\textwidth]{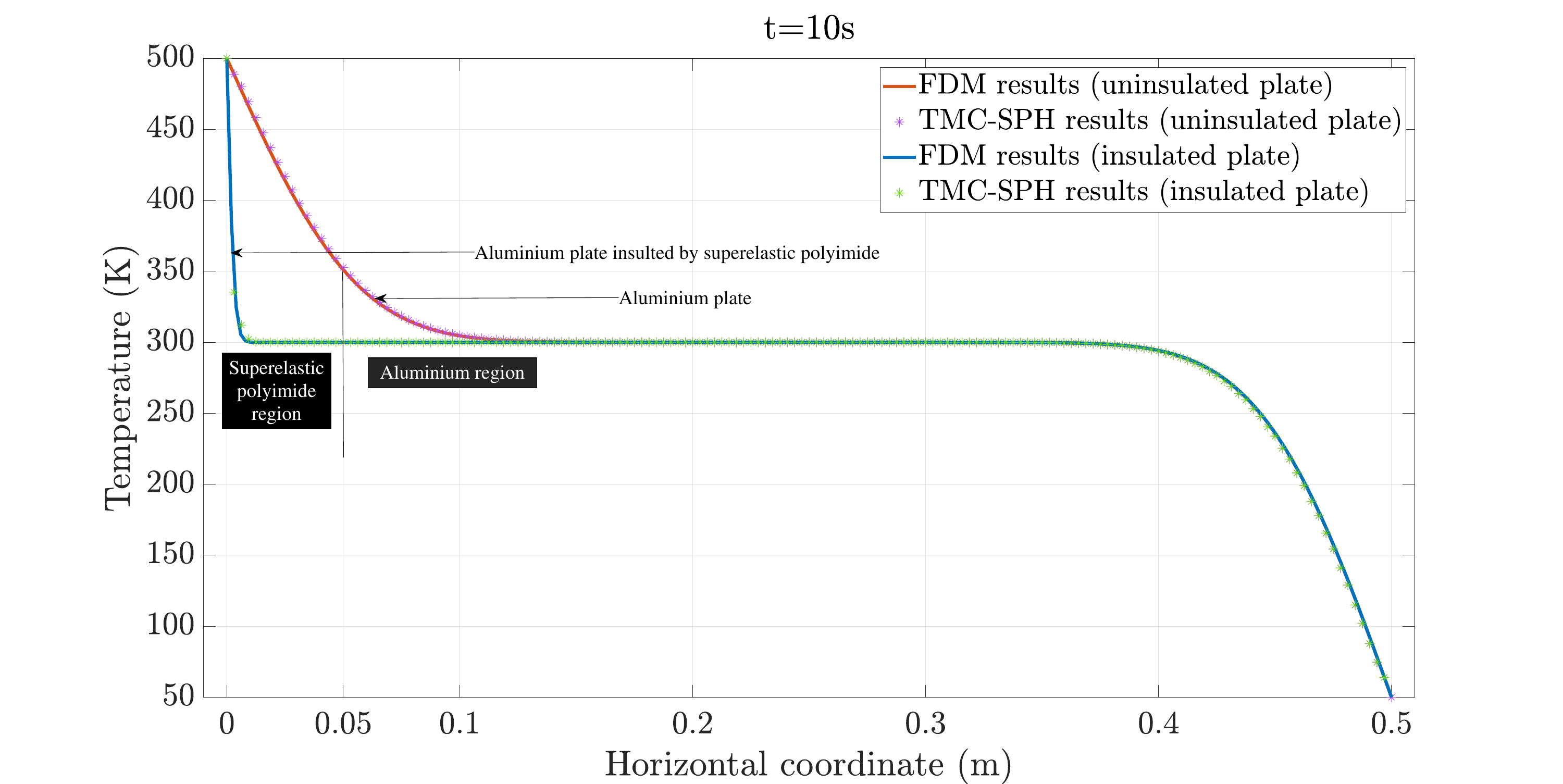}
        \label{fig_temp_10s}
    \end{subfigure}
    \hfill
    \begin{subfigure}[b]{0.49\textwidth}
        \centering
        \includegraphics[trim=3cm 0cm 4cm 0cm, clip,width=\textwidth]{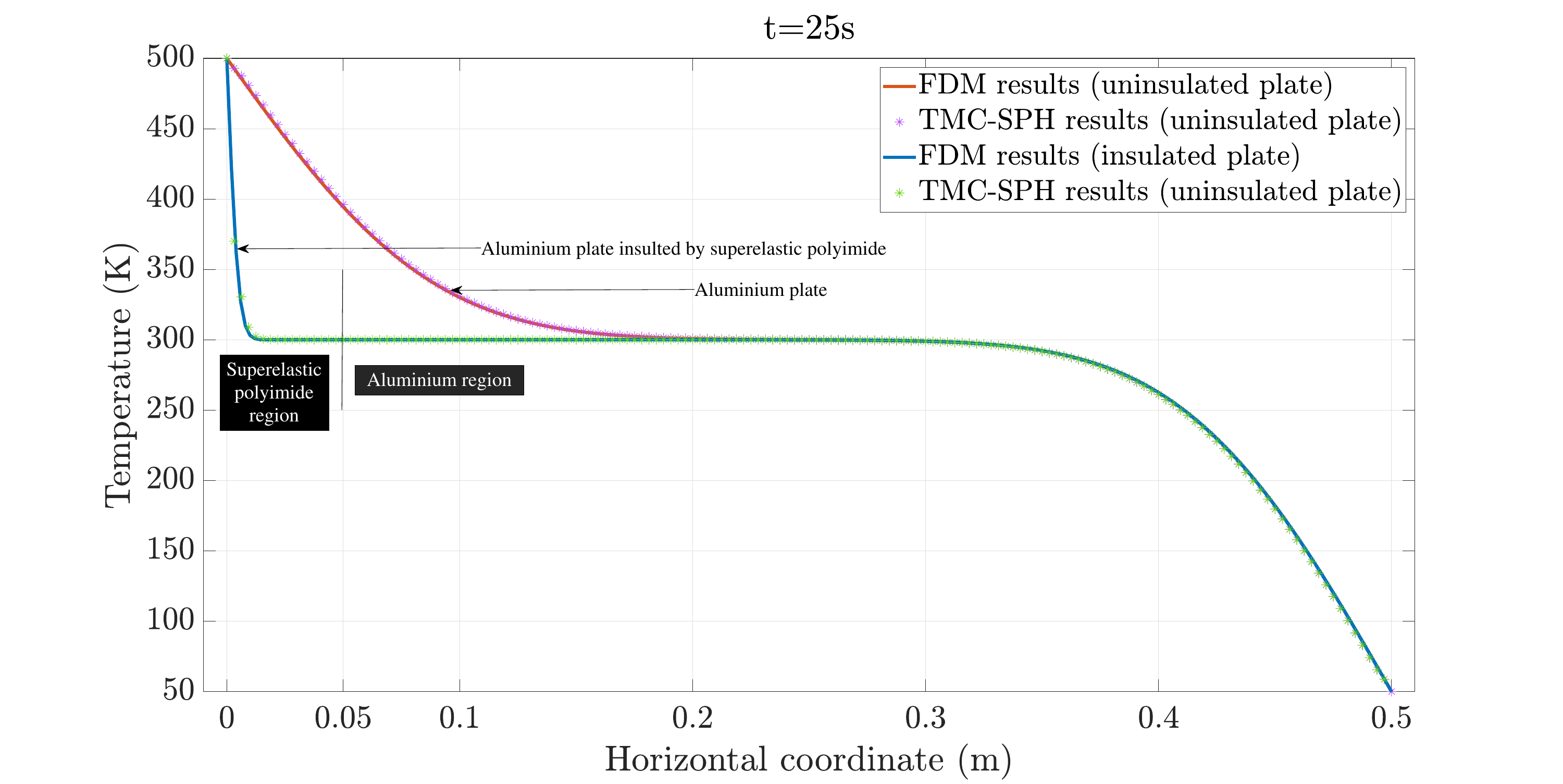}
        \label{fig_temp_25s}
    \end{subfigure}
    \hfill
    \begin{subfigure}[b]{0.49\textwidth}
        \centering
        \includegraphics[trim=3cm 0cm 4cm 0cm, clip,width=\textwidth]{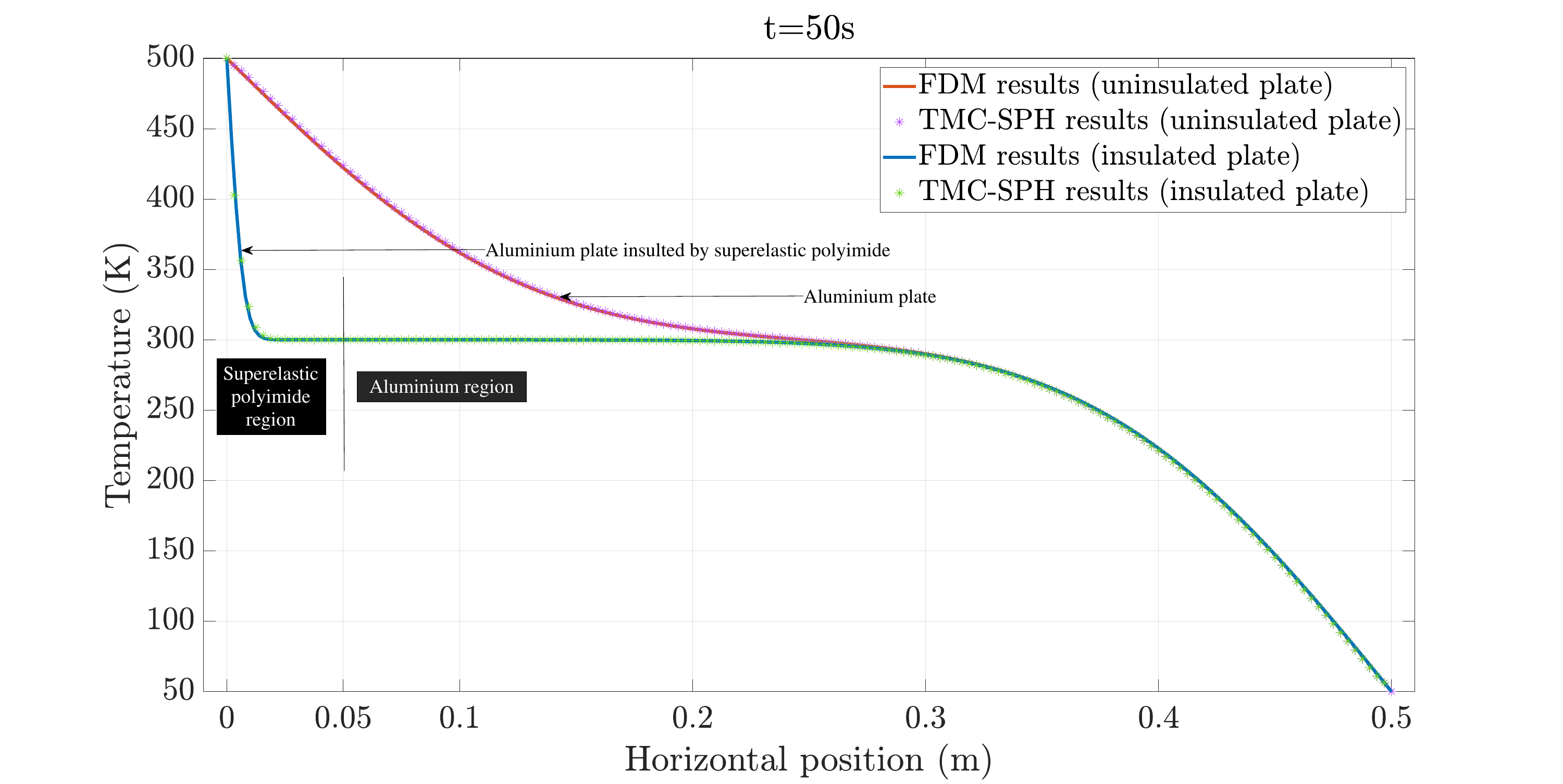}
        \label{fig_temp_50s}
    \end{subfigure}
    \caption{Temperature plot comparing the results of thermomechanical SPH (TMC-SPH) and the finite difference method (FDM) for both insulated and uninsulated aluminum plates at 5, 10, 25, 50 seconds. The TMC-SPH temperature results are taken along the horizontal line through the middle of the plate, while the FDM results are obtained by numerically discretizing and solving the one-dimensional conductive heat transfer equation only.}
    \label{fig_Temp_graph_comparison_for_5_10_25_50}
\end{figure}

A finite difference method (FDM) is used to simulate one-dimensional heat conduction by discretising and time-integrating the temperature field only. \autoref{fig_Temp_graph_comparison_for_5_10_25_50} compares the temperature distributions obtained from SPH and FDM. The one-dimensional conductive heat transfer governing is solved using 151 particles for the uninsulated aluminium plate and 251 particles for the insulated plate, with a timestep of $10^{-3}$. The results show negligible heat transfer from the superelastic polyimide to the aluminium, confirming the polyimide's insulating behavior. Furthermore, the close agreement between the thermomechanically coupled SPH results and the FDM results indicates minimal influence of mechanical coupling on the temperature field.

\section{Conclusions and way forward}\label{conclu}
The preceding sections detail the successful development of a sequential multiscale model, integrating molecular dynamics (MD) and thermo-mechanical coupled smoothed particle hydrodynamics (SPH). This integrated approach is found to be effective in studying thermal stress developed in Aluminium-superelastic polyimide film, which has relevance in the space industry. The mechanical parameters, such as elastic modulus, bulk modulus, Poisson's ratio, and the equation of state, along with thermal parameters like thermal conductivity and thermal expansion coefficient of the superelastic polyimide, are derived entirely from MD simulations. These serve as input to the coupled thermomechanical SPH model.

After validating our SPH model validated against one-dimensional and two-dimensional transient heat transfer cases, we examine the insulating performance of superelastic polyimide. This is achieved by simulating the thermal response of an aluminium plate with and without superelastic polyimide insulation under identical conditions. The results clearly demonstrated the superior insulating capabilities of superelastic polyimide, as evidenced by the thermal stress, strain, and temperature, distributions. The insulated plate experienced very less stress build-up because of the superelastic polyimide, owing to its lower thermal conductivity and bulk modulus, resisting the quick flow of the sudden heat flux generated.


A significant extension of this work would involve incorporating the fracture behavior of superelastic polyimide into the simulations, and modify its molecular topology to introduce defects and irregularities such as chain-end defects. The study also opens avenues for the experimental validation of the computational findings. Advanced and precise testing techniques, such as micro-indentation, micro-tensile testing, and the laser flash method, can be employed to further investigate and characterize the material properties of superelastic polyimide. These experimental approaches would provide critical data to refine the computational model and enhance its predictive accuracy.

\section{Acknowledgments}
MRII acknowledges the computational support provided as a part of the SRG DST-SERB (Grant No. SRG/2023/001266) and the IIT Delhi, India NFS grant on which the simulations have been run.

\FloatBarrier



\begin{thebibliography}{10}
\expandafter\ifx\csname url\endcsname\relax
  \def\url#1{\texttt{#1}}\fi
\expandafter\ifx\csname urlprefix\endcsname\relax\def\urlprefix{URL }\fi
\expandafter\ifx\csname href\endcsname\relax
  \def\href#1#2{#2} \def\path#1{#1}\fi

\bibitem{gong2024application}
Y.~Gong, Y.~Kou, Q.~Yue, X.~Zhuang, F.~Qin, Q.~Wang, T.~Rabczuk, The application of thermomechanically coupled phase-field models in electronic packaging interconnect structures, International Communications in Heat and Mass Transfer 159 (2024) 108033.

\bibitem{patel2019ferroelectric}
T.~Patel, K.~Co, R.~Hebert, S.~Alpay, Ferroelectric films on metal substrates: The role of thermal expansion mismatch on dielectric, piezoelectric, and pyroelectric properties, Journal of Applied Physics 126~(13) (2019).

\bibitem{pantousa2018thermal}
D.~Pantousa, K.~Tzaros, M.-A. Kefaki, Thermal buckling behaviour of unstiffened and stiffened fixed-roof tanks under non-uniform heating, Journal of Constructional Steel Research 143 (2018) 162--179.

\bibitem{wang2002radiation}
J.-A.~J. Wang, R.~C. Singleterry~Jr, R.~J. Ellis, H.~T. Hunter, Radiation effects on spacecraft structural materials, in: Proceedings of International Conference on Advanced Nuclear Power Plants, American Nuclear Society La Grange Park, IL, 2002.

\bibitem{dever2005degradation}
J.~Dever, B.~Banks, K.~de~Groh, S.~Miller, Degradation of spacecraft materials, in: Handbook of environmental degradation of materials, Elsevier, 2005, pp. 465--501.

\bibitem{abd2022radiation}
A.~M. Abd El-Hameed, Radiation effects on composite materials used in space systems: A review, NRIAG Journal of Astronomy and Geophysics 11~(1) (2022) 313--324.

\bibitem{boley1997theory}
B.~Boley, J.~Weiner, Theory of thermal stresses. mineola (1997).

\bibitem{nowacki2013thermoelasticity}
W.~Nowacki, Thermoelasticity, Elsevier, 2013.

\bibitem{hetnarski2009thermal}
R.~B. Hetnarski, M.~R. Eslami, G.~Gladwell, Thermal stresses: advanced theory and applications, Vol.~41, Springer, 2009.

\bibitem{lee2010postbuckling}
Y.~Lee, X.~Zhao, J.~Reddy, Postbuckling analysis of functionally graded plates subject to compressive and thermal loads, Computer Methods in Applied Mechanics and Engineering 199~(25-28) (2010) 1645--1653.

\bibitem{lu2004overview}
G.~Lu, E.~Kaxiras, An overview of multiscale simulations of materials, arXiv preprint cond-mat/0401073 (2004).

\bibitem{lee2013multiscale}
Y.~Lee, C.~Basaran, A multiscale modeling technique for bridging molecular dynamics with finite element method, Journal of Computational Physics 253 (2013) 64--85.

\bibitem{yamakov2008new}
V.~Yamakov, E.~Saether, E.~H. Glaessgen, A new concurrent multiscale methodology for coupling molecular dynamics and finite element analyses, Tech. rep. (2008).

\bibitem{gu2006concurrent}
Y.~Gu, L.~Zhang, A concurrent multiscale method based on the meshfree method and molecular dynamics analysis, Multiscale Modeling \& Simulation 5~(4) (2006) 1128--1155.

\bibitem{wang2009multiscale}
Q.~Wang, T.~Ng, H.~Li, K.~Lam, Multiscale simulation of coupled length-scales via meshless method and molecular dynamics, Mechanics of Advanced Materials and Structures 16~(1) (2009) 1--11.

\bibitem{li2010multiscale}
S.~Li, N.~Sheng, On multiscale non-equilibrium molecular dynamics simulations, International Journal for Numerical Methods in Engineering 83~(8-9) (2010) 998--1038.

\bibitem{bhattacharyya2022multiscale}
S.~Bhattacharyya, M.~R.~I. Islam, P.~K. Patra, Multiscale modelling of fracture in graphene sheets, Theoretical and Applied Fracture Mechanics 122 (2022) 103617.

\bibitem{monaghan1983shock}
J.~J. Monaghan, R.~A. Gingold, Shock simulation by the particle method sph, Journal of computational physics 52~(2) (1983) 374--389.

\bibitem{monaghan2000sph}
J.~J. Monaghan, Sph without a tensile instability, Journal of computational physics 159~(2) (2000) 290--311.

\bibitem{chen1999corrective}
J.~Chen, J.~Beraun, T.~Carney, A corrective smoothed particle method for boundary value problems in heat conduction, International Journal for Numerical Methods in Engineering 46~(2) (1999) 231--252.

\bibitem{hoover2006smooth}
W.~G. Hoover, Smooth particle applied mechanics: the state of the art (2006).

\bibitem{islam2022equivalence}
M.~R.~I. Islam, K.~V. Ganesh, P.~K. Patra, On the equivalence of eulerian smoothed particle hydrodynamics, total lagrangian smoothed particle hydrodynamics and molecular dynamics simulations for solids, Computer Methods in Applied Mechanics and Engineering 391 (2022) 114591.

\bibitem{ganesh2022multiscale}
K.~V. Ganesh, P.~K. Patra, K.~P. Travis, Multiscale modeling of impact through molecular dynamics and smooth particle hydrodynamics, Physica A: Statistical Mechanics and its Applications 593 (2022) 126903.

\bibitem{ganesh2022pseudo}
K.~V. Ganesh, M.~R.~I. Islam, P.~K. Patra, K.~P. Travis, A pseudo-spring based sph framework for studying fatigue crack propagation, International Journal of Fatigue 162 (2022) 106986.

\bibitem{cheng2021super}
Y.~Cheng, X.~Zhang, Y.~Qin, P.~Dong, W.~Yao, J.~Matz, P.~M. Ajayan, J.~Shen, M.~Ye, Super-elasticity at 4 k of covalently crosslinked polyimide aerogels with negative poisson’s ratio, Nature communications 12~(1) (2021) 4092.

\bibitem{rahnamoun2014reactive}
A.~Rahnamoun, A.~Van~Duin, Reactive molecular dynamics simulation on the disintegration of kapton, poss polyimide, amorphous silica, and teflon during atomic oxygen impact using the reaxff reactive force-field method, The Journal of Physical Chemistry A 118~(15) (2014) 2780--2787.

\bibitem{min2017computational}
K.~Min, A.~R. Rammohan, H.~S. Lee, J.~Shin, S.~H. Lee, S.~Goyal, H.~Park, J.~C. Mauro, R.~Stewart, V.~Botu, et~al., Computational approaches for investigating interfacial adhesion phenomena of polyimide on silica glass, Scientific reports 7~(1) (2017) 10475.

\bibitem{rahnamoun2019chemical}
A.~Rahnamoun, D.~P. Engelhart, S.~Humagain, H.~Koerner, E.~Plis, W.~J. Kennedy, R.~Cooper, S.~G. Greenbaum, R.~Hoffmann, A.~C. van Duin, Chemical dynamics characteristics of kapton polyimide damaged by electron beam irradiation, Polymer 176 (2019) 135--145.

\bibitem{marque2008molecular}
G.~Marque, S.~Neyertz, J.~Verdu, V.~Prunier, D.~Brown, Molecular dynamics simulation study of water in amorphous kapton, Macromolecules 41~(9) (2008) 3349--3362.

\bibitem{lu2015reaxff}
X.~Lu, X.~Wang, Q.~Li, X.~Huang, S.~Han, G.~Wang, A reaxff-based molecular dynamics study of the pyrolysis mechanism of polyimide, Polymer Degradation and Stability 114 (2015) 72--80.

\bibitem{LAMMPS}
A.~P. Thompson, H.~M. Aktulga, R.~Berger, D.~S. Bolintineanu, W.~M. Brown, P.~S. Crozier, P.~J. in~'t Veld, A.~Kohlmeyer, S.~G. Moore, T.~D. Nguyen, R.~Shan, M.~J. Stevens, J.~Tranchida, C.~Trott, S.~J. Plimpton, {LAMMPS} - a flexible simulation tool for particle-based materials modeling at the atomic, meso, and continuum scales, Comp. Phys. Comm. 271 (2022) 108171.
\newblock \href {https://doi.org/10.1016/j.cpc.2021.108171} {\path{doi:10.1016/j.cpc.2021.108171}}.

\bibitem{nose1984unified}
S.~Nos{\'e}, A unified formulation of the constant temperature molecular dynamics methods, The Journal of chemical physics 81~(1) (1984) 511--519.

\bibitem{hoover1985canonical}
W.~G. Hoover, Canonical dynamics: Equilibrium phase-space distributions, Physical review A 31~(3) (1985) 1695.

\bibitem{martyna1994constant}
G.~J. Martyna, D.~J. Tobias, M.~L. Klein, Constant pressure molecular dynamics algorithms, J. chem. Phys 101~(4177) (1994) 10--1063.

\bibitem{van2001reaxff}
A.~C. Van~Duin, S.~Dasgupta, F.~Lorant, W.~A. Goddard, Reaxff: a reactive force field for hydrocarbons, The Journal of Physical Chemistry A 105~(41) (2001) 9396--9409.

\bibitem{heinz2013thermodynamically}
H.~Heinz, T.-J. Lin, R.~Kishore~Mishra, F.~S. Emami, Thermodynamically consistent force fields for the assembly of inorganic, organic, and biological nanostructures: the interface force field, Langmuir 29~(6) (2013) 1754--1765.

\bibitem{senftle2016reaxff}
T.~P. Senftle, S.~Hong, M.~M. Islam, S.~B. Kylasa, Y.~Zheng, Y.~K. Shin, C.~Junkermeier, R.~Engel-Herbert, M.~J. Janik, H.~M. Aktulga, et~al., The reaxff reactive force-field: development, applications and future directions, npj Computational Materials 2~(1) (2016) 1--14.

\bibitem{zhao2022dependence}
W.~Zhao, Q.~Wei, C.~Huang, Y.~Zhu, N.~Hu, Dependence of incidence angle and flux density in the damage effect of atomic oxygen on kapton film, Polymers 14~(24) (2022) 5444.

\bibitem{muller1997simple}
F.~M{\"u}ller-Plathe, A simple nonequilibrium molecular dynamics method for calculating the thermal conductivity, The Journal of chemical physics 106~(14) (1997) 6082--6085.

\bibitem{pishkenari2016molecular}
H.~N. Pishkenari, E.~Mohagheghian, A.~Rasouli, Molecular dynamics study of the thermal expansion coefficient of silicon, Physics Letters A 380~(48) (2016) 4039--4043.

\bibitem{hu2017thermomechanically}
H.~Hu, P.~Eberhard, Thermomechanically coupled conduction mode laser welding simulations using smoothed particle hydrodynamics, Computational Particle Mechanics 4 (2017) 473--486.

\bibitem{gingold1977smoothed}
R.~A. Gingold, J.~J. Monaghan, Smoothed particle hydrodynamics: theory and application to non-spherical stars, Monthly notices of the royal astronomical society 181~(3) (1977) 375--389.

\bibitem{monaghan1994simulating}
J.~J. Monaghan, Simulating free surface flows with sph, Journal of computational physics 110~(2) (1994) 399--406.

\bibitem{libersky1993high}
L.~D. Libersky, A.~G. Petschek, T.~C. Carney, J.~R. Hipp, F.~A. Allahdadi, High strain lagrangian hydrodynamics: a three-dimensional sph code for dynamic material response, Journal of computational physics 109~(1) (1993) 67--75.

\bibitem{liu2010smoothed}
M.~Liu, G.~Liu, Smoothed particle hydrodynamics (sph): an overview and recent developments, Archives of computational methods in engineering 17 (2010) 25--76.

\bibitem{liu2003smoothed}
G.-R. Liu, M.~B. Liu, Smoothed particle hydrodynamics: a meshfree particle method, World scientific, 2003.

\bibitem{eringen1967mechanics}
A.~Eringen, Mechanics of continua. jhon wiley and sons, New York (1967).

\bibitem{cleary1999conduction}
P.~W. Cleary, J.~J. Monaghan, Conduction modelling using smoothed particle hydrodynamics, Journal of Computational Physics 148~(1) (1999) 227--264.

\bibitem{monaghan2005solidification}
J.~J. Monaghan, H.~E. Huppert, M.~G. Worster, Solidification using smoothed particle hydrodynamics, Journal of computational Physics 206~(2) (2005) 684--705.

\bibitem{monaghan2005smoothed}
J.~J. Monaghan, Smoothed particle hydrodynamics, Reports on progress in physics 68~(8) (2005) 1703.

\bibitem{shaw2009heuristic}
A.~Shaw, S.~R. Reid, Heuristic acceleration correction algorithm for use in sph computations in impact mechanics, Computer methods in applied mechanics and engineering 198~(49-52) (2009) 3962--3974.

\bibitem{zhou2024thermal}
Z.~Zhou, J.~Bi, Y.~Zhao, C.~Wang, Y.~Zhang, Thermal-mechanical coupling smooth particle hydrodynamics-phase field method modelling of cracking in rocks, Computers and Geotechnics 173 (2024) 106476.

\bibitem{mu2022coupled}
D.~Mu, Z.~Li, A.~Tang, Q.~Liu, D.~Huang, A coupled thermo-mechanical bond-based smoothed particle dynamics model for simulating thermal cracking in rocks, Engineering Fracture Mechanics 265 (2022) 108364.

\bibitem{yan2017coupled}
C.~Yan, H.~Zheng, A coupled thermo-mechanical model based on the combined finite-discrete element method for simulating thermal cracking of rock, International Journal of Rock Mechanics and Mining Sciences 91 (2017) 170--178.

\bibitem{d2017thermal}
P.~D’Antuono, M.~Morandini, Thermal shock response via weakly coupled peridynamic thermo-mechanics, International Journal of Solids and Structures 129 (2017) 74--89.

\bibitem{sun2021pd}
W.~Sun, W.~Lu, F.~Bao, P.~Ni, A pd-fem coupling approach for modeling thermal fractures in brittle solids, Theoretical and Applied Fracture Mechanics 116 (2021) 103129.

\bibitem{heinze2016systematic}
T.~Heinze, G.~Jansen, B.~Galvan, S.~A. Miller, Systematic study of the effects of mass and time scaling techniques applied in numerical rock mechanics simulations, Tectonophysics 684 (2016) 4--11.

\end{thebibliography}

\end{document}